\documentclass[12pt]{iopart}
\usepackage{iopams}
\usepackage{graphics}
\usepackage[next]{inputenc}
\usepackage[dvips]{epsfig}
\begin{document}
\title{Charge and Spin Transport in Superconducting Weak Links}
\author{Gholamreza Rashedi}
\address{ Institute for Advanced Studies in Basic Sciences, 45195-159,
Zanjan, Iran}
\date{\today}
\begin{abstract}
The coherent mixing of the current states in the superconducting
weak link subject to a Josephson phase difference $\phi$ and
subject to an external transport current in the banks is one of
the aims of this work. At $\phi=\pi $ the nonlocal mixing of
current states produces two vortices close to the point-contact
between superconducting bulks. The effect of point-contact
reflection in an impenetrable interface and effect of temperature
on the vortices have been studied. It is obtained that increasing
the reflection of the point-contact destroys the vortices while
increasing the temperature restore these vortices. The vortex
state is a new version of the interference between the macroscopic
states and quantum tunnelling. Also, the weak link between unitary
triplet superconductors which have $f-$wave and $p+h-$wave pairing
symmetry has been studied from the spin and charge current-phase
relation point of view. The main result in the second part of this
thesis, is the polarization of the spin transport when a junction
between triplet superconductors is used. It is observed that the
spin current is the result of the misorientation between the gap
vectors of two supercondcutors. In addition, the weak link between
two bipolar non-unitary triplet superconductors is studied
mathematically. The current-phase relations obtained in third part
of this thesis are totally different from the junctions between
the unitary spin-triplet superconductors and between the
spin-singlet superconductors. The current phase diagrams which
have been obtained in this work can be used to distinguish the
symmetry of the order parameter in the crystals.
\end{abstract}
\maketitle
\section{Introduction}
Superconductor is a substance that below a certain temperature
(critical temperature $T_c$) conducts charge current without
resistance and also repels the magnetic field. Superconductivity
was discovered by Kammerlingh Onnes in a mercury wire at
temperatures less than $T=4.2 K$ in 1911 \cite{Onnes}. In 1957,
Bardeen, Cooper and Schrieffer (BCS), using the concept of Cooper
pairs, proposed a microscopic theory for superconductivity
\cite{BCS}. Cooper had shown earlier that the ground state of a
metal, at sufficient low temperatures, will be stable when its
electrons are collected as the pairs \cite{Cooper}. BCS theory
suggests that Cooper pairs at low temperatures condense into the
same quantum state defined by a macroscopic wave function and can
travel together without dissipation. Also, the pair amplitude of
this wave function as an order parameter in the Ginzburg-Landau
(GL) theory of superconductivity, defines the transition between
normal and superconducting states \cite{Ginzburg}. In the limit of
$T\rightarrow T_c$, the GL order parameter , is directly
proportional to the BCS energy gap, $\Delta$, which is the amount
of energy for breaking a Cooper pair. This energy gap carries the
information about the pairing symmetry of the electrons.

There are different types of superconductivity. According to the
spin states of two electrons in the Cooper pair, superconductor is
either spin-singlet or spin-triplet. Since electrons are fermion,
the total wave function of a cooper pair should be antisymmetric
with respect to electrons:
$\Psi_{cooper-pair}=\psi(\mathbf{r_1,r_2})\varphi(\mathbf{s_1,s_2})$
where $\mathbf{r}$, $\mathbf{s}$ are spatial coordinates and
spins, respectively. For the antisymmetric spin-singlet state, the
orbital part of the wave function should be symmetric under the
interchange of the electrons and the orbital momentum should be
even ($l=0$, $s-$wave, $l=2$, $d-$wave and $l=4$,
$g-$wave).\newline The superconductors with simplest spin-singlet
state, $l=0$, are known as conventional superconductors. All the
other type of superconductors including the spin-singlet and
spin-triplet with $l\neq0$, are unconventional superconductors. A
$s-$wave superconductor, has an isotropic order parameter
$\Delta(\mathbf{\hat{k}})=\Delta_{0}$ in the momentum space, where
$\mathbf{\hat{k}}$ is a unit vector pointed on the Fermi surface.
Earlier superconductors which had been found in the elements like
mercury, are conventional superconductors. Unconventional
superconducting compounds which have anisotropic order parameter
can be defined using the relation
$\sum_{\mathbf{k}}\Delta(\mathbf{k})=0,$ with summation over the
Fermi surface. For the symmetric spin-triplet state, the orbital
part of the wave function is antisymmetric and the orbital
momentum takes the odd numbers ($l=1$, $p-$wave, $l=3$, $f-$wave
and $l=5$, $h-$wave). Here, the terms $s-$wave, $p-$wave and
\emph{etc} have been used from the terminology of the Hydrogen
atom. The pairing state of a spin-triplet superconductor in the
spin space is represented by a three dimensional vector
$\mathbf{d(\hat{k})}$, called gap vector. The gap vector
determines the order parameter matrix, ${\hat\Delta}({\mathbf
k})=i({\mathbf d}({\mathbf k}) \cdot{\hat{\sigma }})\hat{\sigma
}_y$, in which $\hat{\sigma }_j$s are Pauli matrices.\newline
Three complex components of gap vector ($d_1$, $d_2$, $d_3$) over
the Fermi surface are corresponding to three possible spin
directions as follow:
$|spin\rangle=(d_1+id_2)|\hspace{-.1cm}\downdownarrows\rangle+
(-d_1+id_2)|\hspace{-.1cm}\upuparrows\rangle+
d_3(|\hspace{-.1cm}\uparrow\hspace{-.08cm}\downarrow\rangle+
|\hspace{-.1cm}\downarrow\hspace{-.08cm}\uparrow\rangle)$. There
are many experimental and theoretical works in the case of
spin-triplet superconductors from which some important properties
of spin-triplet superconductors are listed below:\newline 1) The
spin-triplet superconductors are generally low $T_c$
superconductors, as compared with the high $T_c$ superconductivity
in $d-$wave superconductors
\cite{Maeno,Mahmoodi,Mackenzie}.\newline 2) The relation between
critical temperature and energy gap at zero temperature is
different from that of the BCS relation for a $s-$wave
superconductor $\Delta(T=0)>1.76 T_c$ \cite{Maki3,SM}.\newline 3)
In structures with spin-triplet superconductivity a spin
supercurrent can flow generally, while in the case of the
spin-singlet superconductivity, the spin current may exist only in
the proximity system of the superconductor and a ferromagnet. This
means that in the spin-triplet superconductors not only the charge
but also the spin of electrons can become superfluid
\cite{Barashbob1,Barashbob2,Brataas}.\newline 4) Ferromagnetic
superconductivity is another interesting phenomenon of
spin-triplet state. It had been observed that all conventional
superconductors were non-magnetic materials and it was concluded
that superconductivity and ferromagnetism are incompatible phases.
While, ferromagnetic superconductivity has been observed recently,
in some of the triplet superconductors like: $ZrZn2$, $UGe2$ and
$URhGe2$ \cite{Mineev,Samokhin}.\newline 5) Nonunitary gap vector
is another fingerprint of the spin-triplet superconductivity. In
the nonunitary spin-triplet state, Cooper pairs may carry a finite
averaged intrinsic spin momentum. This nonunitary state is a
candidate for the $B-$phase of superconductivity in the $UPt_{3}$
compound (Fig.\ref{figrev3}). This phase has been observed at the
low temperatures and low magnetic fields
\cite{Machida,Ohmi}.\newline 6) In addition to the temperature and
magnetic field, pressure influences the phase transition of
triplet superconductors, particularly for the ferromagnetic
superconductors \cite{Pfleiderer,Sheikin}.\newline 7) Another
property of some of triplet superconductors in the similarity of
their structures with some of the important high $T_c$
superconductors. For example, the $Sr_2RuO_4$ spin-triplet
superconductor is isostructural to the spin-singlet
high-temperature superconductor $LaBaCuO.$ Here, $Sr$ and $Ru$
atoms are counterpart of $La(Ba)$ and $Cu$ atoms, respectively.
Consequently, investigation of this triplet superconductor helps
to understand the singlet case \cite{Maeno}.\newline We have to
introduce some of unconventional superconducting compounds,
particularly those which will be investigated in the following
chapters of this work. The uranium compound $UPt_{3}$ will be
studied in chapter (\ref{SpinCharge}) and (\ref{Nonunitary}).
Then, $Sr_{2}RuO_{4}$ and $PrOs_4Sb_{12}$ compounds will be
investigated in chapters (\ref{SpinCharge}) and (\ref{PrOsSb}),
respectively.

The first discovered triplet order parameter, $p-$wave pairing
symmetry, has been observed in the Helium superfluid. Also,
$p-$wave order parameter has been considered as a candidate for
the superconducting state in $Sr_{2}RuO_{4}$ by some authors
\cite{Rice,Agterberg}. A famous form of $p-$wave pairing symmetry
in momentum space is
$\mathbf{d}(T,\mathbf{k})=\Delta(T)(k_x+ik_y)\hat{\mathbf{z}}$ in
which, $\hat{\mathbf{z}}$ is a unit vector \cite{Mahmoodi}.
Another important category of triplet superconductors is $f-$wave
superconductivity, proposed for the pairing symmetry in
heavy-fermion compound $UPt_{3}$ in \cite{Muller,Qian} and the
compound $Sr_{2}RuO_{4}$ in \cite{Maeno,Maeno2}. The different
phases of $f-$wave superconductivity have different order
parameter symmetries in momentum space. For instance, the $f-$wave
axial state symmetry is $\mathbf{d}(T,\mathbf{\hat{k}})=\Delta
_{0}(T)\hat{\mathbf{z}}k_{z}\left( k_{x}+ik_{y}\right) ^{2}$
\cite{Graf1} and the planar state has the form
$\mathbf{d}(T,\mathbf{\hat{k}})=\Delta
_{0}(T)k_{z}(\hat{\mathbf{x}}\left( k_{x}^{2}-k_{y}^{2}\right)
+\hat{\mathbf{y}}2k_{x}k_{y})$, where, $\hat{\mathbf{x}}$,
$\hat{\mathbf{y}}$ and $\hat{\mathbf{z}}$ are the unit vectors
\cite{Machida}. In this work, a nonunitary $f-$wave gap vector in
the $B-$phase of superconductivity (low temperature and low
magnetic field) in $UPt_{3}$ compound has been considered. For
this nonunitary bipolar state, a gap vector of the form
$\mathbf{d}(T,\mathbf{v}_{F})=\Delta
_{0}(T)k_{z}(\hat{\mathbf{x}}\left( k_{x}^{2}-k_{y}^{2}\right)
+\hat{\mathbf{y}}2ik_{x}k_{y})$ has been proposed in the momentum
space in \cite{Machida}. A recently proposed spin-triplet state is
the``$(p+h)-$wave" pairing symmetry which, has been considered for
the superconductivity in $PrOs_4Sb_{12}$ compound in \cite{Maki3}.
Of course, two different phases of ``$(p+h)-$wave" have been
observed for superconductivity in the $PrOs_4Sb_{12}$
\cite{Maki1,Nakajima}. Here, $A-$phase is high magnetic field,
high temperature phase but $B-$phase is low field, low temperature
phase \cite{Goryo}. The first model to explain the properties of
the $A$-phase of $PrOs_4Sb_{12}$ is:
$\mathbf{d}(T,\mathbf{\hat{k}})=\Delta
_{0}(T)(k_x+ik_y)\frac{3}{2} (1-\hat{k}_{x}^4 - \hat{k}_{y}^4 -
\hat{k}_{z}^4)\hat{\mathbf{z}}$, where, $\hat{\mathbf{z}}$ is a
unit vector. The function $ \Delta _{0}=$ $\Delta _{0}\left(
T\right) $ describes the dependence of the gap vector $\mathbf{d}$
on the temperature $T$\cite{Maki3}. The second model to describe
the gap vector of the $B$-phase of $PrOs_4Sb_{12}$ is:
$\mathbf{d}(T,\mathbf{\hat{k}})=\Delta _{0}(T)(k_x+ik_y)
(1-\hat{k}_{z}^4)\hat{\mathbf{z}}$\cite{Maki3}. One of the most
important types of unconventional superconductivity has been
observed in the high $T_c$ superconductor in the ceramic of copper
oxide by Bednorz and M\"{u}ller \cite{Bednorz}. This spin-singlet
superconductor has a $d-$wave pairing symmetry, $l=2$, as
$\Delta(\mathbf{\hat{k}})=\Delta(T)({k_x}^2-{k_y}^2)$ or
$\Delta(\mathbf{\hat{k}})=\Delta(T)2k_x k_y,$ in momentum space.
This discovery was a real revolution in the field of
superconductivity.

Conventional and unconventional superconductors are usually
differentiated by several properties such as anisotropicity of the
order parameter in the momentum space, heat capacity and heat
conductance, density of states, junction behavior and particularly
nodes in the momentum space. Node is a direction in the momentum
space where no order parameter and superconductivity is effective
on the scattering electrons.

One of the most interesting concepts in the field of
superconductivity is, superconducting weak link. The weak link
experiments are classified in the categories of $S-I-S$, $S-N-S$
and $S-c-S$, in which two superconducting bulks have been
separated by a nonsuperconducting interface or point-contact.
Here, $S$, $I$, $N$ and $c$ denote superconductor, insulator,
normal metal and point-contact respectively. The weakness of the
link means that the superconducting order parameters are the same
as the disconnected massive superconducting bulks. The first
configuration had been investigated by Josephson, in which a thin
insulator is located between the two superconducting bulks. In the
second case, a normal layer has been sandwiched by two
superconducting bulks. The third experiment is devoted to the
geometry consisting two superconducting bulks which are separated
by an impenetrable interface (strong insulator) in which a contact
has been prepared for the mixing of two superconducting states.
Josephson effect, $S-I-S$, has been investigated in 1962 by
Josephson \cite{Josephson}. He predicted an electrical current
flowing between two superconductors while they are separated by an
insulator. This insulator layer is very thin and electrons can
pass through it, even with the energies less than height of
potential barrier of insulator. The flow of current between the
superconductors in the absence of an applied voltage is called a
Josephson current (there is a phase difference instead of the
applied voltage) and the motion of electrons across the barrier is
called Josephson tunneling. This tunneling phenomenon, Josephson
effect, can be used in the electronic devices such as
superconducting quantum interference device (\textbf{SQUID}), for
detecting the very small magnetic flux (even a flux quantum). The
Josephson effect can be influenced by the external magnetic fields
through the external phase difference. Consequently, the Josephson
junction can be applied to measure the extremely weak magnetic
fields, in \textbf{SQUID}. The current which flows, (quantum
mechanically tunnels through the potential barrier of the
interface, has a form $j(\phi)=j_c \sin{\phi}$ in which, $\phi$ is
the macroscopic phase difference between two superconductors and
the critical current $j_c$ depends on the geometry of system
(junction). The Josephson effect is a weak link of two massive
banks of superconductor, $S_{1}$ and $S_{2}$ with different phases
$\phi_1$ and $\phi_2$ which are separated by an insulator. The
system allows electron to exchange between the two sides of the
interface and then establishes the phase coherence in the system
as a hole. The Josephson junction can be considered as the mixer
between the superconducting quantum macroscopic states. The result
of mixing is the supercurrent which flows from one of the banks to
the other and it depends on the phase difference
$\phi=\phi_{2}-\phi_{1}$ across the weak link. The function
$j(\phi)$ depends on the geometry of the system. In the simplest
case which had been studied by Josephson, the current has a
sinusoidal dependence on the phase but there are many different
kinds of the weak links in which the current phase diagrams are
not sinusoidal. One of the most famous problems in this category
is problem of Kulik-Omelyanchouk ($S-c-S$ junction) which has been
investigated in \cite{Kulik}. They have used the Eilenberger
equation \cite{Eilenberger} and using the Green function they have
obtained the current-phase diagrams analytically. The current
phase relation depends not only on the manner of coupling but also
on the properties of the supercondcuting bulks. Two coupled
unconventional supercondcuting massive bulks have totaly different
current-phase relation from the conventional superconductivity
which was studied by Josephson and later by Kulik and
Omelyanchouk. For instance $D-c-D$ weak link which is a special
type of $S-c-S$ system, in which $D$ denotes the $d-$wave and high
$T_c$ superconducting bulks and $c$ is the contact, has some new
results. The spontaneous current parallel to the junction
interface, mid-gap states resulting from the sign change of the
order parameter, and the changing period of current $j(\phi)$ to
$j(2\phi)$ are different and new results of weak link between
unconventional superconducting bulks. In addition, it is observed
that the Josephson junction depends not only on the external
Josephson phase difference resulting form the external magnetic
flux but also on the misorientation angle between two
superconducting crystals. In this work, because of similarity
between our problem and problem of Kulik-Omelyanchouk, we use the
generalized form of their formalism.

This thesis consists of three parts. At first, we are to
investigate coherent current states in the superconducting weak
link which has been subjected to the Josephson phase difference
$\phi$ and subjected to the external transport supercurrent state
(parallel to the junction interface) in the banks. Earlier it had
been observed that at $\phi $ close to $\pi $ the mixing of
current states produces two vortices in the vicinity of the
point-contact between superconducting bulks \cite{KOSh}. In this
part we study the effect of transparency coefficient (potential
barrier) of the point-contact in an impenetrable interface and
effect of temperature on the vortices obtained in the paper
\cite{KOSh}.\newline The second part of this dissertation is
devoted to the weak link between unitary triplet superconductors.
These superconductors have $f-$wave and $p+h-$wave pairing
symmetry. The former has been proposed for $UPt_{3}$ and
$Sr_{2}RuO_{4}$ compounds and the later case is considered in
$PrOs_4 Sb_{12}$ complex. The spin-current in the junction between
these unitary triplet superconductors is an important part of this
work. The interesting case which is observed in the second part of
the thesis, is the polarization of the spin transport using the
junction between unitary triplet superconducting bulks. The third
and last part of the thesis discusses the case of weak link
between non-unitary triplet superconductors. The idea behind the
first part of the thesis is suitability of this structure
(vortex-like currents) for the investigation of the quantum
macroscopic phenomena. For the real system the finite transparency
coefficient (finite reflection) which we have investigated is more
suitable than ideal transparent point-contact which had been
considered in \cite{KOSh}. The second part of the thesis discusses
the case of the junction between unitary triplet $f-$wave and
$p+h-$wave superconductors. First of all we know that the triplet
superconductors also the high $T_c$ superconductors are created by
a different mechanism of pairing than the phonon-electron
interaction. Secondly, the molecular structure of the Ruthenate
compound $Sr_{2}RuO_{4}$ is the same as that of the high $T_c$
Cupperate superconductors which are important superconductors and
we are interested to understand their properties. Also, the spin
polarized transport has a valuable motivation for physicist in the
field of spintornics because, the sensitivity and accuracy of the
polarized spin-current systems can be used in the measurement
technology. The second part of this work can be used to develop
the theory of spin-polarized transport systems. The spin transport
in the absence of the charge transport is an interesting case to
investigate. The third part of the work has been devoted to
non-unitary weak links which has intrinsic spin and angular
momentum of systems.

The method for investigation of these weak link experiments is the
quasiclassical method of Green function. This method had been used
by Kulik and Omelyanchouk in paper \cite{Kulik}. They have studied
junction through the point-contact between two static conventional
superconducting bulks, but, here we have applied this formalism
for the case of conventional superconducting bulks with the
external transport current in the banks. Also, we have generalized
the Kulik-Omelyanchouk formalism for the case of triplet
supercondcuting bulks. We have calculated the analytical Green's
function and then we have used that to obtain the current density.
The current-phase diagrams are plotted and in some cases the two
dimensional profile of the current has been plotted in the space.

Arrangement of the rest of this thesis is as follows. In Chapter
(\ref{Introduction}), we review of the concepts which we use in
the rest of the thesis. The quasiclassical approach which is
widely used in the field of solid state physics and specially for
the case of superconducting systems, will be reviewed. The
Kulik-Omelyanchouk problem, that is the ballistic point contact
between two superconducting massive bulks will be studied and
their results will be reproduced. The effect of transparency
coefficient for a point contact with finite reflection in the
impenetrable interface between two bulks ( related to our problem
in the chapter (\ref{Vortex})) will be investigated and some
analytical relations for this system will be obtained. The
Josephson junction between two unitary superconductors will also
be studied in chapter (\ref{Introduction}). We will generalize
this approach in the Chapters
 (\ref{SpinCharge}), (\ref{PrOsSb}) and (\ref{Nonunitary}).\\
A Josephson effect in the ballistic point contact with transport
current on the banks, taking into account the reflection of
electrons from the contact, will be investigated in Chapter
(\ref{Vortex}). The contact is subject to the phase difference
$\phi $ and the transport current tangential to the boundary of
the contact. As it was shown in \cite{KOSh}, in the contact with
direct conductivity at $\phi =\pi$ and near the orifice the
tangential current flows in the opposite direction to the
transport current, and there are two vortices. It is found that by
decreasing the transparency, the vortex-like current will be
destroyed. On the other hand, as the temperature is increased the
vortices are restored. They continue to exist for transparencies
as low as $D=\frac{1}{2}$ in the limit of $T\rightarrow T_{c}$.
This anomalous temperature behavior of the
vortices is an interesting result which have been obtained.\\
In Chapter (\ref{SpinCharge}), we have studied the spin and charge
current in the ballistic Josephson junction in the model of an
ideal transparent interface between two misorientated $f$-wave
superconductors subjected to a phase difference $\phi $. Our
analysis has shown that the misorientation and different models of
the gap vectors influence the spin current. The misorientation
changes strongly the critical values of both the spin current and
charge current. It has been shown that the spin current is the
result of the misorientation between the gap vectors. Furthermore,
it is observed that the different models of the gap vectors and
geometries can be applied to the polarization of the spin
transport. In addition, it is observed that in certain values of
the phase difference $\phi $, the charge-current vanishes while
the spin-current flows, despite the fact that although the
carriers of spin and
charge are the same (electrons).\\
 A stationary Josephson junction as a weak link
between $PrOs_{4}Sb_{12}$ triplet superconductors will be
investigated in Chapter (\ref{PrOsSb}). Recently, the
``$(p+h)-$wave'' form of pairing symmetry has been proposed for
the superconductivity in $PrOs_{4}Sb_{12}$ compound \cite{Maki1}.
The quasiclassical Eilenberger equations are analytically solved
for this system. The spin and charge current-phase diagrams are
plotted and the effect of misorientation between crystals on the
spin current, and spontaneous and Josephson currents is studied.
It is found that such experimental investigations of the
current-phase diagrams can be used to test the pairing symmetry in
the above-mentioned superconductors. Also, it is shown that this
apparatus can be
applied as a polarizer for the spin current.\\
In Chapter (\ref{Nonunitary}), a stationary Josephson effect in a
weak link between misorientated nonunitary triplet superconductors
is studied. The non-self-consistent quasiclassical Eilenberger
equation for this system has been solved analytically and the
current-phase diagrams are plotted for the junction between two
nonunitary bipolar $f-$wave superconducting banks. A spontaneous
current parallel to the interface between superconductors has been
observed. Also, the effect of misorientation between crystals on
the Josephson and spontaneous currents is studied. Such
experimental investigations of the current-phase diagrams can be
used to test the pairing symmetry in the above-mentioned
superconductors. In Chapter (\ref{Conclusions}), the thesis will
be finished with some conclusions.
\newpage
\section{Superconducting weak links}
\label{Introduction}
 When two superconducting bulks are connected to each other by an
 insulator from which the electrons can tunnel, we have a weak link.
 The weakness of the link means that the superconducting
order parameters have their value in the bulks and they are the
same as the disconnected massive superconducting reservoirs. The
Josephson effect arises in a weak link of two separated (by
insulator) superconducting bulks with different phases. The
electrons can be exchanged between two superconducting bulks and
the system (two bulks and contact) tends to be the phase coherent.
Mixing the superconducting states through the contact or link
causes the supercurrent from one of the banks to the other bank.
The current is present because of the phase difference between the
bulks and this phenomenon can be observed in the absence of any
voltage. The phase difference between the bulks which plays a
central role in the weak link phenomena can be the result of the
external magnetic field which is surrounded by the junction and
bulks. The Josephson phase is a kind of the Aharonov-Bohm phase
\cite{Aharonov}. This phase is related to the magnetic flux which
flows from the system as $\phi=\frac{2\pi}{\Phi_{0}}\oint
\mathbf{A \cdot dl}$, where, $\Phi_{0}=\frac{\hbar c}{2 e}$ is the
quantum of flux and $\mathbf{A}$ is the vector potential. The
current in the junction which is a supercurrent and is called
Josephson current can be calculated from $j_J=\frac{2e}{\hbar}
\frac{\partial E}{\partial \phi}$, in which, the $E$ is the energy
of the junction. In the next sections we review some of these weak
link experiments analytically. The quasiclassical Eilenberger
equations have been used to investigate these weak link systems in
this thesis. This method will be explained in Sec.\ref{c2s2}. In
Sec.\ref{c2s3} as an application of the quasiclassical Eilenberger
equation, we solve the problem of a conventional superconducting
bulk without any contact. These results are exactly the same of
results of the standard BCS formalism which can be exerted
directly on the uniform bulk system. Sec.\ref{c2s3} is devoted to
the Kulik-Omelyanchouk problem and at the end of this section we
have generalized this method to a system of a contact with finite
transparency, which is the Zaitsev problem \cite{Zaitsev}. This
problem is as same as our problem in Chapter(3). In Sec.\ref{c2s4}
we have a review in the case of the junction between unitary
$f-$wave superconductors which had been done in paper
\cite{Mahmoodi}. \subsection{Quasiclassical approach}\label{c2s2}
 The normal metals and superconductors can be investigated using the Green functions
\cite{Abrikosov}. It has been shown by Eilenberger that Gorkov
equations for the Green function can be transformed to
transport-like equations for a quasiclassical Green function
\cite{Eilenberger}. These are called Eilenberger equations. Two
conditions for applicability of the quasiclassical approach are
that the characteristic length scales should be much larger than
the Fermi wavelength and energies must be much lower than the
Fermi energy $\varepsilon_{\textrm{F}}= T_{\textrm{F}}$, hereafter
$k_B=1$ and $\hbar=1$ for simplicity. The Green functions which
will be used in our work, are Matsubara Green functions written in
Nambu space. They are $4\times 4$ matrices in a direct-product
space of particle-hole and spin spaces. The general energy
integrated Green function in $\mathbf{k}$-space is of the form
$\breve{g}(\mathbf{\hat{k}},\mathbf{r},\varepsilon_m)$. Here
$\varepsilon_m=\pi T(2m+1)$ are the discrete Matsubara energies
$m=0,1,2...$. The odd integer value of $(2m+1)$ is the result of
Fermion behavior of electrons \cite{Fetter}. Finally, the
Eilenberger equation a ballistic case (there is no scattering) is
as follows:
\begin{equation}
\mathbf{v}_{F}\cdot\nabla \breve{g}+\left[ \varepsilon
_{m}\breve{\sigma}_{3}+i\breve{\Delta},\breve{g}\right] =0,
\label{Eilenberger}
\end{equation}
where, $\mathbf{v}_{F}$ is the Fermi velocity and
$\breve{\sigma}_{3}=\hat{\sigma}_{3}\otimes \hat{I}$ in which
$\hat{\sigma}_{j}\left( j=1,2,3\right) $ are Pauli matrices. This
is a first-order differential equation for the Matsubara
propagator
$\breve{g}(\mathbf{\hat{k}},\mathbf{r},{\varepsilon_m})$ along
classical trajectories of quasiparticles. The Eilenberger equation
is not enough to make the solution unique and so, a separate
normalization condition has to be introduced \cite{Eilenberger}.
With a suitable choice of condition satisfied by physical
solutions of Eilenberger equation, normalization is written as
$\breve{g} \breve{g}=\breve{1}$. To give a closed system of
Eilenberger equations and normalization conditions it should be
supplemented by some self-consistency equations for the
self-energy $\breve{\Delta}$ which will be introduced later in
suitable forms. Finally, the Matsubara propagator $\breve{g}$
which satisfies the Eilenberger equation, normalization condition,
continuity across the interfaces and self-consistency condition
can be written in the form \cite{serenerainer}:
\begin{equation}
\breve{g}=\left(
\begin{array}{cc}
g_{1}+\mathbf{g}_{1}\cdot\mathbf{\hat{\sigma}} & \left( g_{2}+\mathbf{g}_{2}\cdot\hat{%
\sigma }\right) i\hat{\sigma}_{2} \\
i\hat{\sigma}_{2}\left( g_{3}+\mathbf{g}_{3}\cdot\hat{\sigma
}\right)  &
i\hat{\sigma}_{2}(-g_{4}+\mathbf{g}_{4}\cdot\hat{\sigma
})i\hat{\sigma}_{2}
\end{array}
\right) ,\label{Green function}
\end{equation}
 where, the matrix structure of the off-diagonal self energy $\breve{\Delta}$ in the
 Nambu space is
\begin{equation}
\breve{\Delta}=\left(
\begin{array}{cc}
0 & (\Delta+\mathbf{d}\cdot\hat{\sigma })i\hat{\sigma}_{2} \\
i\hat{\sigma}_{2}(\Delta^{*}+\mathbf{{d^{\ast
}}\cdot\hat{\sigma}}) & 0
\end{array}
\right) .\label{order parameter}
\end{equation}
The $\Delta(\mathbf{\hat{k}})=\Delta(\mathbf{-\hat{k}})$ refers to
the spin-singlet but the
$\mathbf{d(\hat{k})}=-\mathbf{d(-\hat{k})}$ has been considered
for the case of spin-triplet superconductivity. Fundamentally, the
gap (order parameter) has to be determined numerically from the
self-consistency equation, while in some cases, we use a
non-self-consistent model for the gap which is much more suitable
for an analytical calculation. The solution of Eq.
(\ref{Eilenberger}) allows us to calculate the current densities.
The expression for current is:
\begin{equation}
\mathbf{j}\left( \mathbf{r}\right) =4\pi ieTN\left( 0\right)
\sum_{m>0}\left\langle \mathbf{v}_{F}g_{1}\left( \mathbf{\hat{v}}_{F},\mathbf{r%
},\varepsilon _{m}\right) \right\rangle_{\mathbf{v}_{F}}
\label{charge-current}
\end{equation}
where, $\left\langle ...\right\rangle_{{\mathbf{\hat{v}}}_{F}}$
stands for averaging over the directions of an electron momentum
on the Fermi surface and $N\left( 0\right) $ is the electron
density of states at the Fermi level of energy. For the case of
the $s-$wave superconductors our Green matrix changes to a matrix
as follows:
\begin{equation}
\breve{g}=\left(
\begin{array}{cc}
g_{1} & g_{2} i\hat{\sigma}_{2} \\
i\hat{\sigma}_{2} g_{3} & -g_{1}
\end{array}
\right) ,
\end{equation}
and for the case of the order parameter we have:
\begin{equation}
\hat{\Delta}=\left(
\begin{array}{cc}
0 & \Delta i\hat{\sigma}_{2} \\
i\hat{\sigma}_{2} \Delta^{*} & 0
\end{array}
\right) .
\end{equation}
The selfconsistent equation for the spin-singlet case is
\begin{equation}
\Delta (\mathbf{r},T)=2\pi \lambda T\sum_{m>0}\left\langle
{g_{2}}(\mathbf {v_{F}},\mathbf{r})\right\rangle _{ {\mathbf
v}_{F}} \label{Eq for Delta}
\end{equation}
where $\lambda $ is the electron-phonon constant of interaction
and $\left\langle ...\right\rangle _{{\mathbf v}_{F}}$ is the
averaging over directions of $ {\mathbf v}_{F}$. The other
important physical variable which can be derived form this Green
function is the density of states. It is related to the diagonal
term of the Green matrix as follows:
\begin{equation}
N(E)=N(0)Re[g_1( \varepsilon_{m}\rightarrow -iE+0)] \label{dos}
\end{equation}
There are many applications of quasiclassical method and in this
chapter some of them will be reviewed.
\subsection{Superconducting bulks}\label{c2s3} The simplest case of superconducting system
which can be investigated by Eilenberger equation is a single bulk
of superconductors without any contact or interaction with other
world. In this system all of superconductivity properties are
uniform and spatially constant. Because of the uniform properties
the gradient term in Eilenberger equation (\ref{Eilenberger}) is
zero,
\begin{equation}
\left[
\varepsilon_{m}\breve{\sigma}_{3}+i\breve{\Delta},\breve{g}\right]
=0
\end{equation}
and the bulk solutions are:
\begin{equation}
g_{1}=\frac{\varepsilon_{m}}{\sqrt{{\varepsilon_{m}}^{2}+|\Delta|^{2}}}
\end{equation}
and
\begin{equation}
g_{2}(g_{3})=\frac{i\Delta(i\Delta^{*})}{\sqrt{{\varepsilon_{m}}^{2}+|\Delta|^{2}}},
\end{equation}
respectively. In conclusion of Eq. (\ref{dos}), the density of
states for $|E|\eqslantgtr|\Delta|$ is as follows:
\begin{equation}
N(E)=\frac{E}{\sqrt{E^{2}-|\Delta|^{2}}},
\end{equation}
while for $|E|\leqslant |\Delta|$ is $ N(E)=0$. This expression
has been obtained before from the BCS theory directly and using
the particle-hole analysis of a bulk of superconductor. Also, the
self-consistent equation for this system is:
\begin{equation}
\Delta (T)=2\pi \lambda T\sum_{m>0}
\frac{\Delta(T)}{\sqrt{{\varepsilon_{m}}^{2}+|\Delta(T)|^{2}}}
\end{equation}
which is the same to the BCS self consistent equation. This latter
had been obtained from the second quantization and quantum field
theory method. This simple self-consistent equation only can be
solved numerically, but near the $T=0$ and $T=T_c$ it has been
solved analytically. Close to the zero temperature the gap
function varies in terms of temperature as:
$$\Delta(T)=\Delta(0)-\sqrt{2\pi T
\Delta(0)}\exp{-(\Delta(0)/T)}$$ and near the $T=T_c$ the gap
dependence on the temperature is as follows:
$$\Delta(T)=\left(\frac{8
\pi^2}{3\zeta{(3)}}\right)^{\frac{1}{2}}\sqrt{T_c(T_c-T)}.$$ Also,
for the case of the gap function at the $T=0$ we have
$\Delta(0)\simeq 1.76 T_c$. So, the current density for this
system can be calculated as follows:
\begin{equation}
\left\langle \mathbf{v}_{F}\right\rangle=0\Rightarrow
\mathbf{j}\left(\mathbf{r}\right) =4\pi ieTN\left( 0\right)
\sum_{m>0}\left\langle \mathbf{v}_{F}\right\rangle
\frac{\varepsilon_{m}}{\sqrt{{\varepsilon_{m}}^{2}+|\Delta|^{2}}}=0.
\end{equation}
As it is clear from the above expression, the current density for
the homogenous superconducting bulks is zero.
\subsection{Kulik-Omelyanchouk problem: superconducting weak link
through a point contact}\label{c2s4} 
\begin{figure}
\includegraphics[width=\columnwidth]{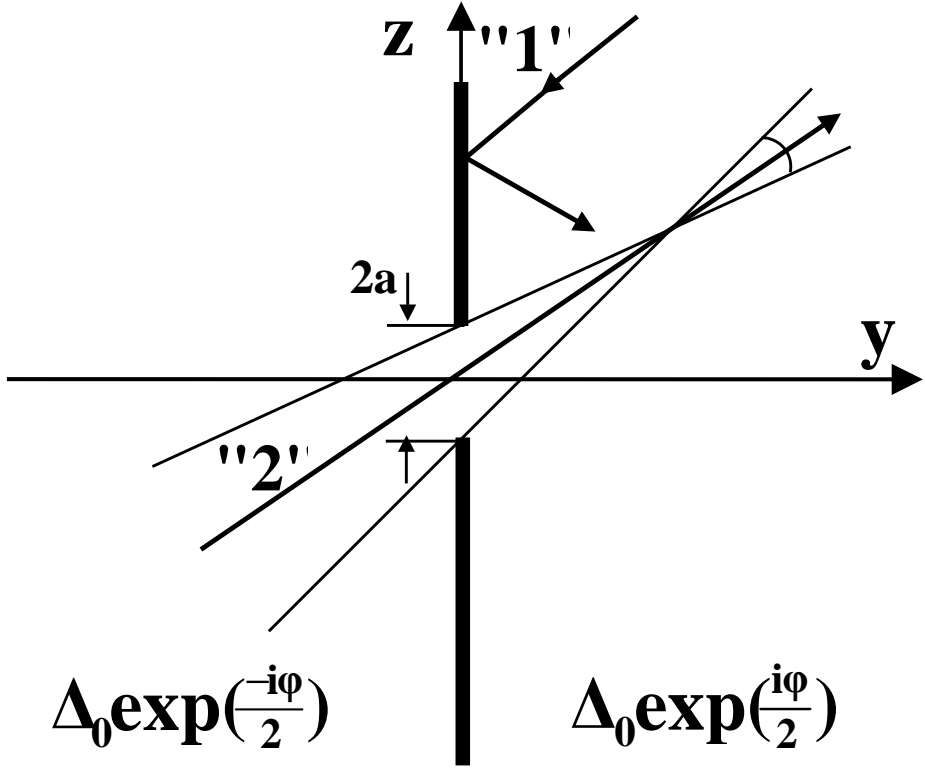}\caption{
Model of an ideal transparent point-contact as an orifice in the
thin impenetrable insulating partition.}\label{figrev1}
\end{figure}

We consider the Josephson
$S-c-S$ weak link as a microbridge between thin superconducting
films of thickness $2a$ (look at Fig.\ref{figrev1}). The length
$L$ and width $d$ of the microbridge, are assumed to be less than
the coherence length $\xi _{0}$. On the other hand, we assume that
$L$ and $2a$ are much larger than the Fermi wavelength $\lambda
_{F}$ and use the quasiclassical approach. The point-contact is an
ideal transparent area for the electrons and there is not any
reflection for the electron. We choose the $\mathbf{z}$-axis along
the interface and the $\mathbf{y}$-axis perpendicular to the
boundary; $y=0 $ is the boundary plane (Fig.\ref{figrev1}).
If the film thickness $d\ll \xi _{0}$ then in the main
approximation in terms of the parameter $d/\xi _{0}$ the
superconducting current depends on the coordinates in the plane of
$\mathbf{\rho}=(z,y)$. The open form of the Eilenberger equation
for the case of spin-singlet superconducting systems is as
follows:
\begin{eqnarray}
\eta \frac{\partial {g_1}_{\left( n\right) }}{\partial
t}+i\Delta^{*}_{n}{g_2}_ {\left( n\right) }-
i\Delta_{n} {g_3}_{\left( n\right)}=0;\\
\eta \frac{\partial {g_2}_{\left( n\right) }}{\partial t}+
2{\varepsilon_m} {g_2}_{\left( n\right) }
-2i\Delta_{n} {g_1}_{\left( n\right) }=0;\\
\eta \frac{\partial {g_3}_{\left( n\right) } }{\partial
t}-2{\varepsilon_m} {g_3}_{\left( n\right) }
+2i\Delta^{*}_{n}{g_1}_{\left( n\right) }=0;
\end{eqnarray}
where, $t=y/|v_y|$ on the Fermi surface, $\eta =\mathbf{sgn}
(v_{y})$ and $n=1,2$ label the left and right hand superconducting
bulks, respectively. Using the quasiclassical approximation, we
select the solution the for this problem  as follows:
\begin{eqnarray}
{g_1}_{\left( n\right) } &=&\frac{\varepsilon_m}{\Omega}+a_{n}\exp \left( -2s{\Omega} t\right) ; \\
{g_2}_{\left( n\right) } &=&\frac{i\Delta_{n}}{\Omega}+b_{n}\exp \left( -2s{\Omega} t\right) ;\\
{g_3}_{\left( n\right) }
&=&\frac{i\Delta^{*}_{n}}{\Omega}+d_{n}\exp \left( -2s{\Omega}
t\right) ;
\end{eqnarray}
 where, $s=\mathbf{sgn}(y)$ and ${\Omega}=\sqrt{{\varepsilon _m}^{2}+|\Delta|^{2}}.$
 By substituting in the Eilenberger equation (\ref{Eilenberger}), we obtain:
\begin{eqnarray}
{g_1}_{\left( n\right) } &=&\frac{\varepsilon_m}{\Omega}+a_{n}\exp
\left( -2s\Omega t\right) ; \\
{g_2}_{\left( n\right) }
&=&\frac{i\Delta_{n}}{\Omega}+a_{n}\left(\frac{i\Delta_{n}}{\varepsilon_m-\eta
s \Omega}\right)\exp \left( -2s\Omega t\right) ;\\
{g_3}_{\left( n\right) }
&=&\frac{i\Delta^{*}_{n}}{\Omega}+a_{n}\left(\frac{i\Delta^{*}_{n}}{\varepsilon_m+\eta
s\Omega} \right)\exp \left( -2s\Omega t\right) ;
\end{eqnarray}
 In the main approximation on the small parameter $a/\xi_{0}\ll1$,
 the self-consistency can been ignored and the model, in which the order parameter
is constant in the two half-spaces $$\Delta (\mathbf{r},T)=\Delta
(T)\exp (\frac{is\phi }{2})$$ in which $\phi $ is the phase
difference between superconductors, can be used.  Solutions of
Eqs. (\ref{Eilenberger}) should satisfy the continuity of
solutions across the contact $y=0,|z|\leq a$ and specular
reflection condition for $y=0,|z|\geq a$. In addition, far from
the contact, solutions should coincide with the bulk solutions.
Consequently, we find the diagonal term of Green functions which
will be used in calculation of the current density, as follows
\cite{Zareyan}:
\begin{equation}
{g_1}{(y=0^{-}) }={g_1}{( y=0^{+})}=\frac{\varepsilon _m
\cos{\frac{\phi}{2}}+i\eta\Omega \sin{\frac{\phi}{2}}}{\Omega
\cos{\frac{\phi}{2}}+i\eta\varepsilon_m \sin{\frac{\phi}{2}}}.
\end{equation}
Because of integration over the fermi surface
(\ref{charge-current}) the antisymmetric and imaginary part of
Green function remains. It is as follows:
\begin{equation}
\mathbf{Im}({g_1}(y=0))=\frac{\eta|\Delta|^{2}
\sin{\frac{\phi}{2}}\cos{\frac{\phi}{2}}}{{\varepsilon_m}^2+|\Delta|^{2}\cos{\frac{\phi}{2}}^{2}},
\end{equation}
and also
\begin{equation}
i T \sum_{m>0}\left\langle \mathbf{\hat{v}}_{F}g_{1}
\right\rangle=\frac{T}{4}\sum_{m>0}\frac{|\Delta|^{2}
\sin{\phi}}{{\varepsilon_m}^2+|\Delta|^{2}\cos{\frac{\phi}{2}}^{2}}=\frac{
 \Delta(T)\sin{\frac{\phi}{2}}}{8}\tanh{\left(\frac{\Delta(T)\cos{\frac{\phi}{2}}}{2T}\right)}
\end{equation}
where we have done the angular integration and we have substituted
$\left\langle \mathbf{v}_{F}\eta \right\rangle=\frac{1}{2}$ and $
\frac{\pi}{4x}\tanh{\frac{\pi
x}{2}}=\sum_{m>0}\frac{1}{x^2+(2m+1)^2}$ in the above-mentioned
relation. Consequently the current density has an expression as
follows \cite{Kulik,Zareyan}:
\begin{equation}
I\left(\phi\right) =\frac{\pi}{2} S e \Delta(T)N\left( 0\right)
v_F\sin{\frac{\phi}{2}}\tanh{\left(\frac{\Delta(T)\cos{\frac{\phi}{2}}}{2T}\right)}
\end{equation}
where, in the case of the low temperatures $T\rightarrow0$ we have
$I\left(\phi\right)=\frac{\pi \Delta}{e R_0}\sin{\frac{\phi}{2}}$
\cite{Kulik} while for $T \rightarrow T_c$ we obtain
$I\left(\phi\right)=\frac{\pi \Delta}{2 e R_0}\sin{\phi}$
\cite{Ambegaokar}, in which, ${R_0}^{-1}=\frac{1}{2}e^2 S N(0)v_F$
is the Sharvin resistance of the junction in the normal state
\cite{Sharvin} and $S$ is the effective square of the contact.
This means that near the critical temperature current-phase
relation is sinusoidal like
 Josephson prediction, while in the low temperatures the
 current-phase relation is non-sinusoidal and we have some unusual
 jumps at $\phi=\pi$.
Now, we want to investigate the effect of transparency coefficient
and reflection of point-contact on the current density which has
been studied before by Zaitsev \cite{Zaitsev}. Using the Zaitsev
quasiclassical boundary conditions for the case of Matsubara Green
function we have obtained:
\begin{equation}
\mathbf{Im}({g_1}(y=0))=\frac{D\eta|\Delta|^{2}
\sin{\frac{\phi}{2}}\cos{\frac{\phi}{2}}}{{\varepsilon_m}^2+|\Delta|^{2}(1-D\sin{\frac{\phi}{2}}^{2})},
\end{equation}
where $D$ is transparency coefficient which is related to the
interface physical properties as the potential barrier against our
tunneling phenomenon. The transparency coefficient usually depends
on the direction of scattering electron velocity, but here for
simplicity we use the constant transparency coefficient formalism.
The current density exactly at the contact is as follows:
\begin{equation}
I\left(\phi\right) =\frac{\pi}{4} S e \Delta(T)N\left( 0\right)
v_F\frac{\sin{\phi}D}{\sqrt{1-D(\sin{\frac{\phi}{2}})^2}}
\tanh{\left(\frac{\Delta(T)\sqrt{1-D(\sin{\frac{\phi}{2}})^2}}{2T}\right)}.
\end{equation}
Obviously, for the low value of transparency coefficient which is
called tunneling limit, $D\rightarrow0$, the current has linear
dependence on the transparency and it is a sinusoidal function of
phase and we have $I\left(\phi\right) =\frac{\pi}{4} S e
\Delta(T)N\left( 0\right) v_F D
\sin{\phi}\tanh{\left(\frac{\Delta(T)}{2T}\right)}.$ So, at the
tunneling limit and $T\rightarrow0$, the current is:
$I\left(\phi\right) =\frac{\pi}{4} S e \Delta(0)N\left( 0\right)
v_F D \sin{\phi}.$ These two former expressions for current are
linear functions of transparency coefficient and sinusoidal
functions of phase $\phi.$ Also in the limit of $T \rightarrow
T_c$, the current has the form of : $I\left(\phi\right) =\frac{\pi
\Delta^2}{8T_c} S e N\left( 0\right) v_F \sin{\phi}D$. This means
that at high temperatures the current is linear function of
transparency and has a sinusoidal dependence on the phase. The
nonsinusoidal dependence of phase will be happened for both the
low temperature and high value of transparency coefficient.
Consequently, nonlinearity and nonsinusoidal current-phase
relation are coupled with each other. For the density of states
using the quasiclassical formalism we obtain:
\begin{equation}
N(E)=N(0)Re[g_1( \varepsilon_{m}\rightarrow -iE+0)]=
\frac{E\sqrt{E^2-\Delta^2}}{E^2-\Delta ^{2}(1-D{\sin \frac{ \phi
}{2}}^{2})}.
\end{equation}
In the limit of $D\rightarrow1$ we obtain the density of states of
the problem of Kulik-Omelyanchouk, $N(E)=N(0)
\frac{E\sqrt{E^2-\Delta^2}}{E^2-\Delta ^{2}{\cos\frac{ \phi
}{2}}^{2}}$, and in the limit of $D\rightarrow0$ we have a
disconnected system and the density of states for the system tends
to the density of states of the bulk,
$N(E)=N(0)\frac{E}{\sqrt{E^2-\Delta^2}},$ as we expect from the
BCS theory \cite{Abrikosov}.
\subsection{Charge transport in the
weak link between unitary and triplet
 supercondcutors}\label{c2s5} In this section we theoretically
study the stationary Josephson effect in a small ballistic
junction between two spin-triplet superconducting bulks with
different orientations of the crystallographic axes. We consider a
model of a ballistic point contact as an orifice with a diameter
$2a$ in an impenetrable for electrons partition between two
superconducting half spaces. We assume that the thickness of
interface, $d$, is much larger than the Fermi wavelength and use
the quasiclassical approach. In order to calculate the charge
current in point contact we use Eilenberger equations
(\ref{Eilenberger}) and the normalization condition $\breve{g}
\breve{g}=\breve{1}$. For the case of the unitary and pure
spin-triplet superconductivity, the Green function matrix
(\ref{Green function}) is written in the form
\cite{Mahmoodi,serenerainer}:
\begin{equation}
\breve{g}=\left(
\begin{array}{cc}
g_{1}+\mathbf{ g}_{1}\cdot \hat{\sigma } & i\mathbf{
g}_{2}\cdot \hat{\mathbf{\sigma }}\hat{\sigma }_{2} \\
i\hat{\sigma }_{2}\mathbf{ g}_{3}\cdot \hat{\mathbf{ \sigma }} &
g_{4}-\hat{\sigma }_{2}\mathbf{ g}_{4}\cdot\hat{\mathbf{ \sigma
}}\hat{\mathbf{\sigma }}_{2}
\end{array}
\right).
\end{equation}
Matrix structure of the off-diagonal self energy $\breve{\Delta}$
in Nambu space is
\begin{equation}
\breve{\Delta }=\left(
\begin{array}{cc}
0 & {\hat\Delta} \\
{\hat\Delta^{\dag}} & 0
\end{array}
\right)= \left(
\begin{array}{cc}
0 & i\mathbf{ d}\cdot\hat{\mathbf{ \sigma }}\hat{\mathbf{ \sigma }}_{2} \\
i\hat{\sigma }_{2}\mathbf{ d}^{\ast }\cdot\hat{\sigma } & 0
\end{array}
\right),
\end{equation}
Also, we have:
\begin{equation} {\hat\Delta}({\mathbf
k})= i{\mathbf d}({\mathbf k}) \cdot{\hat{\sigma
}}\hat{\sigma}_2=\left(
\begin{array}{cc}
{id_2-d_1} & { d_3} \\
{ d_3} & {id_2+d_1}
\end{array}
\right).
\end{equation}
Below we consider a unitary states, for which $\mathbf{ d\times
d}^{\ast }=0.$ Solutions of Eq. (\ref{Eilenberger}) must satisfy
the conditions for Green functions and vector $\mathbf{d}$ in the
bulks of superconductors far from the orifice:
\begin{eqnarray}
\breve{g}\left( \mp \infty \right) &=&\frac{\varepsilon _{m}%
\breve{\tau }_{3}+i\breve{\Delta }_{1,2}}{\sqrt{%
\varepsilon _{m}^{2}+\left| \mathbf{ d}_{1,2}\right| ^{2}}};  \label{g(inf)} \\
\mathbf{ d}\left( \mp \infty \right) &=&\mathbf{ d}_{1,2}\left( \hat{\mathbf{ k}}%
\right) \exp \left( \mp \frac{i\phi }{2}\right) ,  \label{d(inf)}
\end{eqnarray}
where $\phi $ is the external phase difference. Eq.
(\ref{Eilenberger}) has to be supplemented by the continuity of
solutions at the contact plane ($|y|\leq a$) and conditions of
reflection at the interface between superconductors, remainder
part of the interface ($|y|>a$). Below we assume that this
interface is smooth and electron scattering is negligible. The
solution of Eq. (\ref{Eilenberger}) should be used to calculate
the current density. We consider a simple model of the constant
order parameter up to the surface.
\begin{figure}[tbp]
\includegraphics[width=\columnwidth]{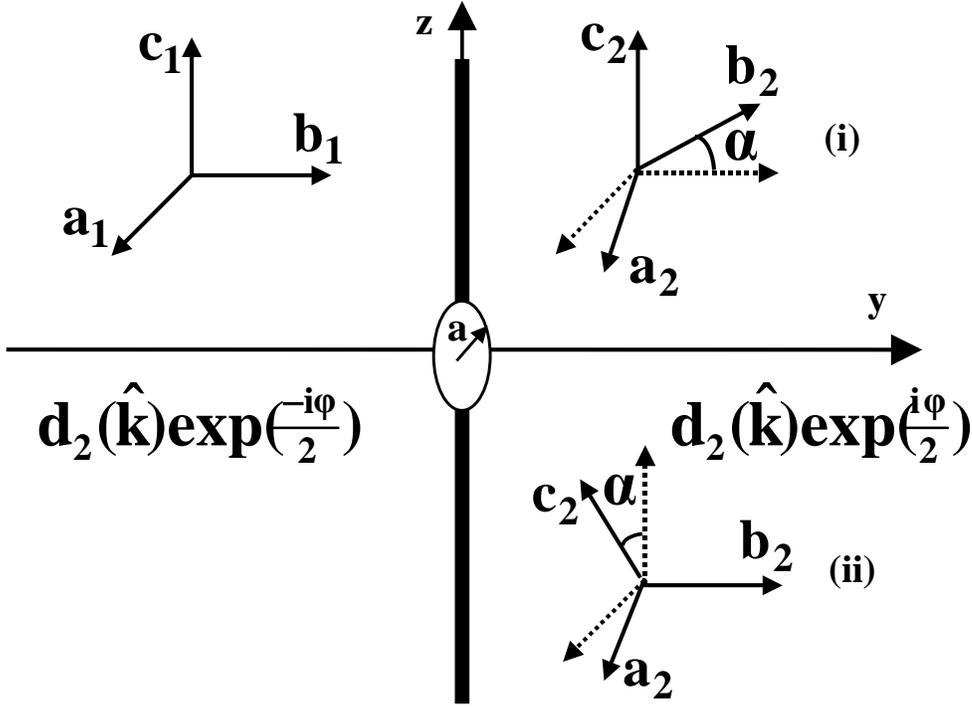} \caption{Scheme
of a ideal transparent point-contact in an impenetrable flat
interface between two superconducting bulks, which are
misorientated as much as $\protect\alpha $.} \label{figrev2}
\end{figure}
We assume that the order parameter does not depend on the
coordinates and in each half-space equals to its value far from
the point contact which is called superconducting massive bulk.
For this non-self-consistent model the current-phase dependence of
a Josephson junction can be calculated analytically. In a
ballistic case, the system of $13$ equations for functions $g_{i}$
and $\mathbf{ g}_{i}$ can be decomposed on independent blocks of
equations. The set of equations which enables us to find the Green
functions are:
\begin{eqnarray}
\eta \frac{\partial g_{1}}{\partial t}+i\left( \mathbf{
g}_{2}\mathbf{ d}^{\ast }-
\mathbf{ g}_{3}\mathbf{ d}\right) &=&0;  \label{a} \\
\eta \frac{\partial \mathbf{ g}_{-} }{\partial t}+\left( \mathbf{
d\times g}_{3}+
\mathbf{d}^{\ast }\mathbf{ \times g}_{2}\right) &=&0;  \label{b} \\
\eta \frac{\partial \mathbf{ g}_{2}}{\partial t}+\varepsilon _{m}\mathbf{ g}%
_{2}-ig_{1}\mathbf{ d}-\mathbf{ d}\times \mathbf{ g}_{-}
&=&0;\label{c}\\
\eta\frac{\partial \mathbf{ g}_{3}}{\partial t}-\varepsilon
_{m}\mathbf{ g}_{3}+ig_{1}\mathbf{ d}^{\ast }-\mathbf{ d}^{\ast
}\times \mathbf{ g}_{-} &=&0;\label{d}
\end{eqnarray}
where $\mathbf{ g}_{-}=\frac{\mathbf{ g}_{1}-\mathbf{ g}_{4}}{2}$,
$t=y/|v_y|$ on the Fermi surface and $\eta =\mathbf{sgn} (v_{y})$.
The Eqs. (\ref{a})-(\ref{d}) can be solved by integrating over the
ballistic trajectories of electron in the right and left
half-spaces. The general solution satisfying the boundary
conditions at infinity is:
\begin{eqnarray}
g_{1}^{\left( n\right) } &=&\frac{\varepsilon _{m}}{\Omega
_{n}}+a_{n}\exp
\left( -2s\Omega _{n}t\right) ;  \label{ei} \\
{\mathbf g}_{-}^{\left( n\right) } &=&{\mathbf C}_{n}\exp \left(
-2s\Omega
_{n}t\right) ;  \label{fi} \\
{\mathbf g}_{2}^{\left( n\right) } &=&\frac{i{\mathbf
d}_{n}}{\Omega _{n}}-\frac{ia_{n}{\mathbf d}_{n}+{\mathbf
d}_{n}\times {\mathbf C}_{n}}{s\eta \Omega _{n}-\varepsilon
_{m}}\exp \left( -2s\Omega
_{n}t\right) ;  \label{gi} \\
{\mathbf g}_{3}^{\left( n\right) } &=& \frac{i {\mathbf
d}_{n}^{\ast
}}{\Omega _{n}}+\frac{ia_{n}{\mathbf d}_{n}^{\ast }-{\mathbf d}%
_{n}^{\ast }\times {\mathbf C}_{n}}{s\eta \Omega _{n}+\varepsilon
_{m}}\exp \left( -2s\Omega _{n}t\right); \label{hi}
\end{eqnarray}
where $t$ is the time of the flight along the trajectory,
$sgn\left( t\right) =\mathbf{sgn}\left( y\right) =s,$ $\eta
=\mathbf{sgn}\left( v_{y}\right)$ and $\Omega
_{n}=\sqrt{\varepsilon _{m}^{2}+\left| \mathbf{ d}_{n}\right|
^{2}}.$ Index $n$ numbers left $ \left( n=1\right) $ and right
$\left( n=2\right) $ half-spaces. By matching the solutions
(\ref{ei}-\ref{hi}) at the orifice plane $\left( y=0\right) $, we
find constants $a_{n}$ and $\mathbf{ C_{n}}$. Exactly at the
orifice plane we obtain:
\begin{eqnarray}
\frac{\varepsilon _{m}}{\Omega _{1}}+a_{1}&=&\frac{\varepsilon _{m}}{\Omega _{2}}+a_{2} ; \\
{\mathbf C}_{1}&=&{\mathbf C}_{2}; \\
 \frac{i{\mathbf d}_{1}}{\Omega _{1}}+\frac{ia_{1}{\mathbf d}_{1}+{\mathbf d}_{1}\times
 {\mathbf C}_{1}}{\eta \Omega_{n}+\varepsilon_{m}}&=&\frac{i{\mathbf d}_{2}}{\Omega _{2}}-\frac{ia_{2}
 {\mathbf d}_{2}+{\mathbf d}_{2}\times{\mathbf C}_{2}}{\eta \Omega_{2}-\varepsilon_{m}}; \\
\frac{i {\mathbf d}_{1}^{\ast }}{\Omega_{1}}-\frac{ia_{1}{\mathbf
d}_{1}^{\ast }-{\mathbf d}_{1}^{\ast }\times {\mathbf C}_{1}}{\eta
\Omega_{1}-\varepsilon _{m}}&=&\frac{i {\mathbf d}_{2}^{\ast
}}{\Omega_{2}}+\frac{ia_{2}{\mathbf d}_{2}^{\ast}-{\mathbf
d}_{2}^{\ast }\times {\mathbf C}_{2}}{\eta
\Omega_{2}+\varepsilon_{m}};
\end{eqnarray}
Consequently, the function $g_{1}\left( 0\right)$ which determines
the current density at the contact is as follows:
\begin{equation}
g_{1}\left( 0\right)= \frac{\eta\left[\mathbf{d_1\cdot d_1}\left(
\eta\Omega _{2}+\varepsilon_{m}\right)^{2}-\mathbf{d_2\cdot
d_2}\left( \eta\Omega
_{1}-\varepsilon_{m}\right)^{2}\right]}{\mathbf{d_1\cdot
d_1}\left( \eta\Omega
_{2}+\varepsilon_{m}\right)^{2}+\mathbf{d_2\cdot d_2}\left(
\eta\Omega _{1}-\varepsilon_{m}\right)^{2}+2\mathbf{d_1\cdot
d_2}\left( \eta\Omega _{1}-\varepsilon_{m}\right)\left( \eta\Omega
_{2}+\varepsilon_{m}\right)}. \label{greenterm}
\end{equation}
Using the $g_{1}\left( 0\right)$ one can calculate the current
density at the orifice plane $\mathbf{ j}(0)$:
\begin{equation}
\mathbf{ j}(0)=4\pi eN(0)v_{F}T\sum_{m=0}^{\infty }\int
d^{3}k\hat{\mathbf{ k}}g_{1}(0). \label{jJos}
\end{equation}
Misorientation of the crystals produces a spontaneous current
along the interface \cite{Barash,AmOmZ} generally, as can be
calculated by projecting vector $\mathbf{ j}$ at the corresponding
direction. To illustrate the results obtained by computing the
formula (\ref{greenterm}), we can plot the current-phase diagrams
for the different models of the pairing symmetry and for two
different geometries. These geometries are corresponding to the
different orientations of the crystals in the right and left sides
of the interface (Fig.\ref{figrev2}). In the right hand side of
the interface, the $\mathbf{ab-}$plane has been rotated around the
$\mathbf{c-}$axis and the $\mathbf{c-}$axis has been rotated
around the $\mathbf{b-}$axis by $\alpha $ in geometries (i) and
(ii), respectively. Also for the further calculations we need to a
certain model of the gap vector $ \mathbf{d}$ which is called
order parameter vector.
\begin{figure}[tbp]
\includegraphics[width=\columnwidth]{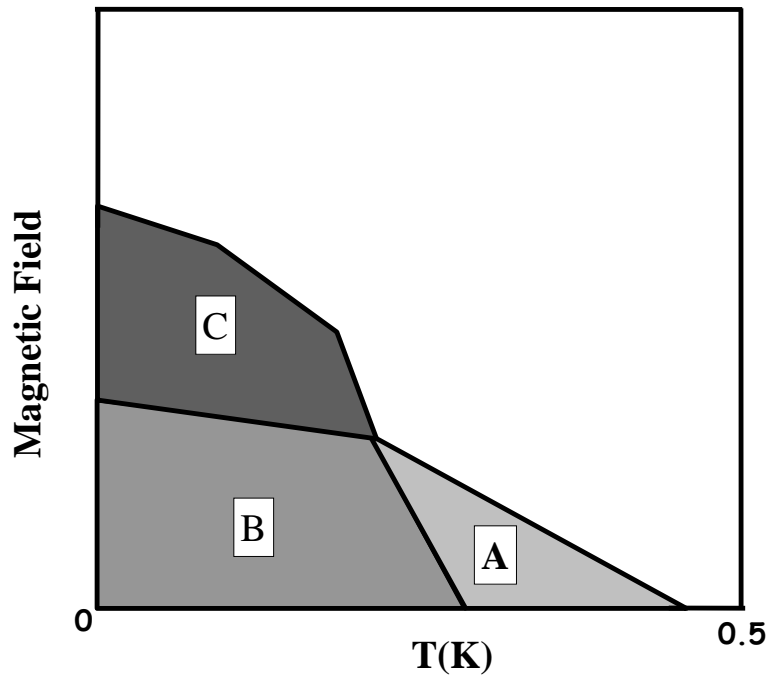} \caption{Superconducting phase diagram of
 heavy fermion $UPt_3$ compound, in the magnetic field and temperature space.} \label{figrev3}
\end{figure}
There are three models which have been successful to explain
properties of the three phases of triplet superconductivity in
$UPt_{3}$ compound (Fig.\ref{figrev3}). For the high-temperature
and low field phase, A-phase, of superconductivity in $UPt_{3}$
the order parameter (gap vector) has an equatorial line node and
two longitudinal line nodes \cite{Mahmoodi,Machida,Ohmi}. This
state which is known as the polar state, is as follows:
\begin{equation}
\mathbf{ d=}\Delta _{0}\hat{z}k_{z}\left( k_{x}^2-k_{y}^2\right).
\label{aphase}
\end{equation}
The gap vector dependence in momentum space for the
low-temperature and low field phase, B-phase, or the axial state
is as follows \cite{Mahmoodi,Machida,Ohmi}:
\begin{equation}
\mathbf{ d=}\Delta _{0}\hat{z}k_{z}\left( k_{x}+ik_{y}\right)
^{2}. \label{bphase}
\end{equation}
Here, the longitudinal line nodes are closed and there is a pair
of point nodes. The coordinate axes $\mathbf{\hat{x}},
\mathbf{\hat{y}}$ and $\mathbf{\hat{z}}$ are chosen along the
crystallographic axes $\hat{\mathbf{ a}}, \hat{\mathbf{ b}}$ and
$\hat{\mathbf{ c}}$ as the left of Fig.\ref{figrev2}. The function
$\Delta _{0}\left( T\right) $ describes the dependence of the
order parameter $\mathbf{ d}$ on the temperature $T$. Other
candidate to describe the orbital states implying the weak
effective spin-orbital coupling in the $c-$phase of
superconductivity in $UPt_{3}$ compound, is the unitary planar
state \cite{Mahmoodi,Machida,Ohmi}:
\begin{equation}
\mathbf{ d=}\Delta _{0}k_{z}(\hat{x}\left( k_{x}^{2}-k_{y}^{2}\right) +%
\hat{y}2k_{x}k_{y}).  \label{planar}
\end{equation}
Using these forms of the triplet order parameters, we can plot the
Josephson current-phase relation $j_{J}(\phi )=j_{y}(y=0)$
calculated from Eq.\ref{jJos} for a particular value of
misorientation angle $ \alpha $ under the rotation of
$\mathbf{ab}$-plane, the geometry (i), and rotation around the
normal axis $\mathbf{\hat{y}}$ or geometry (ii). For simplicity we
use the spherical model of the Fermi surface. The different gap
vectors have different current-phase diagrams and such a different
behavior can be a criterion for distinguish between the different
phases. In some cases, the Josephson current formally does not
equal to zero at $\phi =0.$ This state corresponds to a
spontaneous phase difference, which depends on the misorientation
angle $\alpha$. The tangential components of current, $x$ and $z$,
as the functions of $\phi$ are not zero when the Josephson current
is zero. This spontaneous tangential current is due to the
specific ''proximity effect'' similar to spontaneous current in
contacts between ''$d$-wave'' superconductors \cite{AmOmZ,LSW00}.
The total current is determined by the Green function, which
depends on the order parameters in both superconductors. As a
result of this, for nonzero misorientation angles the current
parallel to the surface can be generated. It can be shown that the
current-phase relations are totally different for different models
of the gap vector. Because the order parameter phase depends on
the momentum direction on the Fermi surface, the misorientation of
the superconductors leads to spontaneous phase difference that
corresponds to the zero Josephson current and to the minimum of
the weak link energy. This phase difference depends on the
misorientation angle and can possess any values. It is observed
that, in the "$f-$wave" superconductors the spontaneous current
can be generated in a direction parallel to the plane of contact.
Generally speaking this current is not equal to zero in the
absence of the Josephson current. Finally, study of current-phase
diagrams of Josephson junction for different misorientations can
be used to distinguish or demonstrate the different phases and
different triplet gap vectors of superconductivity $UPt_{3}$.
\newpage
\section{Coherent mixing of Josephson and transport supercurrents}
\label{Vortex}
\subsection{Introduction} The investigations of Josephson effect
manifestations in different systems are continuing due to it's
importance both for basic science and industry. A point contact
between two massive superconductors (S-c-S junction) is one of the
possible Josephson weak links. A microscopic theory of the
stationary Josephson effect in ballistic point contacts between
conventional superconductors was developed in \cite{Kulik}. Later,
this theory was generalized for a pinhole model in $^{3}He$
\cite{Kurk,Yip}, for point contacts between ''$d$-wave''
\cite{AmOmZ,Yip1}, and triplet superconductors \cite{Mahmoodi}.
The Josephson effect is the phase sensitive instrument for the
analysis of an order parameter in novel (unconventional)
superconductors, where current-phase dependencies $I_{J}\left(
\phi \right) $ may differ essentially from those in conventional
superconductors \cite{AmOmZ,Yip1}. In some cases the model with
total transparency of the point contact does not quite adequately
correspond to the experiment, and the electron reflection should
be taken into account. The influence of electron reflection on the
Josephson current in ballistic point contacts was first considered
by Zaitsev \cite{Zaitsev}. He had shown that reflection from the
contact not only changes the critical value of current, but also
the current-phase dependence $I_{J}\left( \phi \right) \sim \sin
\left( \phi /2\right) $ at low temperature which has been
predicted in \cite{Kulik}. The current-phase dependence for small
values of transparency, $D\ll 1$, is transformed to the
$I_{J}\left( \phi \right) \sim \sin \phi $, similar to the planar
tunnel junction. The effect of transparency for point contact
between unconventional (d-wave) superconductors is studied in the
papers \cite{Amin,Coury,Galaktionov,LSW00}. The non-locality of
Josephson current in point contacts was investigated in
\cite{Heida}. The authors of \cite{Heida} concentrated on the
influence of magnetic field on the zero voltage supercurrent
through the junction. They found an periodic behavior in terms of
magnetic flux and demonstrated that this anomalous behavior is a
result of a non-locality supercurrent in the junction. This
observation was explained theoretically in \cite{Zagoskin}. In
this chapter we want to investigate theoretically the influence of
electron reflection on dc Josephson effect in a ballistic point
contact with transport current in the right and left banks which
are separated by an interface (look at Fig.\ref{figa1}). In Ref.
\cite{KOSh} for an ideal transparent point-contact in the
impenetrable interface, it has been observed that at the phase
differences close to the $\phi=\pi$ two antisymmetric vortex-like
currents appear (see Fig.\ref{figa5}). we want to study the effect
of finite transparency (reflection) of the point-contact in the
interface, on these vortex-like currents near the contact and at
the phase difference $\phi =\pi$. We show that at low temperatures
even a small reflection on the contact destroys the mentioned
vortex-like current states, which can be restored by increasing of
the temperature. In our system which will be investigated in this
chapter, a point-contact between two massive superconductors
(S-c-S junction) is considered as a possible Josephson weak links.
For such systems it is convenient to use Kulik-Omelyanchouk method
\cite{Kulik} for the ballistic point-contact. The microscopic
theory of the stationary Josephson effect in the ballistic point
contacts between conventional superconductors was developed in
\cite{Kulik}. Later, this theory was generalized for
point-contacts between ''$d$-wave'' high-$T_c$ supercondcutors in
Ref. \cite{AmOmZ}. The Josephson effect is a phase sensitive
instrument for the analysis of an order parameter in novel
(unconventional) superconductors, where current-phase dependencies
$I_{J}\left( \phi \right) $ may differ essentially from those in
conventional superconductors \cite{AmOmZ,Mahmoodi}. In some cases
the model with ideal transparent point-contact does not correspond
to the experiment totally, and the electron reflection should be
taken into account. The influence of electron reflection on the
Josephson current in ballistic point contacts was first considered
by Zaitsev \cite{Zaitsev}. He had shown that reflection from the
contact not only changes the critical value of current, but also
the current-phase dependence $I_{J}\left( \phi \right) \sim \sin
\left( \phi /2\right) $ at low temperature which has been
predicted in \cite{Kulik}. The current-phase dependence for small
values of transparency, $D\ll 1$, is transformed to the
$I_{J}\left( \phi \right) \sim \sin \phi $, similar to the planar
tunnel junction. In addition, the non-locality of Josephson
current in point contacts was investigated in Ref. \cite{Heida}.
The authors of Ref. \cite{Heida} concentrated on the influence of
magnetic field on the zero voltage supercurrent through the
junction. They found an periodic behavior in terms of magnetic
flux and demonstrated that this anomalous behavior is a result of
a non-locality supercurrent in the junction. This observation was
explained theoretically in \cite{Zagoskin,Lederman,M.H.S.Amin}.
Recently an influence of transport supercurrent, which flows in
the contacted banks and is parallel to the interface, to the
Josephson effect in point contacts has been analyzed
theoretically\cite{KOSh}. It was found that a non-local mixing of
two superconducting currents results in the appearance of two
vortex-like current states in vicinity of the contact, when the
external phase difference is $\phi \simeq \pi $. The Josephson
current through superconducting weak link is a result of quantum
interference between order
parameters with phase difference $\phi $. Obviously, the finite reflection $%
R=1-D$ of electrons from the Josephson junction suppresses this
interference and it must influence the vortex-like current states,
which are predicted in \cite{KOSh}. In this chapter, we study the
effect of finite transparency on the current-phase dependence and
distribution of the superconducting current near the ballistic
point contact in the presence of homogeneous current states far
from the contact. We show that at low temperatures $\left(
T\rightarrow 0\right) $ the electron reflection destroys the
mentioned vortex-like current states even for a very small value
of reflection coefficient $R\ll 1.$ On the other hand we have
found that, as the temperature increases the vortices are restored
and they exist for transparency as low as $D=\frac{1}{2}$ in the
limit of $T\rightarrow T_{c}$. The arrangement of the rest of this
chapter is as follows. In Sec.(\ref{suba}) we describe the model
of the point contact, quasiclassical equations for Green functions
and boundary conditions. The analytical formulas for the Green
functions are derived for a ballistic point contact with arbitrary
transparency. In Sec.(\ref{subb}) we apply them to analyze a
current state in the ballistic point contact. The influence of the
transport current on the Josephson current and vice versa at the
contact plane is considered. In Sec.(\ref{subc}) we present the
numerical results for the distribution of the current in the
vicinity of the contact. \subsection{Model and equations}
\label{suba} 
\begin{figure}
 \hspace{2.5cm}\includegraphics[width=\columnwidth]{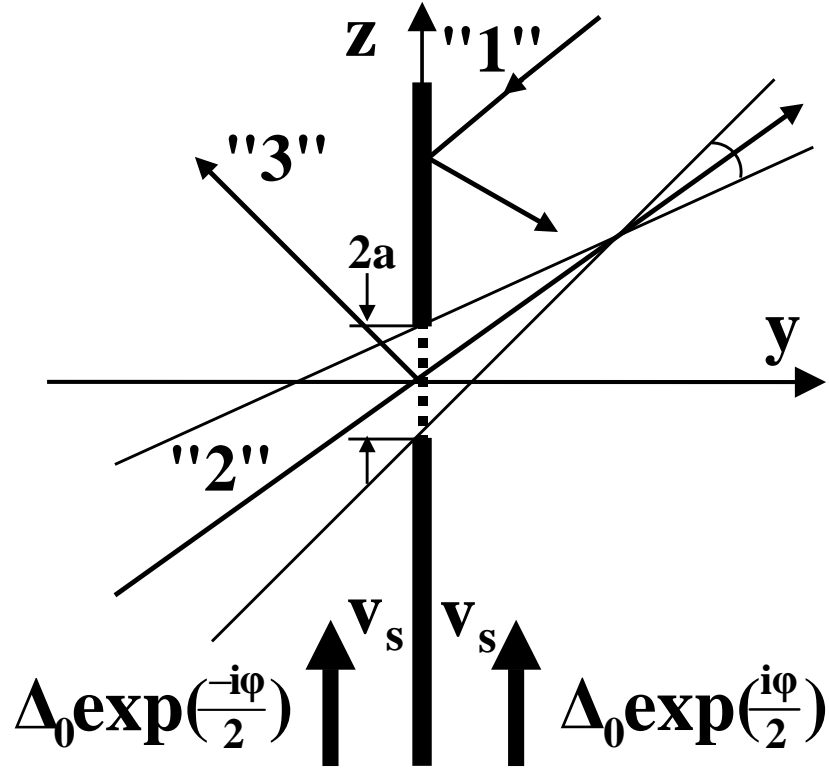}\caption{\label{figa1}
Model of the point-contact with finite reflection coefficient as a
slit in the thin impenetrable insulating partition.}
\end{figure}

We consider the Josephson weak link as a microbridge
between thin superconducting films of thickness $d$. The length
$L$ and width $2a$ of the microbridge, are assumed to be less than
the coherence length $\xi _{0}$. On the other hand, we assume that
$L$ and $2a$ are much larger than the Fermi wavelength $\lambda
_{F}$ and use the quasiclassical approach. There is a potential
barrier in the contact, resulting in a finite probability for the
electron that is to be reflected back. In the banks of
superconductors a homogeneous current with a superconducting
velocity $\mathbf{{v}_{s}}$ flows parallel to the partition. We
choose the $\mathbf{z}$-axis along $\mathbf{{v}_{s}}$ and the
$\mathbf{y}$-axis perpendicular to the boundary; $y=0$ is the
boundary plane (see Fig.\ref{figa1}). If the film thickness $d\ll
\xi _{0}$ then in the main approximation in terms of the parameter
$d/\xi _{0}$ the superconducting current depends on the
coordinates in the plane of the film $ \mathbf{\rho} =(y,z)$ only.
The superconducting current in the quasiclassical approximation
\begin{equation}
\mathbf{j}(\mathbf{\rho} ,\mathbf {v_{s}})=4\pi
ieN(0)T\sum_{m>0}\left\langle \mathbf{v_{F}}{g_{1}}(\mathbf
{v_{F}},\mathbf{\rho}, \mathbf{v_{s}}) \right\rangle _{\mathbf
{v_{F}}} \label{Current}
\end{equation}
is defined by the energy integrated Green matrix for the case of
singlet superconductors which is following:
\begin{equation}
\breve{g}(\tilde{\varepsilon},{\mathbf v}_{F},{\mathbf \rho},
{\mathbf v}_{s})=\left(
\begin{array}{cc}
g_{1} & g_{2} i\hat{\sigma}_{2} \\
i\hat{\sigma}_{2} g_{3} & -g_{1}
\end{array}
\right) ,
\end{equation}
and for the case of the order parameter we have:
\begin{equation}
\hat{\Delta}=\left(
\begin{array}{cc}
0 & \Delta i\hat{\sigma}_{2} \\
i\hat{\sigma}_{2} \Delta^{*} & 0
\end{array}
\right) .
\end{equation}
This Green matrix in the ballistic case satisfies the Eilenberger
equations as follows \cite{Eilenberger,Belzig}:
\begin{eqnarray}
\eta \frac{\partial {g_1}_{\left( n\right) }}{\partial
t}+i\Delta^{*}_{n}{g_2}_ {\left( n\right) }-
i\Delta_{n} {g_3}_{\left( n\right)}=0;\\
\eta \frac{\partial {g_2}_{\left( n\right) }}{\partial t}+
2{\tilde{\varepsilon}} {g_2}_{\left( n\right) }
-2i\Delta_{n} {g_1}_{\left( n\right) }=0;\\
\eta \frac{\partial {g_3}_{\left( n\right) } }{\partial
t}-2{\tilde{\varepsilon}} {g_3}_{\left( n\right) }
+2i\Delta^{*}_{n}{g_1}_{\left( n\right) }=0;
\end{eqnarray}
where, $t=y/|v_y|$ on the Fermi surface, $\eta =\mathbf{sgn}
(v_{y})$ and $n=1,2$ label the left and right hand superconducting
bulks, respectively. Using the quasiclassical approximation, we
select the solution the for this problem  as follows:
\begin{eqnarray}
{g_1}_{\left( n\right) } &=&\frac{\tilde{\varepsilon}}{\Omega}+a_{n}\exp \left( -2s{\Omega} t\right) ; \\
{g_2}_{\left( n\right) } &=&\frac{i\Delta_{n}}{\Omega}+b_{n}\exp \left( -2s{\Omega} t\right) ;\\
{g_3}_{\left( n\right) }
&=&\frac{i\Delta^{*}_{n}}{\Omega}+d_{n}\exp \left( -2s{\Omega}
t\right) ;
\end{eqnarray}\label{vortex}
 where, $s=\mathbf{sgn}(y)$ and ${\Omega}=\sqrt{\tilde{\varepsilon}^{2}+|\Delta|^{2}}.$ Here $N(0)$ is the density of states at the Fermi level,
$\tilde{{\varepsilon}}=\varepsilon _{m}+i{\mathbf p}_{F}\cdot
{\mathbf v}_{s}$, ${\mathbf v}_{F}$ and ${\mathbf p}_{F}$ are the
electron velocity and momentum on the Fermi surface,
$\varepsilon_{m}=(2m+1)\pi T$ are the Matsubara frequencies, $m$
is an integer number, $ {\mathbf v}_{s}$ is the superfluid
velocity and $T$ is the temperature. Eqs. (\ref{vortex}) should be
supplemented by the equation for the superconducting order
parameter $\Delta $
\begin{equation}
\Delta ({\mathbf \rho},{\mathbf v}_{s},T)=2\pi \lambda
T\sum_{m>0}\left\langle g_{2}({\mathbf v}_{F},{\mathbf \rho}
,{\mathbf v}_{s})\right\rangle _{ {\mathbf v}_{F}} \label{Eq for
Delta}
\end{equation}
where $\lambda $ is the constant of pairing interaction and
$\left\langle ...\right\rangle _{{\mathbf v}_{F}}$ is the
averaging over directions of $ {\mathbf v}_{F}$. After substitute
in the Eilenberger equation (\ref{Eilenberger}), we obtain:
\begin{eqnarray}
{g_1}_{\left( n\right) } &=&\frac{\tilde{\varepsilon}}{\Omega
_{n}}+a_{n}\exp
\left( -2s\Omega _{n}t\right) ; \\
{g_2}_{\left( n\right) } &=&\frac{\Delta_{n}}{\Omega
_{n}}+a_{n}\left(\frac{\Delta_{n}}{\tilde{\varepsilon}-\eta s
\Omega_{n}}\right)\exp \left( -2s\Omega
_{n}t\right) ;\\
{g_3}_{\left( n\right) } &=&\frac{\Delta^{*}_{n}}{\Omega
_{n}}+a_{n}\left(\frac{\Delta^{*}_{n}}{\tilde{\varepsilon}+\eta s
\Omega_{n}}\right)\exp \left( -2s\Omega _{n}t\right) ;
\end{eqnarray}
 As it was shown in \cite{Kulik} in the zero approximation
in terms of the small parameter $a/\xi _{0}\ll 1$ for a
self-consistent solution of the problem it is not necessary to
consider Eq. (\ref{Eq for Delta}). The model, in which the order
parameter is constant in the two half-spaces $\Delta ({\mathbf
\rho},{\mathbf v}_{s},T)=\Delta ({\mathbf v} _{s},T)\exp
($sgn$\left( y\right) \frac{i\phi }{2})$ ($\phi $ is the phase
difference between superconductors), can be used.

\begin{figure}
 \includegraphics[width=\columnwidth]{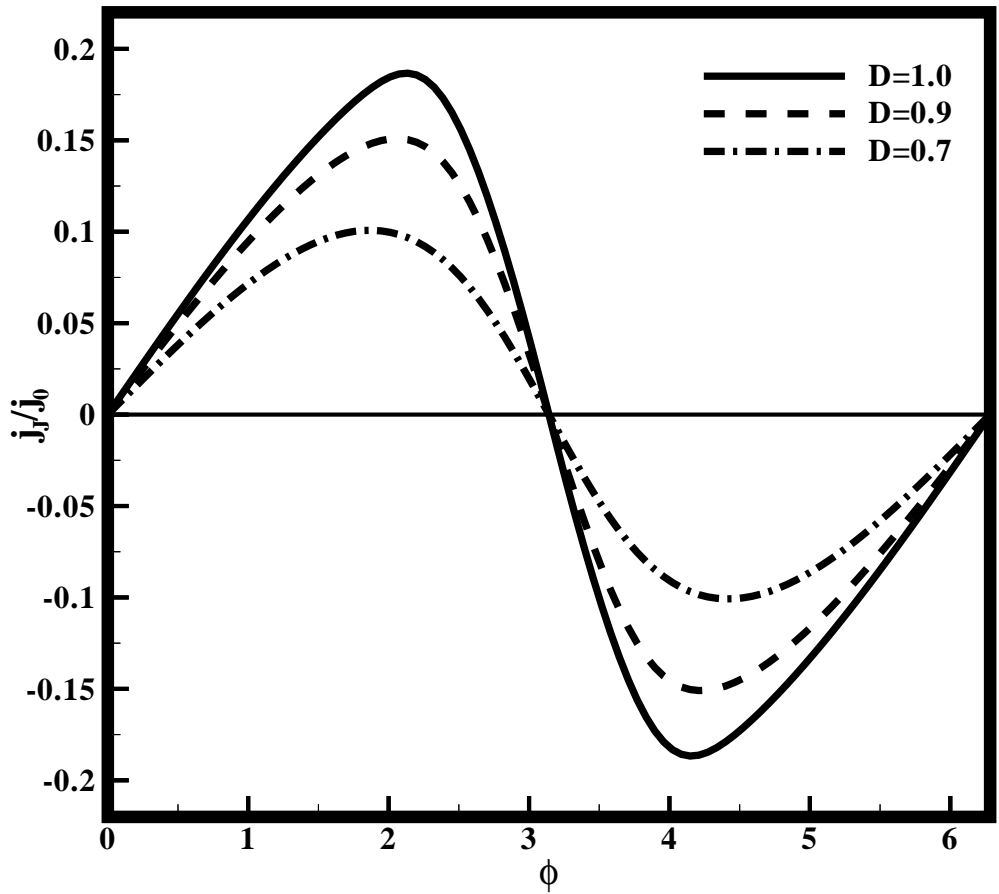}\caption{\label{figa2}Josephson
current $j_{J}$ versus phase $\protect\phi$ for $T/T_c=0.1$,
$q=0.5$ and $j_{0}=4\protect\pi \left| e\right| N(0)v_{F}T_{c}$.}
\end{figure}
In the same approximation the velocity ${\mathbf v}_{s}$ does not
depend on the coordinates. The Eq. (\ref{Eq for Delta}) enables us
to calculate a spatial distribution of the order parameter $\Delta
({\mathbf \rho} )$ in the next order approximation in terms of the
parameter $a/\xi _{0}.$ Solutions of Eqs. (\ref{Eilenberger})
should satisfy Zaitsev's boundary conditions (\cite{Zaitsev})
across the contact $y=0,|z|\leq a$ and specular reflection
condition for $y=0,|z|\geq a$. In addition, far from the contact,
solutions should coincide with the bulk solutions. The Zaitsev
boundary conditions \cite{Zaitsev} have been considered in
\cite{Belzig}, but some improvements are necessary for using these
boundary conditions. These improvements have been done in
\cite{Rashedi1}. The Zaitsev boundary conditions at the contact
can be written as \cite{Zaitsev,Belzig,Rashedi1}
\begin{equation}
\widehat{d}^{~l}=\widehat{d}^{~r}\equiv \widehat{d}
\label{Zaitsev1}
\end{equation}
\begin{equation}
{\frac{D}{2-D}}\left[
(1+{\frac{\widehat{d}}{2}})\widehat{s}^{~r},\widehat{s}
^{~l}\right] =\widehat{d}\ \widehat{s}^{~l2}  \label{Zaitsev2}
\end{equation}
where
\begin{equation}
\widehat{s}^{~r}=\breve{g}^{r}({\mathbf
v}_{F},y=0)+\breve{g}^{r}({{\mathbf v}_{F}}^{\prime},y=0)
\end{equation}
\begin{equation}
\widehat{d}^{~r}=\breve{g}^{r}({\mathbf
v}_{F},y=0)-\breve{g}^{r}({{\mathbf v}_{F}}^{\prime },y=0)
\end{equation}
with ${{\mathbf v}_{F}}^{\prime }$ being the reflection of
${\mathbf v}_{F}$ with respect to the boundary and $D$ is the
transparency coefficient of point contact. Indexes $l$ and $r$
denote that the Green function are taken at the left $\left(
y=-0\right) $ or right $\left( y=+0\right) $ hand from the
barrier.\ Similar relations also hold for $\widehat{s}^{~l}$ and $
\widehat{d}^{~l}$. The first boundary condition implies that the
antisymmetric part of Green function is continuous, this is a form
of charge conservation, because the antisymmetric part of Green
function is related to the current directly. But the second
boundary condition means discontinuity in symmetric part of Green
function. This discontinuity is the result of potential barrier as
the interface. In general, $D$ could be be momentum dependent. For
simplicity in our calculations we assumed that $D$ is independent
of the Fermi velocity direction. \subsection{Current-phase
dependencies for Josephson and tangential currents} \label{subb}
Making use of the solution of Eilenberger equations
(\ref{Eilenberger}), we obtain the following expression for the
current density (\ref{Current}) at the slit \cite{Rashedi2}:
\begin{equation}
\mathbf{ j}(y=0,\left| z\right| <a)= 4\pi
eN(0)Tv_{F}\sum\limits_{m>0}\left\langle \widehat{\mathbf{ v}}_F
\left( \frac{{\tilde{\varepsilon}}\Omega -i\eta D\Delta ^{2}\sin
\frac{\phi }{2}\cos \frac{\phi }{2}}{{\tilde{\varepsilon}}
^{2}+\Delta ^{2}\left[1-D{(\sin \frac{ \phi
}{2})}^{2}\right]}\right) \right\rangle_{\mathbf{ \hat{v}}}
\label{j1}
\end{equation}
 where, $\Omega =\sqrt{{\tilde{\varepsilon}
}^{2}+\Delta ^{2}}$, $\mathbf{ \hat{v}}=\mathbf{v}_{F}/v_{F}$ is
the unit vector and $\eta =\mathbf{sgn} (v_{y})$. In the case,
$\mathbf{ v}_{s}\neq 0$, the current (\ref{j1}) has both $\mathbf{
j}_{J} $ and $\mathbf{ j}_{z}$ components.
\begin{figure}
 \includegraphics[width=\columnwidth]{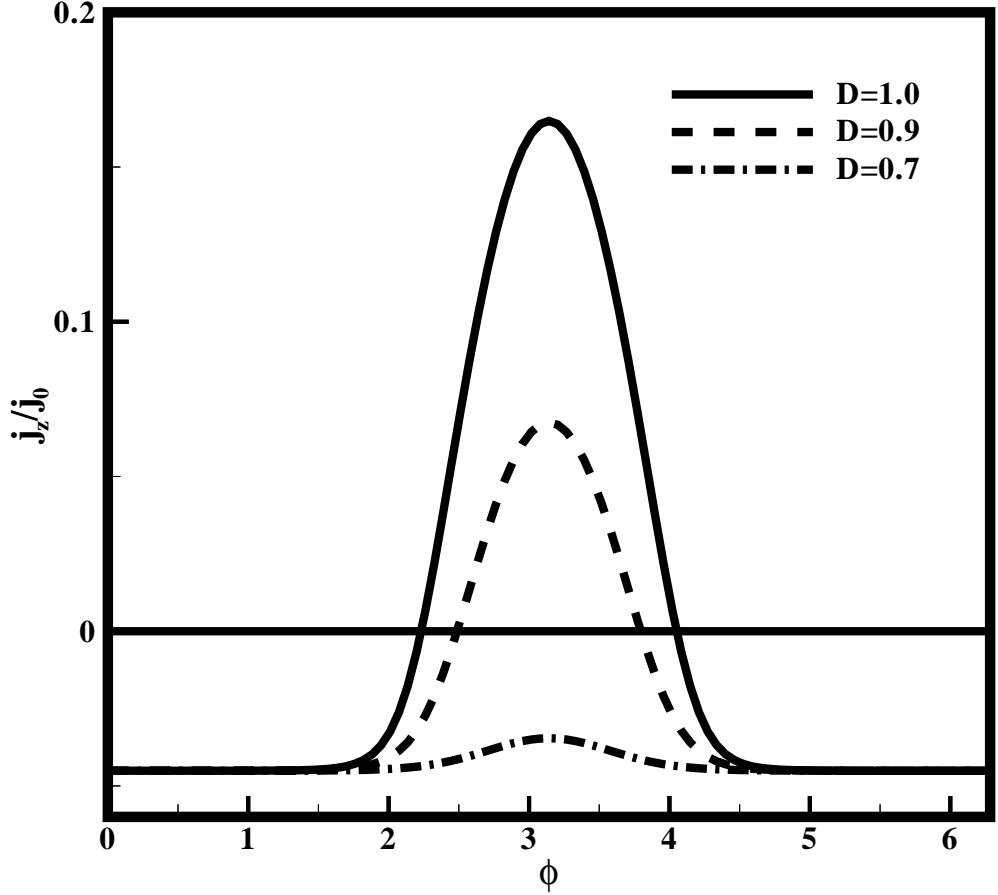}\caption{\label{figa3}Tangential
current $j_{z}$ versus phase $\protect\phi $ for $T/T_c=0.1$ and
$q=0.5$.}
\end{figure}
\begin{figure}
 \includegraphics[width=\columnwidth]{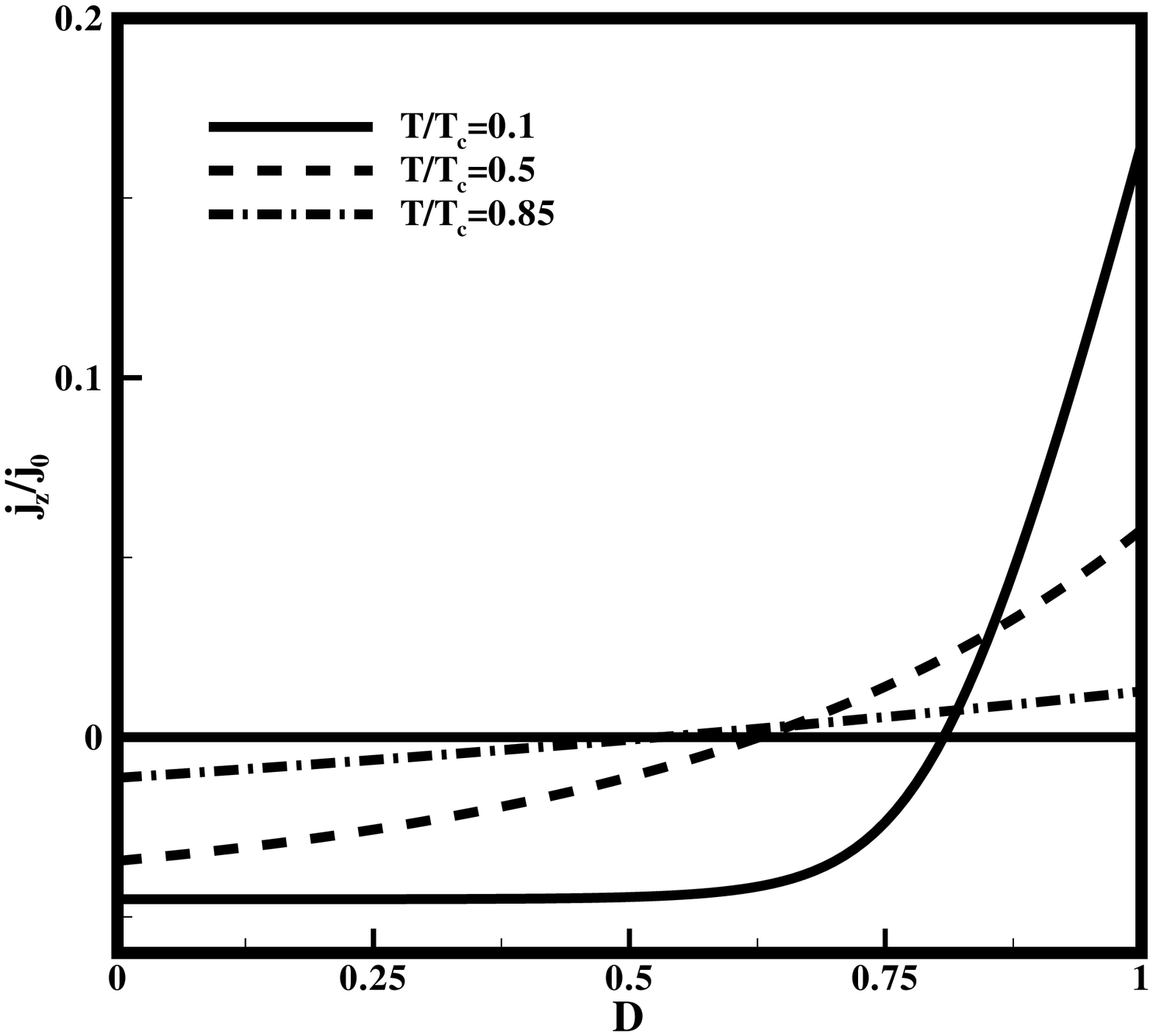}\caption{\label{figa4}Tangential
current $j_z$ versus the transparency $D$ at $\protect\phi=\pi$
and $q=0.5$.}
\end{figure}
The tangential current $\mathbf{ j}_{z}$ depends on the order
parameters phase difference $\phi $ and is not equal to the
transport current $\mathbf{ j}_{T}$ on the banks, in other words
the total current is not equal to the vector sum of Josephson and
transport currents. For the case $\mathbf{ v}_{s}=0$, at the
contact the tangential current is zero and the normal component,
i.e. the Josephson current is as found for the finite transparent
contact in \cite{Zaitsev}. Detaching explicitly the Josephson
current $\mathbf{ j}_{J}$ and the spatially homogeneous
(transport) current $\mathbf{ j}_{T}$ that is produced by the
superfluid velocity $\mathbf{ v}_{s}$, we can write the current as
the sum of three terms: $\mathbf{ j}_{J}$, $\mathbf{ j}_{T}$, and
the ''interference'' current $\mathbf{ j}_{int}$. Also we have
\begin{equation}
\mathbf{ j}=\mathbf{ j}_{J}(\phi ,D,\mathbf{ v}_{s})+\mathbf{
j}_{T}( \mathbf{ v}_{s})+\mathbf{ j}_{int}(\phi ,D,\mathbf{
v}_{s}) \label{detaching}
\end{equation}
The ''interference'' current takes place in the vicinity of the
contact, where both coherent currents $\mathbf{ j}_{J}(\phi )$ and
$\mathbf{ j}_{T}(\mathbf{ v}_{s})$ exist (see also the next
subsection). At first we consider the current density (\ref{j1})
for temperatures close to the critical temperature ($T_{c}-T\ll
T_{c}$). From Eqs. (\ref{j1}) at the contact and for the
temperatures close to the critical temperature we have:
\begin{equation}
\hspace{-.4cm}\mathbf{ j}= j_{0}\sum\limits_{m>0}\left\langle
\widehat{\mathbf{ v}}_F \mathbf{Im}\left(
\frac{\tilde{\varepsilon}\Omega -i\eta D\Delta ^{2}\sin \frac{\phi
}{2}\cos \frac{\phi }{2}}{\tilde{\varepsilon}^{2}+\Delta
^{2}\left[1-D{(\sin \frac{ \phi
}{2})}^{2}\right]}\times\frac{(\tilde{\varepsilon}^{*})
^{2}+\Delta ^{2}\left[1-D{(\sin \frac{ \phi
}{2})}^{2}\right]}{(\tilde{\varepsilon}^{*}) ^{2}+\Delta
^{2}\left[1-D(\sin \frac{\phi}{2})^{2}\right]}\right)
\right\rangle_{\widehat{\mathbf{v}}}
\end{equation}
where, $j_{0}=4\protect\pi \left| e\right| N(0)v_{F}T_{c}$ and
numerator and denominator of the Green function fraction has been
multiplied by the expression ${{(\tilde{\varepsilon}^{*}})
^{2}+\Delta ^{2}\left[1-D{(\sin \frac{ \phi }{2})}^{2}\right]}$,
to escape of the complexity problems. It is well-known that for
the temperatures close to the critical temperature,
$\Delta(T\rightarrow T_c) << T_c$. Consequently, we use the Taylor
expansion in terms of the small parameter
$\frac{\Delta(T\rightarrow T_c)}{T_c}$. So at the contact we have:
\begin{equation}
\mathbf{ j}= j_{0}\sum\limits_{m>0}\left\langle \widehat{\mathbf{
v}}_F
\mathbf{Im}\left((\tilde{\varepsilon}{\tilde{\varepsilon}^{*}})^{2}+\tilde
{\varepsilon}^{2}\Delta^{2}[1-D{(\sin \frac{\phi
}{2})}^{2}]+\frac{1}{2}(1-i\eta
D\sin{\phi})\Delta^{2}({\tilde{\varepsilon}^{*}})^{2}\right)/{\varepsilon_{m}}^{4}
\right\rangle_{\widehat{\mathbf{ v}}}
\end{equation}
we obtain \cite{Rashedi2}:
\begin{equation}
\mathbf{ j}_{J}(\phi ,D,\mathbf{ v}_{s})=\frac{1}{2}AD\sin \phi
\mathbf{ e}_{y}
\end{equation}
\begin{equation}
\mathbf{ j}_{T}(\mathbf{ v}_{s})=-\frac{1}{3}Ak\mathbf{e}_{z},
\label{j1tc}
\end{equation}
\begin{equation}
\mathbf{ j}_{int}(\phi ,D,\mathbf{ v}_{s})=\frac{1}{3}AkD(1-\cos
\phi) \mathbf{ e}_{z}.\label{j2tc}
\end{equation}
where $A=\frac{1}{16}j_{0}\frac{\Delta ^{2}}{T_{c}^{2}}$,\
$k=\frac{ 14\varsigma (3)}{\pi ^{3}}\frac{v_{s}p_{F}}{T_{c}}$,
$\mathbf{ e}_{i}$ is the unit vector in the $i-$direction. This
consideration shows how the current is affected by the interplay
of Josephson and transport currents. At the contact the
''interference'' current $\mathbf{j}_{int}$ is anti-parallel to $
\mathbf{j}_{T}$ and if the phase difference $\phi =\pi $,
$\mathbf{j} _{int}=-2D\mathbf{ j}_{T}$. When there is no phase
difference (at $\phi =0)$, we obtain $j_{int}=0$. So at
transparency values $D$ up to ${\frac{1}{2}}$ the total tangential
current at the contact flows in the opposite direction to the
transport current. Thus, for such $D$ in the vicinity of the
contact, two vortices should exist. At arbitrary temperatures
$T<T_{c}$ the current-phase relations can be analyzed numerically.
In our calculations we
define the parameter, $q$, in which $q=\frac{p_{F}v_{s}}{\Delta _{0}}$ and ${%
\Delta _{0}}=\Delta (T=0,v_{s}=0)$. The value of $q$ can be in the range $%
0<q<q_{c}$ and it's critical value $q_{c}$, corresponds to the
critical current in the homogeneous current state \cite{Bardeen}.
At $T=0$, $q_{c}=1$ and the gap $\Delta $ does not depend on $q$.
In Fig.\ref{figa2} and Fig.\ref{figa3}, we plot the Josephson and
tangential currents at the contact as functions of $\phi $ at
temperatures far from the critical (namely, $ T=0.1T_{c}$) and for
$q=0.5$ and for different values of transparency $D$. Far from
$\phi =\pi $, the tangential current is not disturbed by the
contact, it tends to its value on the bank. The Josephson
current-phase relation is the same as when the transport current
is absent. However, when $ \phi $ tends to $\pi $, for the highly
transparent contact\ $\left( D=1,0.9\right) $ the tangential
current becomes anti-parallel to the bulk current. But for $D=0.7$
the ''interference'' current is strongly suppressed and the
tangential current flows parallel to the bulk current. In
Fig.\ref{figa4}, we plot $%
j_{z}\left( D\right) =j_{T}+j_{int}$ at $\phi =\pi $ for different
temperatures. These plots show that by increasing the temperature
a counter-flow $j_{z}\left( D\right) <0$ exists in a wider
interval of transparency $D_{c}\left( T\right) <D\leq 1$ and
$D_{c}\left( T\rightarrow T_{c}\right) \rightarrow {\frac{1}{2}}.$
This numerical result coincides with analytical results
(\ref{j1tc},\ref{j2tc}).
\begin{figure}
 \includegraphics[width=\columnwidth]{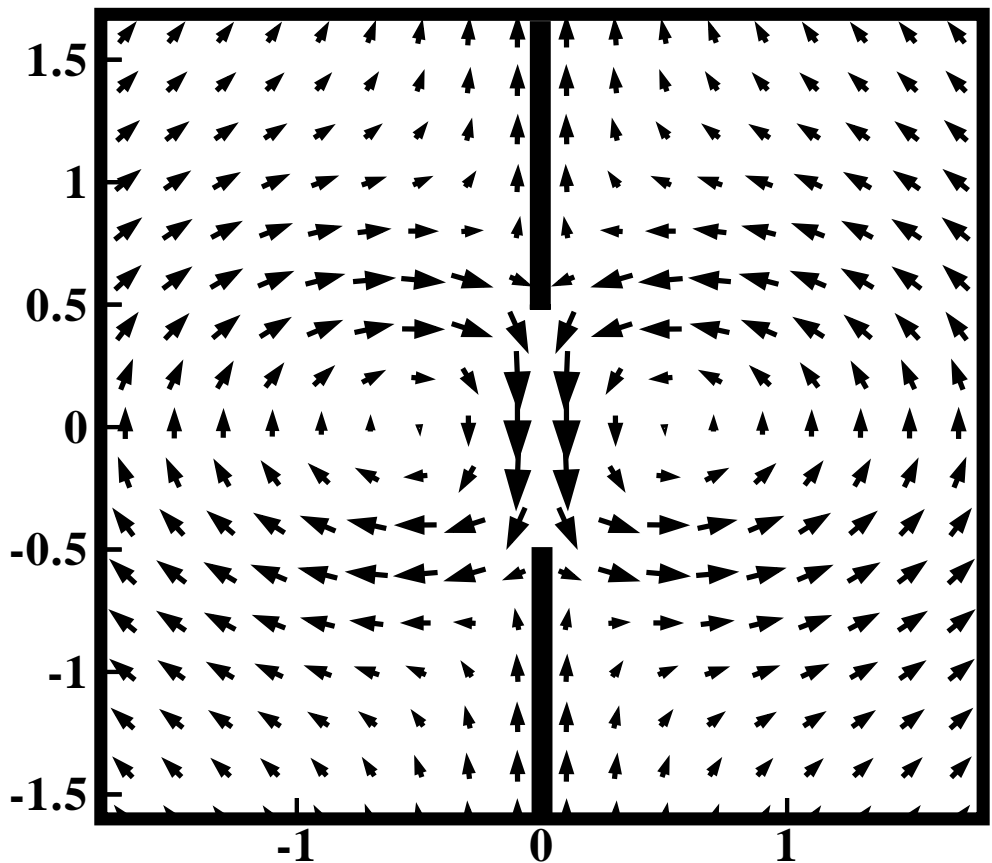}\caption{\label{figa5} Vector plot of
the current
 for $\protect\phi=\protect\pi$, $q=0.5$, $T/T_c=0.1$ and $D=0.95$.
  Axes are marked in units of the contact size $a$.}
\end{figure}
\begin{figure}
 \includegraphics[width=\columnwidth]{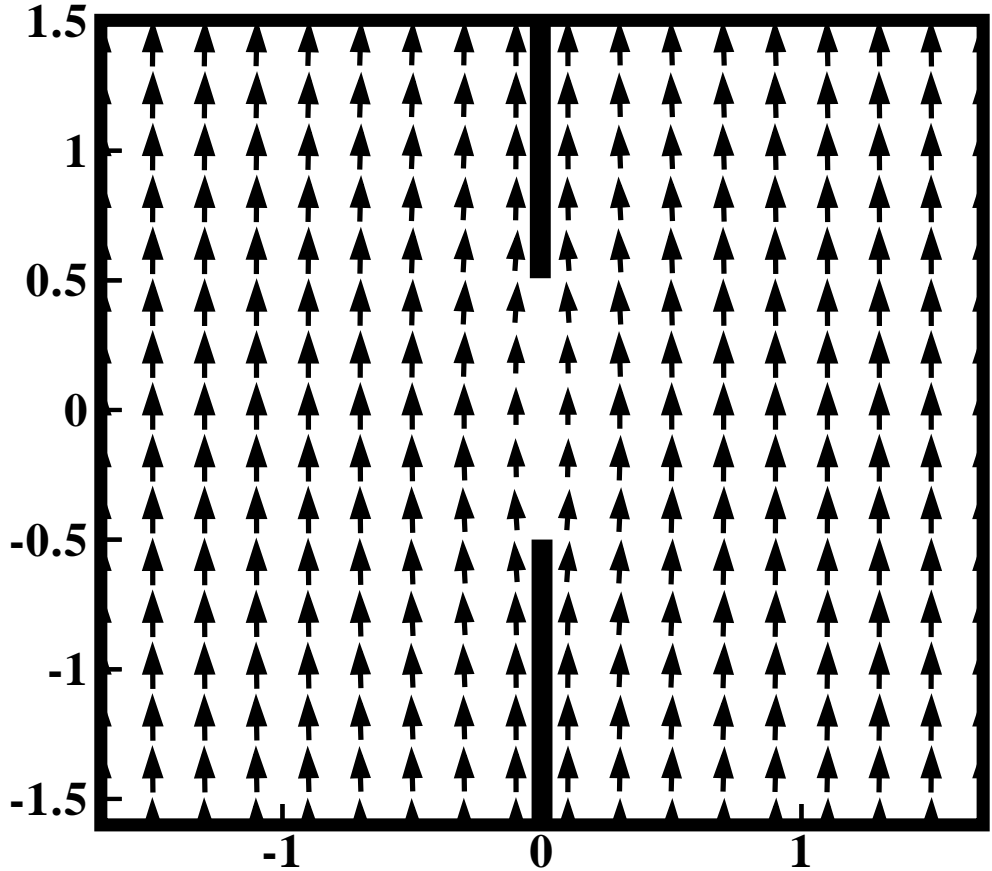}\caption{\label{figa6}Vector plot of
the current for $\protect\phi=\protect\pi$, $q=0.5$, $T/T_c=0.1$
and $D=0.7$.}
\end{figure}
\subsection{Spatial distribution of the current near the contact}
\label{subc} In this subsection we consider the spatial
distribution of the current near the orifice. The superconducting
current (\ref{Current}) can be written as
\begin{equation}
\mathbf{j}(\mathbf{\rho},\mathbf{ v_{s}})=j_{0}\frac{T}{T_{c}}
\sum\limits_{m>0}\left \langle \widehat{\mathbf{ v}}{g_1}(\mathbf{
\rho },\mathbf{ v_{s}})\right\rangle _{\widehat{\mathbf{v_{F}}}},
\label{j}
\end{equation}
where, $\ j_{0}=4\pi \left| e\right| N(0)v_{F}T_{c}$. We should
note that although the current (\ref{j}) depends only on the
coordinates in the film plane, the integration over velocity
directions $\widehat{\mathbf{v}}$ is carried out over all of the
Fermi sphere as in a bulk sample. This method of calculation is
correct only for specular reflection from the film surfaces when
there is no back scattering after electron interaction with them.
 At a point, $\mathbf{ \rho }=(y,z)$, all ballistic trajectories can
be categorized as transit and non-transit trajectories
(see,Fig.\ref{figa1}). For the transit trajectories ''1'' (their
reflected counterparts marked by ''3'' in Fig.\ref{figa1}) a
projection $\widehat{\mathbf{ v}}_{\Vert }$ of the vector
$\widehat{\mathbf{v}}$ to the film plane belongs to the angle at
which the slit is seen from the point $\mathbf{ \rho },$
$\widehat{\mathbf{ v}}_{\Vert }\in \alpha (\mathbf{\rho }),$ and
for non-transit(marked by ''2'' in Fig.\ref{figa1})
$\widehat{\mathbf{{v}}}_{\Vert }\notin \alpha ( \mathbf{\rho )}$.
For transit trajectories the Green functions satisfy boundary
conditions on both banks and at the contact. The non-transit
trajectories should satisfy the specular reflection condition [or
Zaitsev's boundary conditions (\ref{Zaitsev1})-(\ref{Zaitsev2})
for $D=0$ at $y=0,|z|\geq a$]. Then for the current at $T_{c}-T\ll
T_{c}$ we obtain an analytical formula \cite{Rashedi2}:
\begin{equation}
\mathbf{j}(\rho ,\phi ,D,\mathbf{v_s})= j_{c}D\left \langle \sin
\phi \widehat{\mathbf{{v}}}\mathbf{sgn}(v_{y})+k(1-\cos \phi
)\widehat{\mathbf{{v}}} \widehat{{v}_{z}}\right\rangle
_{{\widehat{\mathbf{{v}}}_{\Vert }}\in \alpha }-j_{c}k\left\langle
\widehat{\mathbf{v}}\widehat{{v} _{y}}\right\rangle
_{\widehat{\mathbf{v}}} \label{j_T_c}
\end{equation}
 where, $j_{c}\left( T,\mathbf{v}_{s}\right)
=\frac{\pi |e|N(0)v_{F}}{8}\frac{ \Delta ^{2}\left(
T,\mathbf{v}_{s}\right) }{T_{c}}$. To illustrate how the current
flows near the contact, we plot the Fig.\ref{figa5} and
Fig.\ref{figa6}, for $ \phi =\pi $ and temperatures much smaller
than critical ($T/T_{c}=0.1$), and for different values of
transparency. At such value of the phase $\phi $ there is no
Josephson current and at the large $D=0.95$ the current is
disturbed in such a way that there are two anti-symmetric vortices
close to the orifice (see Fig. \ref{figa5}). For the such
temperature at $D=0.7$ the vortices are absent in Fig.
\ref{figa6}.
\begin{figure}
 \includegraphics[width=\columnwidth]{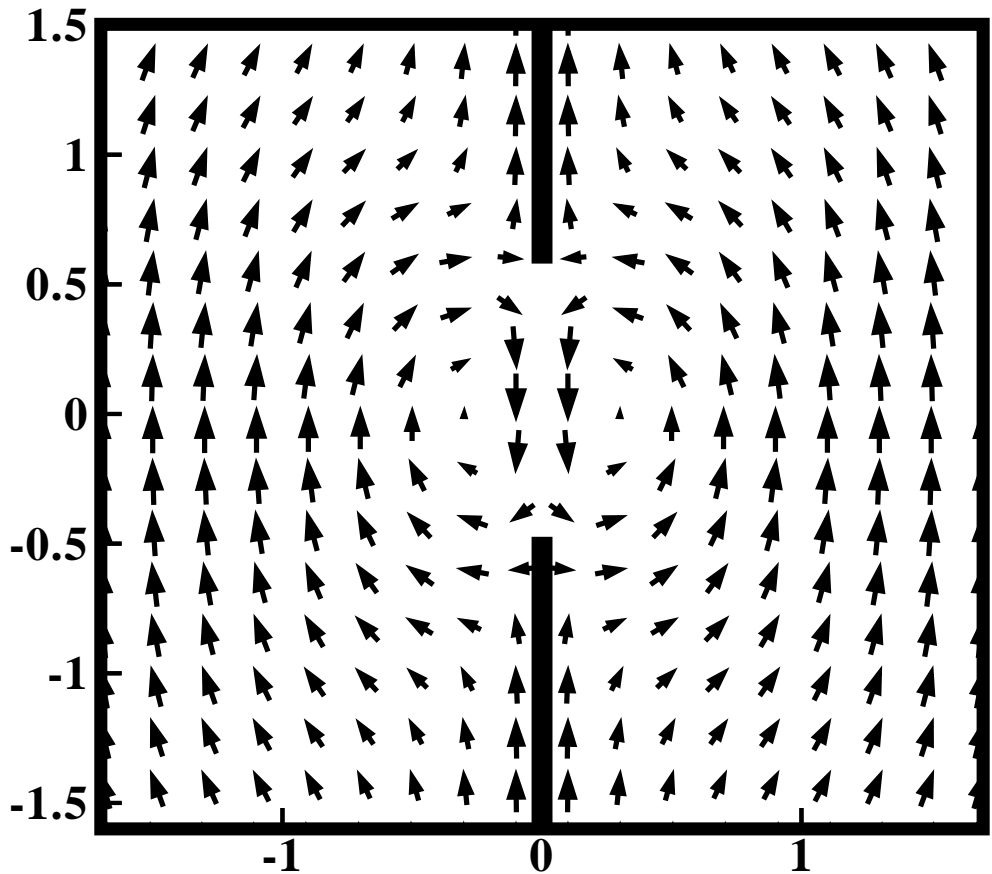}\caption{\label{figa7}Vector plot of
the current for $\protect\phi%
=\protect\pi$, $q=0.5$, $D=0.7$, and $T/T_c=0.85$.}
\end{figure}
Near the critical temperature $\left( T/T_{c}=0.85\right) $ the
vortex-like currents are restored for $D=0.7$ (see
Fig.\ref{figa7}). Far from the orifice (at the distances $l\sim \xi _{0}\gg a$%
) the Josephson current is spread out and the current is equal to
its value at infinity. Considering the current distributions and
current-phase diagrams, we observed that:\newline 1). For fixed
values of temperature and superfluid velocity, by decreasing the
transparency the vortex-like current disappears at $D=D_{c}\left(
T\right) ;$ $0.5\leq D_{c}\left( T\right) <1$\newline 2). For
intermediate values of transparency $D$ $(D_{c}\left( T\right)
<D<1)$ by increasing the temperature the vortex-like currents,
which were destroyed by the effect of electron reflection at the
contact, may be restored.\newline It is clear that both Josephson
and ''interference'' currents are the result of the quantum
interference between two coherent states. By decreasing the
transparency the interference effect will be weaker and these two
currents will decrease, while the transport current will remain
constant. On the other hand, the presence of vortices depends on
the result of competition between transport and ''interference''
current.
\begin{figure}
 \includegraphics[width=\columnwidth]{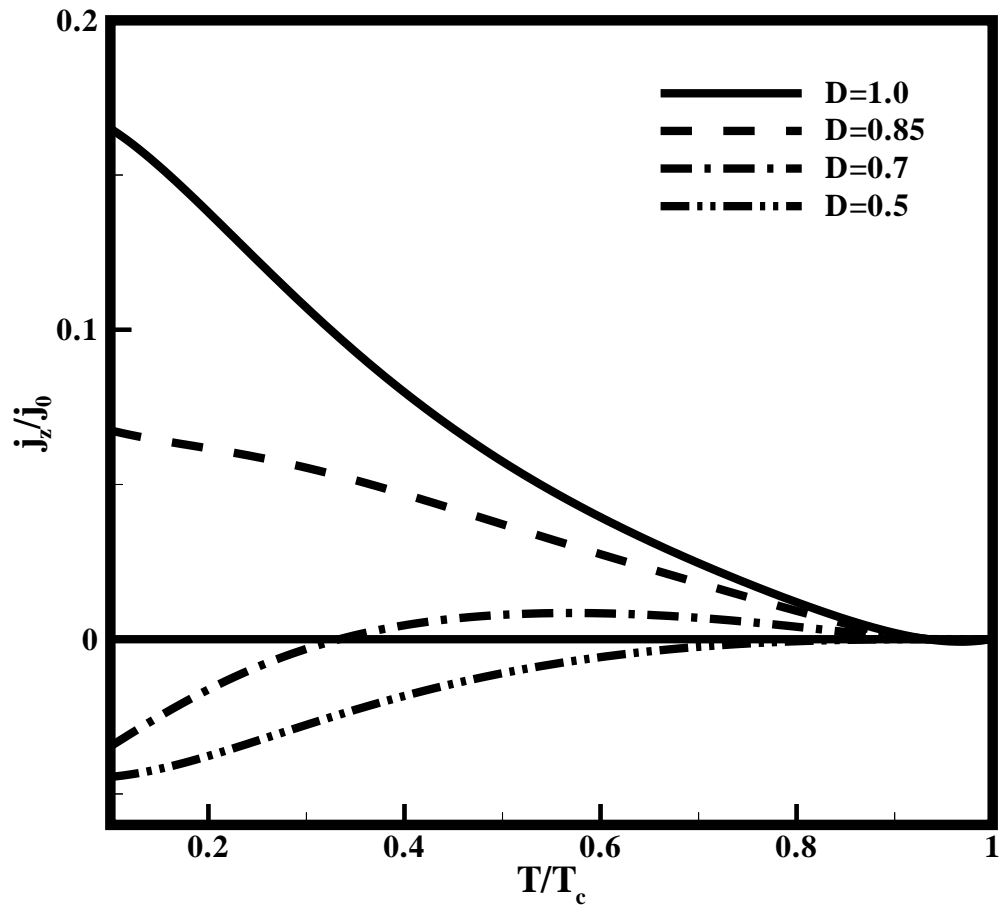}\caption{\label{figa8}Tangential
current $j_z$ versus the temperature $T$ at
$\protect\phi=\protect\pi$ and $q=0.5$.}
\end{figure}
 Thus, by decreasing the transparency the tunneling and
consequently the ''interference'' current will decrease and
vortices may be destroyed (\ref{j1tc},\ref{j2tc}). Similar to the
case $D=1$ in \cite{KOSh}, at high values of transparency, the
''interference'' current can dominate the transport current and
tangential current can be anti-parallel to the transport current,
thus the vortices appear. But for low transparency the tangential
current will be parallel to the transport current and the vortices
disappear.\\
The second point is an anomalous temperature behavior of the
effect. The vortices are the result of the coherent current
mixing. One could expect that by increasing the temperature the
vortices would disappear whereas, for intermediate values of
transparency, by increasing the temperature the vortices will be
restored. As considered in Fig.\ref{figa6} and Fig.\ref{figa7} for
the transparency $D=0.7$ the vortices at low temperature are
absent but at high temperature they are present. In the plots for
tangential current versus transparency, Fig. \ref{figa4} we can
observe this phenomenon (appearance of the counter-flow near the
contact at high temperatures).\\
Usually superconducting currents are monotonic and descendant
functions of temperature. Josephson and transport currents have
this property, but about the tangential current $j_{y}$, the
situation is totally different. At high values of transparency the
$j_{y}$ has similar behavior to the two other currents, but at low
and intermediate values of transparency at $\phi =\pi $ it has a
non-monotonic dependence on the temperature and this is the origin
of the anomalous temperature behavior of vortices. As the
temperature increases, the tangential current first increases and
then decreases. In Fig.\ref{figa8} we plotted the tangential
current ( ''interference''+ transport current) versus the
temperature for different values of transparency. We observed that
for intermediate values of transparency $0.5<D<1$, at low
temperatures and $\phi =\pi $ the tangential current has anomalous
dependence on the temperature. The reason for this dependence is
that the "interference" current flows in the opposite direction to
the transport current. This current is suppressed by the
reflection, but with increasing of the temperature it decreases
slowly than the transport current. As a consequence of that with
increasing of $T$ the tangential current can change its sign and
vortices appear. We found that for low values of transparency
$0<D<0.5$ the ''interference'' current cannot dominate the
transport current and in addition the tangential current has the
same direction as the transport current for any temperature
$T<T_{c}.$
\newpage
\section{Weak link between unitary $f$-wave superconductors}
\label{SpinCharge}
\subsection{Introduction} In this chapter the spin current in the
Josephson junction as a weak-link (interface) between
misorientated triplet superconductors will be investigated
theoretically for the models of the order parameter in $UPt_{3}$.
Green functions of the system will be obtained from the
quasiclassical Eilenberger equations. The analytical results for
the charge and spin currents will be illustrated by numerical
calculations for the certain
misorientation angles of gap vector of superconductors.\\
Triplet superconductivity has become one of the most interesting
topics of condensed matter physics \cite{Maeno,Mackenzie},
particularly in view of the recently discovered ferromagnetic
superconductivity \cite{Mineev,Samokhin}. The mechanism of
pairing, physics of interaction and gap structure in this type of
superconductors have been the subject of many experimental and
theoretical works \cite{Ishida1,Deguchi}. The Cooper pairing in
the triplet superconductors has been investigated, for example,
using the thermal conductivity in papers \cite{Izawa,Graf} and
Knight shift experiments in papers
 \cite{Tou1,Ishida2}. Also, the Josephson effect in the point
contact between triplet superconductors has been studied in paper
\cite{Mahmoodi}. These weak-link structures have been used to
demonstrate the order parameter symmetry in Ref.
\cite{Stefanakis}. Eventually, the $f$-wave
symmetry of order parameter has been proposed for $UPt_{3}$ and $%
Sr_{2}RuO_{4}$ compounds. In addition, the spin polarized
transport through the systems consisting of superconductors,
normal metals, ferromagnetic layers and other structures as one of
the modern topics of mesoscopic physics, has attracted much
attention recently \cite{Imamura,Chtchelkatchev,Sun,Maekawa}. In
this chapter, the ballistic Josephson weak-link as the interface
between two bulk of $f$-wave superconductors with different
orientations of the crystallographic axes has been investigated.
It is shown that the current-phase dependencies are totally
different from the current-phase dependencies of the junction
between conventional ($s$-wave) superconductors \cite{Kulik} and
high $T_{c}$ ($d$-wave) superconductors \cite{Coury}. It is found
that for the certain values of the misorientation, the
spin-current in the both directions, tangential and perpendicular
to the interface, may exist and it has totally unusual dependence
on the external phase difference. The effect of misorientation on
the spin current is investigated. It is observed that the
misorientation between gap vectors is the origin of the spin
current. As the important result of this chapter, it is obtained
that, at some of certain values of phase difference, at which the
charge current is zero, the spin current has the finite value.
Another result of this chapter is the capability of this proposed
experiment for polarization of the spin transport using the
junction between $f$-wave superconductors. Eventually, one of the
states and geometries of our system can be used as a switch which
is able to divide the spin and charge currents into two
parts: parallel and perpendicular to the interface.\\
The arrangement of the rest of this chapter is as follows. In
Sec.(\ref{subd}) we describe our configuration, which is
investigated. For a non-self-consistent model of the order
parameter, the quasiclassiacl Eilenberger equations
\cite{Eilenberger} are solved and suitable Green functions are
obtained analytically. In Sec.(\ref{sube}) the obtained formulas
for the Green functions are used for calculation the charge and
spin current densities at the interface. An analysis of numerical
results will be done in Sec.(\ref{subf}).
 \subsection{Basic Equations}
\label{subd}
\begin{figure}[tbp]
\includegraphics[width=\columnwidth]{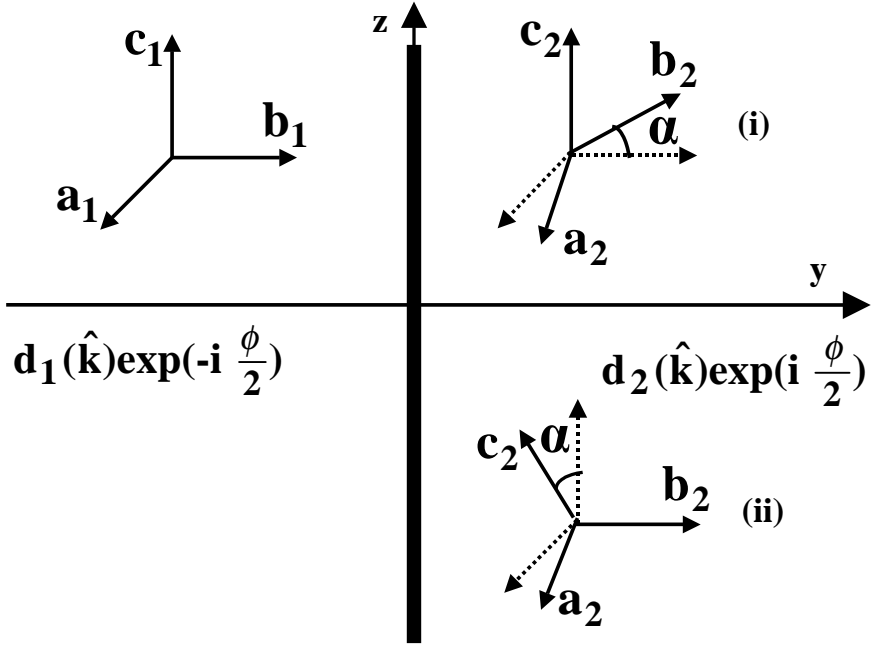}
\caption{Scheme of a flat interface between two superconducting
bulks, which are misorientated as much as $\protect\alpha $.}
\label{figb1}
\end{figure}

We consider a model of a flat interface $y=0$ between
two misorientated $f-$wave superconducting half-spaces
(Fig.\ref{figb1}) as a ballistic Josephson junction. In the
quasiclassical ballistic approach, in order to calculate the
charge and spin current, we use ``transport-like'' equations
\cite{Eilenberger}
for the energy integrated Green functions $\breve{g}\left( \mathbf{\hat{v}}_{F},%
\mathbf{r},\varepsilon _{m}\right) $
\begin{equation}
\mathbf{v}_{F}\nabla \breve{g}+\left[ \varepsilon _{m}\breve{\sigma}_{3}+i%
\breve{\Delta},\breve{g}\right] =0,  \label{Eilenberger-unitary}
\end{equation}
and the normalization condition
\begin{equation}
\breve{g}\breve{g}=\breve{1},  \label{Normalization}
\end{equation}
where $\varepsilon _{m}=\pi T(2m+1)$ are discrete Matsubara energies $%
m=0,1,2...$, $T$ is the temperature and $\mathbf{v}_{F}$ is the Fermi velocity and $\breve{\sigma}_{3}=%
\hat{\sigma}_{3}\otimes \hat{I}$ in which $\hat{\sigma}_{j}\left(
j=1,2,3\right) $ are Pauli matrices.
The Matsubara propagator $\breve{g}$ can be written in the
standard form:
\begin{equation}
\breve{g}=\left(
\begin{array}{cc}
g_{1}+\mathbf{g}_{1}\mathbf{\hat{\sigma}} & \left( g_{2}+\mathbf{g}_{2}\hat{%
\mathbf{\sigma }}\right) i\hat{\sigma}_{2} \\
i\hat{\sigma}_{2}\left( g_{3}+\mathbf{g}_{3}\hat{\mathbf{\sigma
}}\right)  &
g_{4}-\hat{\sigma}_{2}\mathbf{g}_{4}\hat{\mathbf{\sigma
}}\hat{\sigma}_{2}
\end{array}
\right) ,
\end{equation}
\label{Green-function-unitary} where, the matrix structure of the off-diagonal self energy $\breve{%
\Delta}$ in the Nambu space is
\begin{equation}
\breve{\Delta}=\left(
\begin{array}{cc}
0 & \mathbf{d}\hat{\mathbf{\sigma }}i\hat{\sigma}_{2} \\
i\hat{\sigma}_{2}\mathbf{{d^{\ast }}\hat{\sigma}} & 0
\end{array}
\right) .
\end{equation}
\label{order parameter} In this chapter, the unitary states, for which $\mathbf{%
d\times d}^{\ast }=0,$ is investigated. Also, the unitary states
vectors $\mathbf{d}_{1,2}$ can be written as
\begin{equation}
\mathbf{d}_{n}=\mathbf{\Delta }_{n}\exp i\psi _{n},
\end{equation}
where $\mathbf{\Delta }_{1,2}$ are the real vectors in the left
and right sides of the junction.\ The gap (order parameter) vector
$\mathbf{d}$ has to be determined from the self-consistency
equation, near the Fermi surface:
\begin{equation}
\mathbf{d}\left( \mathbf{\hat{v}}_{F},\mathbf{r}\right) =\pi
TN\left(
0\right) \sum_{m}\left\langle V\left( {\mathbf{\hat{v}}}_{F},{\mathbf{\hat{v}%
}}_{F}^{\prime }\right) \mathbf{g}_{2}\left(
{\mathbf{\hat{v}}}_{F}^{\prime },\mathbf{r},\varepsilon
_{m}\right) \right\rangle   \label{self-consistent}
\end{equation}
where $V\left(
{\mathbf{\hat{v}}}_{F},{\mathbf{\hat{v}}}_{F}^{\prime }\right) $,
is a potential of pairing interaction, $\left\langle
...\right\rangle $ stands for averaging over the directions of an
electron momentum on the Fermi surface
${\mathbf{\hat{v}}}_{F}^{\prime }$ and $N\left( 0\right) $ is the
electron density of states at the Fermi level of energy.
\begin{figure}[tbp]
\includegraphics[width=\columnwidth]{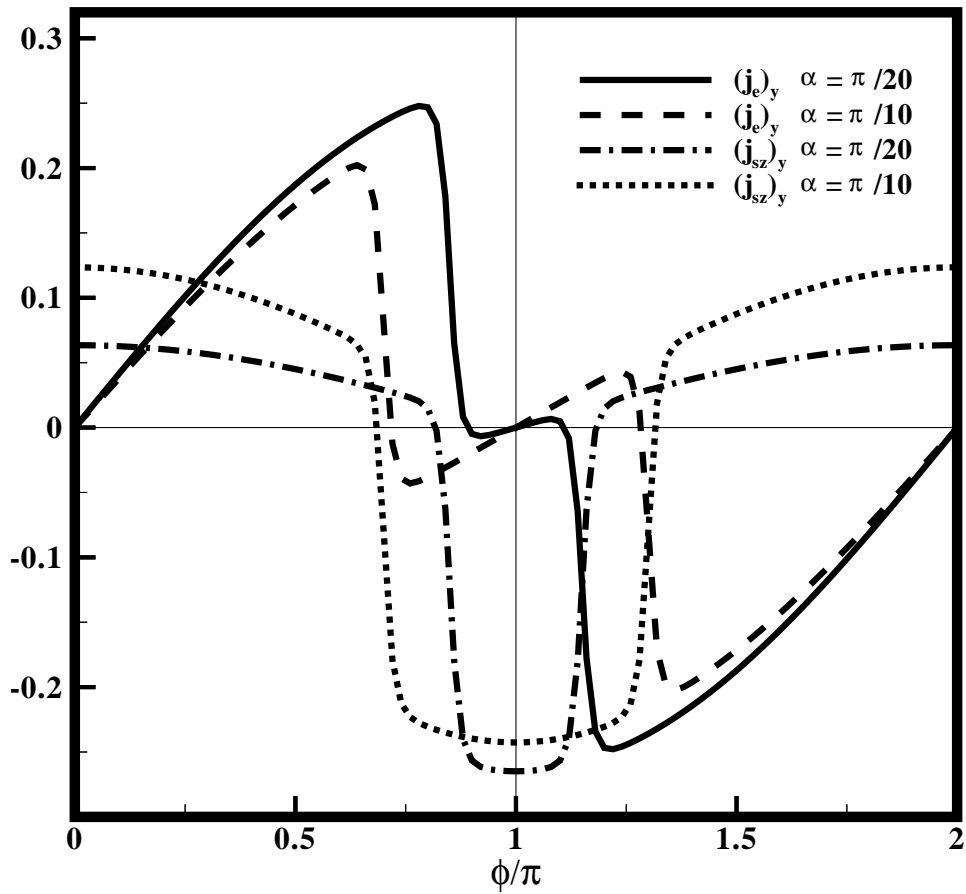}
\caption{Charge and spin current ($s_{z}$) versus the phase
difference $\protect\phi $ for the planar state (\ref{planar}),
geometry (i) and the different misorientations. Currents are given
in units of $j_{0,e}=\frac{\protect\pi }{2}eN(0)v_{F}\Delta
_{0}(0) $ and $j_{0,s}=\frac{\protect\pi }{4}\hbar N(0)v_{F}\Delta
_{0}(0)$ respectively.} \label{figb2}
\end{figure}
Solutions to Eqs. (\ref{Eilenberger-unitary}) and
(\ref{self-consistent}) must satisfy the conditions for Green
functions and vector $\mathbf{d}$ in the bulks of the
superconductors far from the interface as follow:
\begin{eqnarray}
\breve{g}\left( \pm \infty \right)  &=&\frac{\varepsilon _{m}\breve{\sigma}%
_{3}+i\breve{\Delta}_{2,1}}{\sqrt{\varepsilon _{m}^{2}+\left| \mathbf{d}%
_{2,1}\right| ^{2}}};  \label{Bulk solution} \\
\mathbf{d}\left( \pm \infty \right)  &=&\mathbf{d}_{2,1}\left( \mathbf{\hat{v%
}}_{F}\right) \exp \left( \mp \frac{i\phi }{2}\right) ,
\label{Bulk order parameter}
\end{eqnarray}
where $\phi $ is the external phase difference between the order
parameters of the bulks. Eqs. (\ref{Eilenberger-unitary}) and
(\ref{self-consistent}) have to be supplemented by the continuity
conditions at the interface between superconductors. For all
quasiparticle trajectories, the Green functions satisfy the
boundary conditions both in the right and left bulks as well as at
the interface.\\
The set of equations (\ref{Eilenberger-unitary}) and
(\ref{self-consistent}) can be solved only numerically. For
unconventional superconductors such solution
requires the information of the function $V\left( {\mathbf{\hat{v}}}_{F},{%
\mathbf{\hat{v}}}_{F}^{\prime }\right)$. This information, as that
of the nature of unconventional superconductivity in novel
compounds, in most cases is unknown. Usually, the spatial
variation of the order parameter and its dependence on
the momentum direction can be separated in the form of $\Delta ({\mathbf{%
\hat{v}}}_{F},y)=\Delta ({\mathbf{\hat{v}}}_{F})\Psi (y)$. It has
been shown that the absolute value of a self-consistent order
parameter and $\Psi (y)$ are suppressed near the interface and at
the distances of the order of the coherence length, while its
dependence on the direction in the momentum space ($\Delta
({\mathbf{\hat{v}}}_{F})$) remains unaltered \cite{Barash}.
Consequently, this suppression doesn't influence the Josephson
effect drastically. This suppression of the order parameter keeps
the current-phase dependence unchanged but, it changes the
amplitude value of the current. For example, it has been verified
in Ref. \cite{Coury} for the junction between unconventional
$d$-wave, in Ref. \cite{Barash} for the case of ``$f$-wave''
superconductors and in Refs. \cite{Thuneberg,Viljas} for pinholes
in $^{3}He$ that, there is a good qualitative agreement between
self-consistent and non-self-consistent results. Also, it has been
observed that the results of the non-self-consistent investigation
of ferromagnet-d-wave proximity structure in Ref. \cite{Faraii}
are coincident with the experimental results of the paper
\cite{Freamat} and the results of the non-self-consistent model in
paper \cite{Yip} are similar to the superfluid weak-link
experiment \cite{Backhaus}. In the paper \cite{Faraii}, they have
investigated the proximity effect between a ferromagnet and a
high$-T_c$ superconductor.
\begin{figure}[tbp]
\includegraphics[width=\columnwidth]{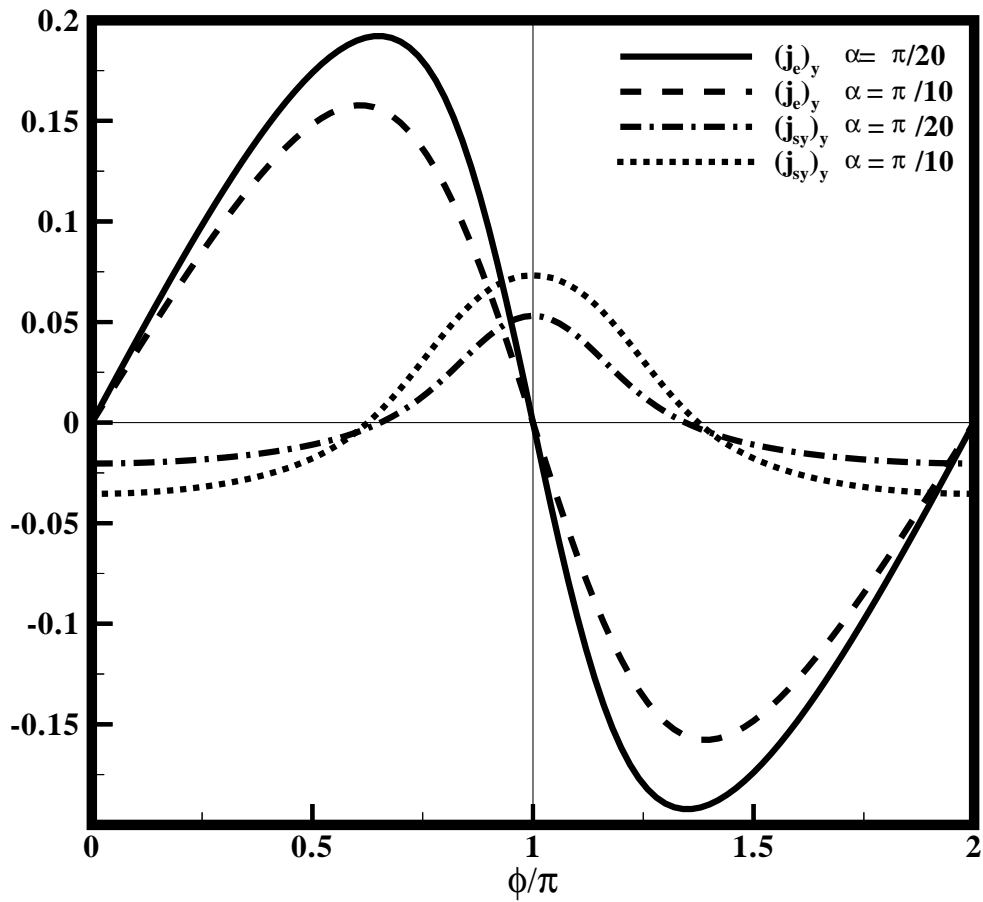}
\caption{Charge and spin current ($s_{y}$) versus the phase
difference $\protect\phi $ for the axial state (\ref{axial}),
geometry (ii) and the different misorientations ($y$-component).}
\label{figb3}
\end{figure}
They have solved the Eilenberger equation and using the obtained
Green function investigated the Andreev bound states. The density
of states in this system has been studied and the spatial
oscillations have been observed in this nonselfconsistent paper.
The results of this paper had been observed before in an
experimental report in paper \cite{Freamat}. In addition, there
are many published papers \cite{Thuneberg,Viljas,Faraii} and
\cite{Beasley,Hirai,Inoue,Asano} in which, such approximation has
been used for different systems containing unconventional
superconductors and important analytical results have been
obtained. In Refs. \cite{Thuneberg,Viljas,Faraii} Eilenberger
equation has been solved and Bogoliobov- Degennes equation has
been considered in papers \cite{Beasley,Hirai,Inoue,Asano}
non-self-consistently. Consequently, despite the fact that this
estimation cannot be applied directly for a quantitative analyze
of the real experiment, only a qualitative comparison of
calculated and experimental current-phase relations is possible.
In our calculations, a simple model of the constant order
parameter up to the interface is considered and the pair breaking
and the scattering on the interface are ignored. We believe that
under these strong assumptions our results describe the real
situation qualitatively. In the framework of such model, the
analytical expressions for the charge and spin current can be
obtained for an arbitrary form of the order parameter. Also, we
have done our calculations for the small misorientations
$\alpha=\frac{\pi}{20}$, $\alpha=\frac{\pi}{15}$ and
$\alpha=\frac{\pi}{10}$. As the results of paper \cite{Coury}, for
the small misorientations, selfconsistent and nonselfconsistent
calculations have the same results approximately. Consequently,
authors of paper \cite{Coury} concluded that the nonselfconsistent
formalism can be used for the junction between unconventional
superconducting bulks with small misorientations.

\subsection{Analytical Green functions} \label{sube}The solution of
Eqs. (\ref{Eilenberger-unitary}) and (\ref{self-consistent})
allows us to calculate the charge and spin current densities. The
expression for the charge current is:
\begin{equation}
\mathbf{j}_{e}\left( \mathbf{r}\right) =2i\pi eTN\left( 0\right)
\sum_{m}\left\langle \mathbf{v}_{F}g_{1}\left( \mathbf{\hat{v}}_{F},\mathbf{r%
},\varepsilon _{m}\right) \right\rangle
\label{charge-current-unitary}
\end{equation}
and for the spin current we have:
\begin{equation}
\mathbf{j}_{s_{i}}\left( \mathbf{r}\right) =2i\pi (\frac{\hbar
}{2})TN\left(
0\right) \sum_{m}\left\langle \mathbf{v}_{F}\left( \mathbf{{\hat{e}}}_{i}%
\mathbf{g}_{1}\left( \mathbf{\hat{v}}_{F},\mathbf{r},\varepsilon
_{m}\right) \right) \right\rangle  \label{spin-current}
\end{equation}
where, $\mathbf{{\hat{e}}}_{i}\mathbf{=}\left( \hat{\mathbf{x}},\hat{%
\mathbf{y}},\hat{\mathbf{z}}\right) $. We assume that the order
parameter
does not depend on coordinates and in each half-space it equals to its value (%
\ref{Bulk order parameter}) far from the interface in the left or
right bulks. For such a model, the current-phase dependence of a
Josephson junction can be calculated analytically. It enables us
to analyze the main features
of current-phase dependence for the different models of the order parameter of ``%
$f$-wave'' superconductivity. The Eilenberger equations
(\ref{Eilenberger-unitary}) for Green functions $\breve{g}$, which
are supplemented by the condition of continuity of solutions
across the interface, at $y=0$, and the boundary conditions at the
bulks, should be solved for a non-self-consistent model of the
order parameter analytically.
In a ballistic case the system of equations for functions $g_{i}$ and $%
{\mathbf g}_{i}$ can be decomposed on independent blocks of
equations. The set of equations which enables us to find the Green
function $g_{1}$ is:
\begin{eqnarray}
v_{F}\hat{{\mathbf k}}\nabla g_{1}=i\left({\mathbf d}\cdot{\mathbf
g}_{3}-{\mathbf d}^{\ast }\cdot{\mathbf g}_{2}
\right);  \label{au} \\
v_{F}\hat{{\mathbf k}}\nabla {\mathbf g}_{-}=-2\left( {\mathbf
d\times g}_{3}+{\mathbf d
}^{\ast }{\mathbf \times g}_{2}\right);  \label{bu} \\
v_{F}\hat{{\mathbf k}}\nabla {\mathbf g}_{2}=-2\varepsilon
_{m}{\mathbf g}
_{2}+2ig_{1}{\mathbf d}+{\mathbf d}\times {\mathbf g}_{-};  \label{cu}\\
v_{F}\hat{{\mathbf k}}\nabla {\mathbf g}_{3}=2\varepsilon
_{m}{\mathbf g} _{3}-2ig_{1}{\mathbf d}^{\ast }+{\mathbf d}^{\ast
}\times {\mathbf g}_{-}; \label{du}
\end{eqnarray}
where ${\mathbf g}_{-}={\mathbf g}_{1}-{\mathbf g}_{4}.$ The Eqs.
(\ref{au})-(\ref{du}) can be solved by integrating over ballistic
trajectories  of electrons in the\ right and left half-spaces. The
general solution satisfying the boundary conditions (\ref{Bulk
solution}) at infinity is
\begin{eqnarray}
g_{1}^{\left( n\right) } &=&\frac{\varepsilon _{m}}{\Omega
_{n}}+a_{n}\exp
\left( -2s\Omega _{n}t\right) ;  \label{eu} \\
{\mathbf g}_{-}^{\left( n\right) } &=&{\mathbf C}_{n}\exp \left(
-2s\Omega
_{n}t\right) ;  \label{fu} \\
{\mathbf g}_{2}^{\left( n\right) } &=&\frac{i{\mathbf
d}_{n}}{\Omega _{n}}-\frac{2ia_{n}{\mathbf d}_{n}+{\mathbf
d}_{n}\times {\mathbf C}_{n}}{2s\eta \Omega _{n}-2\varepsilon
_{m}}\exp \left( -2s\Omega
_{n}t\right) ;  \label{gu} \\
{\mathbf g}_{3}^{\left( n\right) } &=& \frac{i {\mathbf
d}_{n}^{\ast
}}{\Omega _{n}}+\frac{2ia_{n}{\mathbf d}_{n}^{\ast }-{\mathbf d}%
_{n}^{\ast }\times {\mathbf C}_{n}}{2s\eta \Omega
_{n}+2\varepsilon _{m}}\exp \left( -2s\Omega _{n}t\right);
\label{hu}
\end{eqnarray}
where $t$ is time of flight along the trajectory, $sgn\left(
t\right) =sgn\left( y\right) =s$ and $\eta =sgn\left(
v_{y}\right).$  By matching the solutions (\ref{eu}-\ref{hu}) at
the interface $\left(y=0, t=0\right) $, we find constants $a_{n}$
and ${\mathbf C_{n}}.$ Indices $n=1,2$ label the left and right
half-spaces respectively. The function $g_{1}\left( 0\right)
=g_{1}^{\left( 1\right) }\left( -0\right) =g_{1}^{\left( 2\right)
}\left( +0\right) ,$ which is a diagonal term of Green matrix and
determines the current density at the interface, $y=0$, is as
follows: Two diagonal terms of Green matrix which determine the
current densities at the interface, $y=0$, are following.
\begin{figure}[tbp]
\includegraphics[width=\columnwidth]{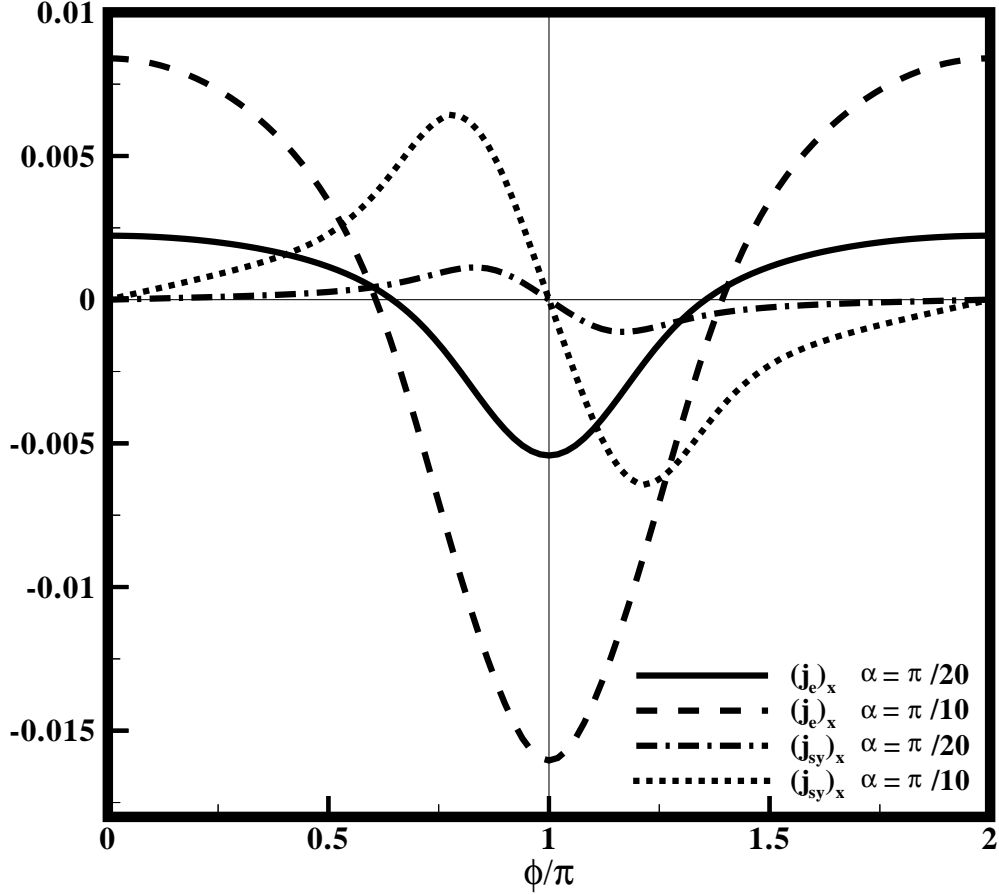}
\caption{Charge and spin current ($s_{y}$) versus the phase
difference $\protect\phi $ for
the axial state (\ref{axial}), geometry (ii) and the different misorientations ($x$%
-component)} \label{figb4}
\end{figure}
For the relative term to the charge current we obtain:
\begin{equation}
g_{1}\left( 0\right) =\frac{\varepsilon _{m}(\Omega _{1}+\Omega
_{2})\cos\beta+i\eta\sin\beta(\Omega _{1} \Omega _{2}+\varepsilon
_{m}^2)}{i\eta\sin\beta\varepsilon _{m}(\Omega _{1}+\Omega
_{2})+\cos\beta(\Omega _{1}\Omega _{2}+ \varepsilon
_{m}^2)+\mathbf{\Delta}_{1}\mathbf{\Delta}_{2}}
\label{charge-term}
\end{equation}
and for the case of spin current we have:
\begin{equation}
\mathbf{g_{1}}\left( 0\right) =\label{spin-term}
\end{equation}
$$\mathbf{M}[(B-1)^{2}\exp(i\beta)(\eta \Omega _{1}+\varepsilon
_{m})(\eta \Omega _{2}+\varepsilon _{m})
-(B+1)^{2}\exp(-i\beta)(\eta \Omega _{2}-\varepsilon _{m})(\eta
\Omega _{1}-\varepsilon _{m})]
$$
where $\eta =sgn\left( v_{y}\right) $, $\Omega
_{n}=\sqrt{\varepsilon _{m}^{2}+\left| \mathbf{d}_{n}\right|
^{2}}$, $\beta=\psi_1-\psi_2+\phi$,
\begin{equation}
B=\frac{\eta\varepsilon _{m}(\Omega _{1}+\Omega
_{2})\cos\beta+i\sin\beta(\Omega _{1} \Omega _{2}+\varepsilon
_{m}^2)}{i\eta\sin\beta\varepsilon _{m}(\Omega _{1}+\Omega
_{2})+\cos\beta(\Omega _{1}\Omega _{2}+ \varepsilon
_{m}^2)+\mathbf{\Delta}_{1}\mathbf{\Delta}_{2}},
\label{mathematics-term}
\end{equation}
\begin{equation}
A=\frac{\mathbf{\Delta}_1\mathbf{\Delta}_2(B-1)\exp(i\beta)}{(\eta
\Omega _{1}-\varepsilon _{m})(\eta \Omega _{2}-\varepsilon
_{m})}+\frac{\mathbf{\Delta}_1\mathbf{\Delta}_2(B+1)\exp(-i\beta)}{(\eta
\Omega _{1}+\varepsilon _{m})(\eta \Omega _{2}+\varepsilon _{m})}
\end{equation}
and
\begin{equation}
\mathbf{M}=\frac{\eta\mathbf{\Delta}_{1}\times\mathbf{\Delta}_{2}}{(A+2B)\left|
\mathbf{d}_{1}\right| ^{2}\left| \mathbf{d}_{2}\right| ^{2}}
\end{equation}
  Also, $n=1,2$ label the left and right
half-spaces respectively.
\begin{figure}[tbp]
\includegraphics[width=\columnwidth]{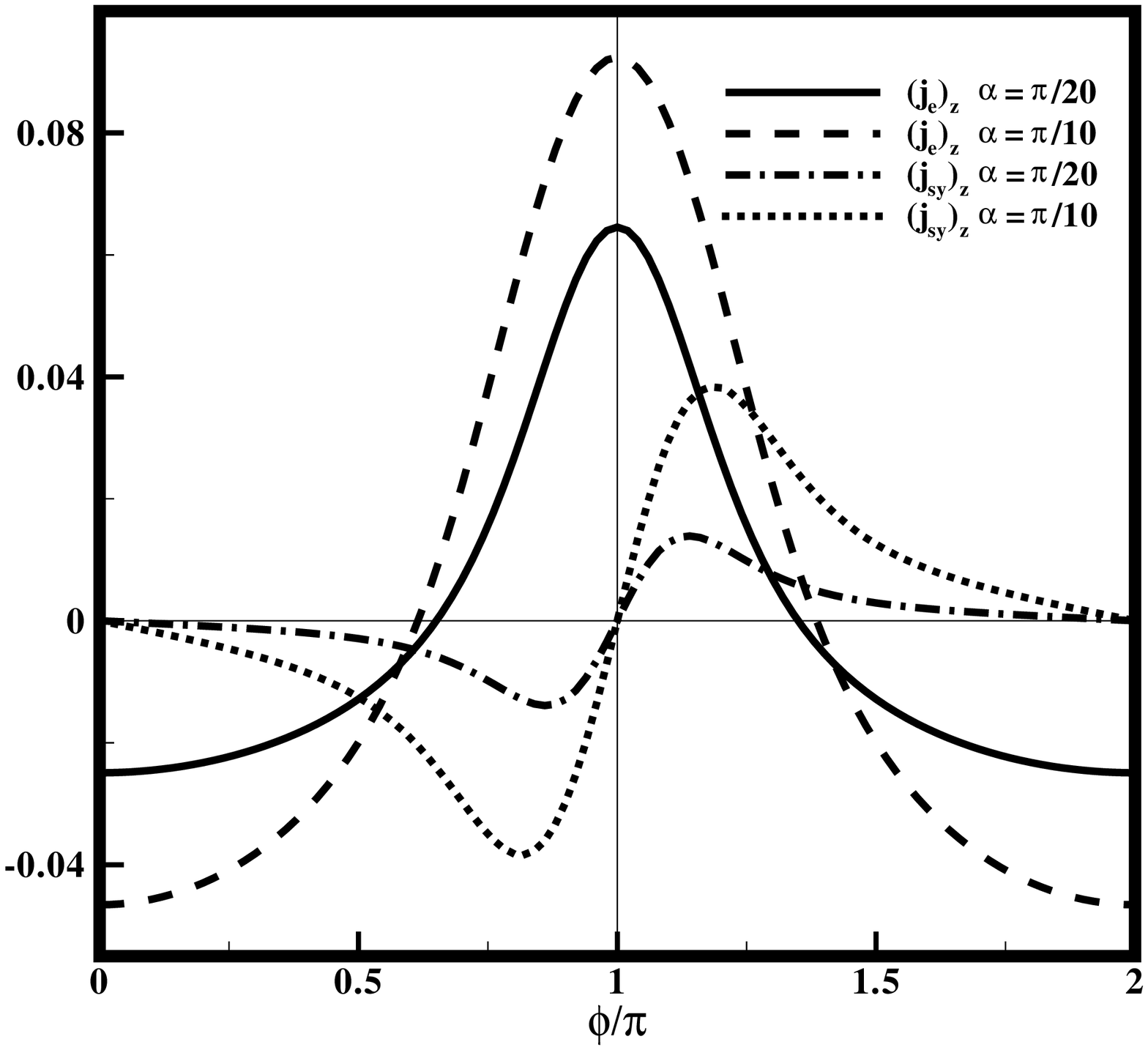}
\caption{Charge and spin current ($s_{y}$) versus the phase
difference $\protect\phi $ for
the axial state (\ref{axial}), geometry (ii) and the different misorientations ($z$%
-component)} \label{figb5}
\end{figure}
We consider a rotation $\breve{R}$ only in the right
superconductor (see, Fig.\ref{figb1}), (i.e., $\mathbf{d}_{2}(\hat{\mathbf{k}}) =\breve{R}\mathbf{d}%
_{1}( \breve{R}^{-1}\hat{\mathbf{k}}),$ $\hat{\mathbf{k}}$ is the
unit vector in the momentum space). The crystallographic
$\mathbf{c}$-axis in the left half-space is selected parallel to
the partition between the superconductors (along $\mathbf{z}$-axis
in Fig.\ref{figrev2}). To illustrate the results obtained by
computing the formula (\ref{charge-term},\ref{spin-term}), we plot
the
current-phase diagrams for the different models of the ``$f$-wave'' pairing symmetry (%
\ref{axial},\ref{planar}) and for two different geometries. These
two geometries are corresponding to the different orientations of
the crystals in the right and left sides of the interface (see,
Fig.\ref{figb1}):\newline (i) The basal $ab$-plane in the right
side has
been rotated around the $c$-axis by $\alpha $; $\hat{\mathbf{c}}_{1}\Vert \hat{\mathbf{c}}_{2}$.%
\newline
(ii) The $c$-axis in the right side has been rotated around the axis perpendicular to the interface ($y$%
-axis in Fig.\ref{figb1}) by $\alpha $; $\hat{\mathbf{b}}_{1}\Vert
\hat{\mathbf{b}}_{2}$.\newline
Further calculations require a certain model of the gap vector (vector of order parameter) $%
\mathbf{d}$. \subsection{Numerical results} \label{subf}
 In this chapter, two most possible forms of the
$f$-wave order parameter vector in $UPt_{3}$ are considered. The
first model which is successful to explain the properties of the
$B$-phase of $UPt_{3}$ is the axial state. This sate describes the
strong spin-orbital coupling with vector \textbf{d} directed along
the \textbf{c} axis of the lattice and it is:
\begin{equation}
\mathbf{d}(T,\mathbf{v}_{F})=\Delta
_{0}(T)\hat{\mathbf{z}}k_{z}\left( k_{x}+ik_{y}\right) ^{2}.
\label{axial}
\end{equation}
The coordinate axes
$\hat{\mathbf{x}},\hat{\mathbf{y}},\hat{\mathbf{z}}$
here and below are chosen along the crystallographic axes $\hat{\mathbf{a}},%
\hat{\mathbf{b}},\hat{\mathbf{c}}$ in the left side of
Fig.\ref{figb1}. The function $ \Delta _{0}=$ $\Delta _{0}\left(
T\right) $ in Eq. (\ref{axial})
and below describes the dependence of the order parameter $\mathbf{d}$ on the temperature $%
T$ (our numerical calculations have been done at the temperatures
close to the $T=0$). The second model of the order parameter which
describes the weak spin-orbital coupling in $UPt_{3}$ states, is
the unitary planar state. The planar model of gap vector is:
\begin{equation}
\mathbf{d}(T,\mathbf{v}_{F})=\Delta
_{0}(T)k_{z}(\hat{\mathbf{x}}\left( k_{x}^{2}-k_{y}^{2}\right)
+\hat{\mathbf{y}}2k_{x}k_{y}). \label{planar}
\end{equation}
Using these two models of order parameters
(\ref{axial},\ref{planar}) and solutions to the Eilenberger
equations (\ref{charge-term}) and (\ref{spin-term}), we have
calculated the spin current and charge current densities at the
interface numerically.
\begin{figure}[tbp]
\includegraphics[width=\columnwidth]{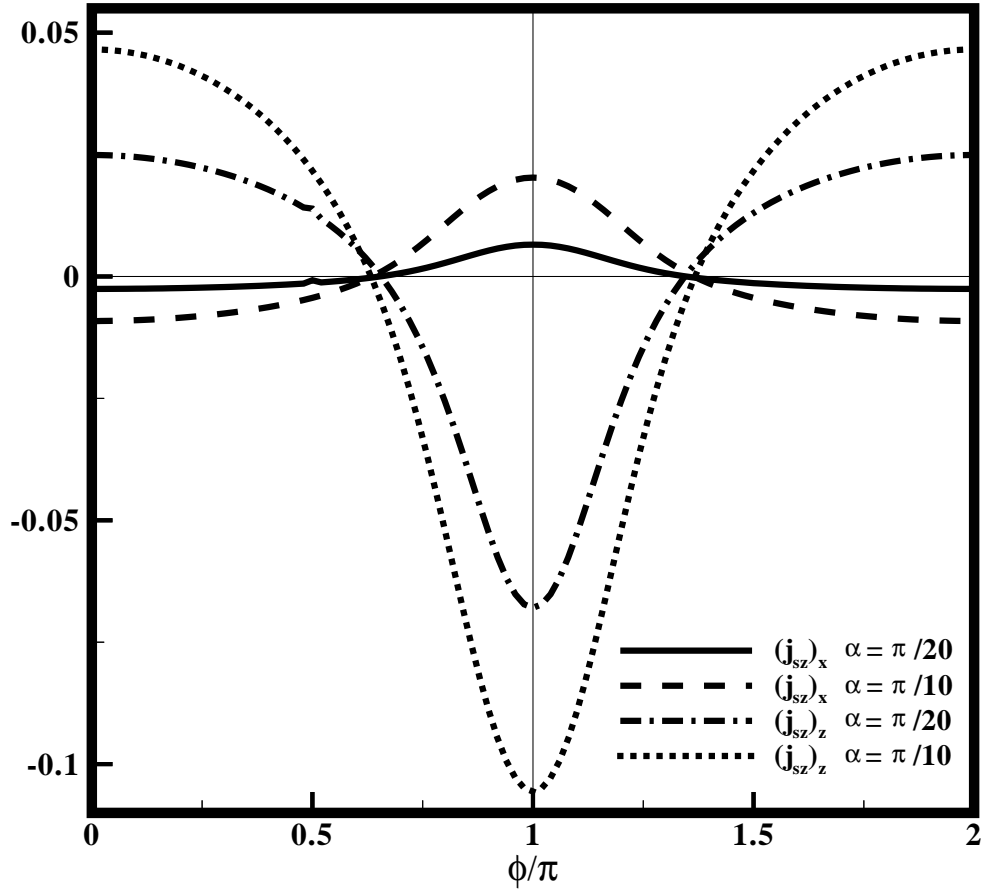}
\caption{Tangential spin current ($s_{z}$) versus the phase
difference $\protect\phi $ for the planar state (\ref{planar}),
geometry (ii) and the different misorientations. The perpendicular
component ($y$-direction) of the spin current is absent.}
\label{figb6}
\end{figure}
These numerical results are listed below:\newline 1) The spin
current can be present, only when misorientation between gap
vectors exists. Because in our Green function (\ref{spin-term}),
the spin current is proportional to the ``cross product'' between
the left and right gap vectors. For instance, the spin current for
the case of the axial state (\ref{axial}) and
geometry (i) is zero, because both of the gap vectors are in the same direction (%
$\mathbf{{\hat{z}}}$). (Geometry (i) is a rotation as much as
$\alpha $, around the $z$ axis).\newline 2) In Fig.\ref{figb2} it
is shown that for the planar state and geometry (i), it is
possible to observe the current of $s_{z}$ in the direction
perpendicular to the interface, but in
Figs.\ref{figb3},\ref{figb4} and \ref{figb5}, it is demonstrated
that, for the axial state and geometry (ii), only the current of
$s_{y}$ can be observed.
\begin{figure}[tbp]
\includegraphics[width=\columnwidth]{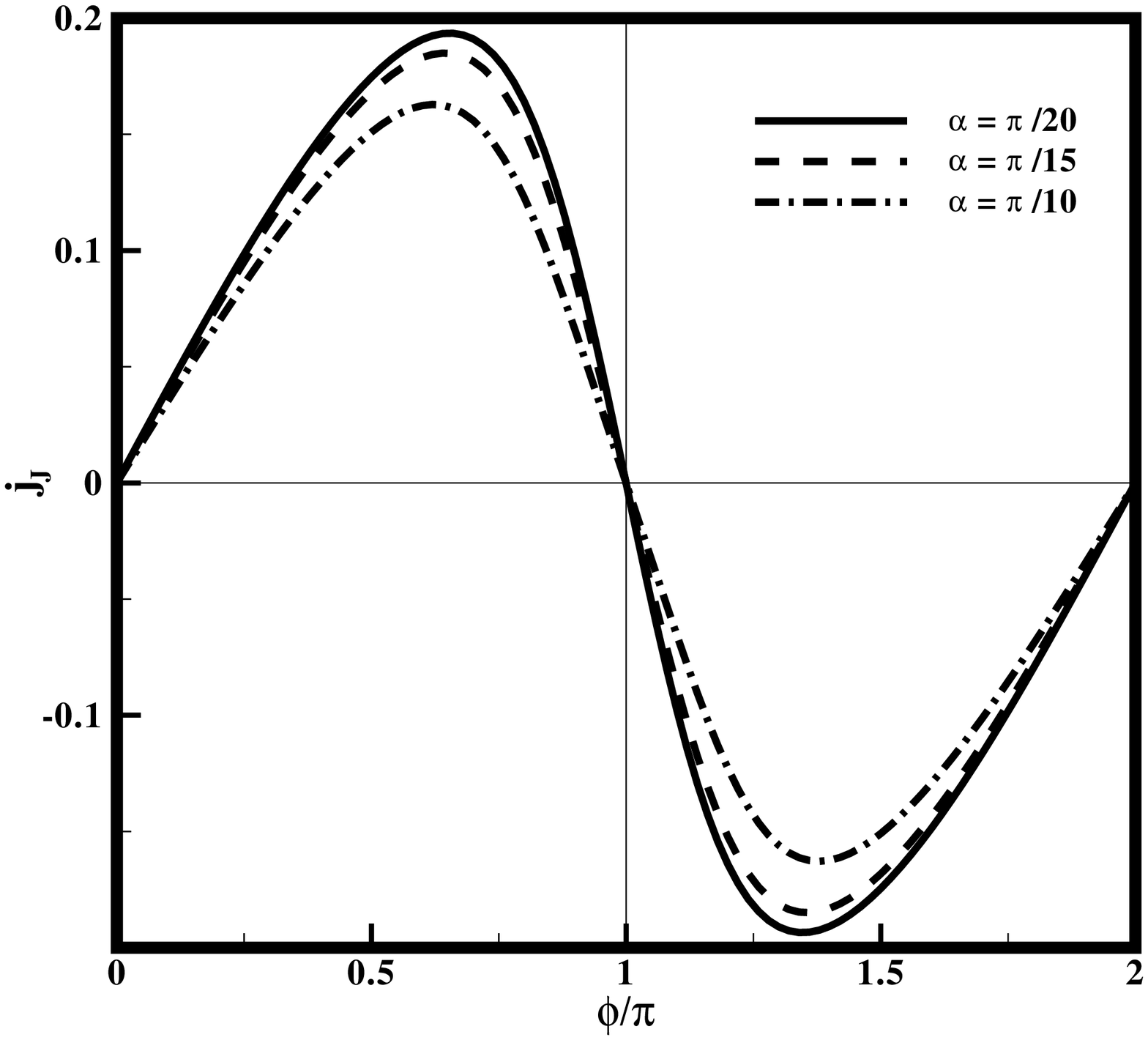}
\caption{The perpendicular component of the charge current (Josephson) versus the phase difference $%
\protect\phi $ for the planar state (\ref{planar}), geometry (ii)
and the different misorientations. The tangential components ($x$
and $z$-directions) are absent.} \label{figb7}
\end{figure}
Consequently, this kind of junction can be applied as a polarizer
or filter for the spin currents.\\ However, for the planar state
and geometry (ii), all terms of the spin current ($s_{x}$, $s_{y}$
and $s_{z}$) can be observed (see Eq.\ref{spin-term}).\newline 3)
In Figs.\ref{figb3}-\ref{figb9} (planar states), it is shown that
the value of the phase differences in which the currents are in
the maxima, minima and zero values, are not very sensitive to the
misorientation angle $\alpha $, while the amplitude of maxima and
minima, are strongly dependent on the value of misorientation
$\alpha $.\\
4) In the Figs.\ref{figb2},\ref{figb3} and Fig.\ref{figb6},
\ref{figb7}, while the charge currents are the odd functions of
$\phi $ with respect to the line of $\phi =\pi $, the spin
currents are even functions of the phase difference; $\mathbf{j}_{e}(\phi =\pi +\delta \phi )=-\mathbf{j}%
_{e}(\phi =\pi -\delta \phi )$ and for the spin current $\mathbf{j}%
_{s_{i}}(\phi =\pi +\delta \phi )=\mathbf{j}_{s_{i}}(\phi =\pi
-\delta \phi ) $. On the contrary, in the Figs.\ref{figb4} and
\ref{figb5}, the
charge and spin currents are even and odd functions of $\phi $ with respect to the line of $%
\phi =\pi $, respectively; $\mathbf{j}_{e}(\phi =\pi +\delta \phi )=\mathbf{j%
}_{e}(\phi =\pi -\delta \phi )$ and $\mathbf{j}_{s_{i}}(\phi =\pi
+\delta \phi )=-\mathbf{j}_{s_{i}}(\phi =\pi -\delta \phi
)$.\newline 5) In Fig.\ref{figb2}, the perpendicular component of
the spin and charge current in terms of the external phase
difference $\phi $ for the case of the planar state
(\ref{planar}), geometry (i) and for two different misorientations
are plotted. The solid line is the charge current-phase dependence
\cite{Mahmoodi}. Also, at the $\phi =0$, $\phi =\pi $ and $\phi
=2\pi $, the charge current (Josephson current) is zero while the
spin current has the
finite value.\\
6) The perpendicular component of the charge (Josephson current)
and spin current for the case of the axial state (\ref{axial}) and
geometry (ii) are plotted in Fig.\ref{figb3}, and the tangential
components of them, are plotted in Figs.\ref{figb4},\ref{figb5}.
The charge current-phase diagrams have been obtained before in
paper \cite{Mahmoodi} and they are totally different from the case
of conventional superconductors in the paper \cite{Kulik}.
\begin{figure}
\includegraphics[width=\columnwidth]{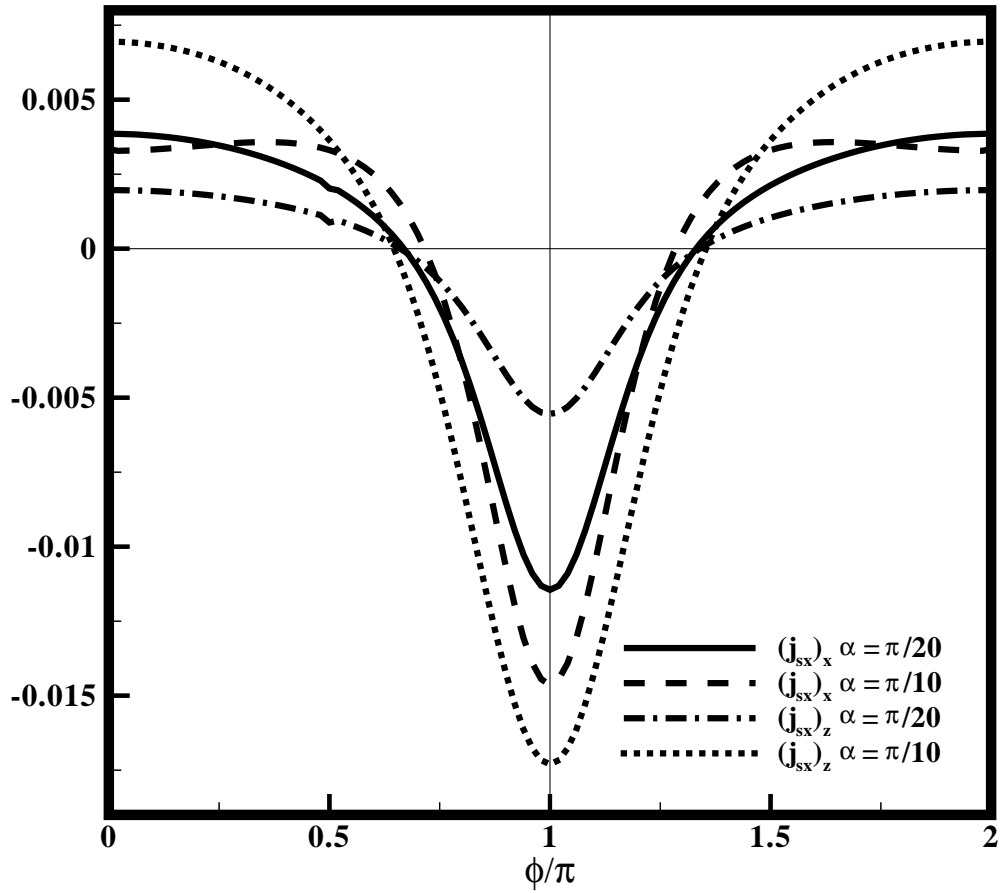}
\caption{Tangential spin currents ($s_{x}$) versus the phase
difference $\protect\phi $ for the planar state (\ref{planar}),
geometry (ii) and the different misorientations. The perpendicular
component ($y$-direction) of the spin current is absent.}
\label{figb8}
\end{figure}
At the phase values of $\phi =0$, $\phi =\pi $ and $\phi =2\pi $,
in which the charge current is exactly zero, the spin current has
the finite values and may select its maximum value. In
Figs.\ref{figb2},\ref{figb3} and specially Fig.\ref{figb9}, for a
small value of misorientation we have a very
long but narrow peak in the spin current phase diagram, close to the $\phi =\pi$.\\
7) Both the planar state with geometry (i) and the axial state
with geometry (ii) can be applied as the filter for polarization
of the spin transport (see Figs\ref{figb2}-\ref{figb5}), the
former transports only the $s_z$ but the latter case flows the
$s_y$ (see .\ref{spin-term}). In addition, the planar states with
geometry (ii) can be used as a switch for the spin and charge
current into two directions: parallel and perpendicular to the
interface. In this case, the spin and charge currents select only
one of the directions parallel or perpendicular to the interface.
Namely, it is impossible to observe the tangential and
perpendicular components of the currents at the same time for
planar state with geometry (ii)(Figs.\ref{figb6}-\ref{figb9}).
\label{subh}
\begin{figure}[tbp]
\includegraphics[width=\columnwidth]{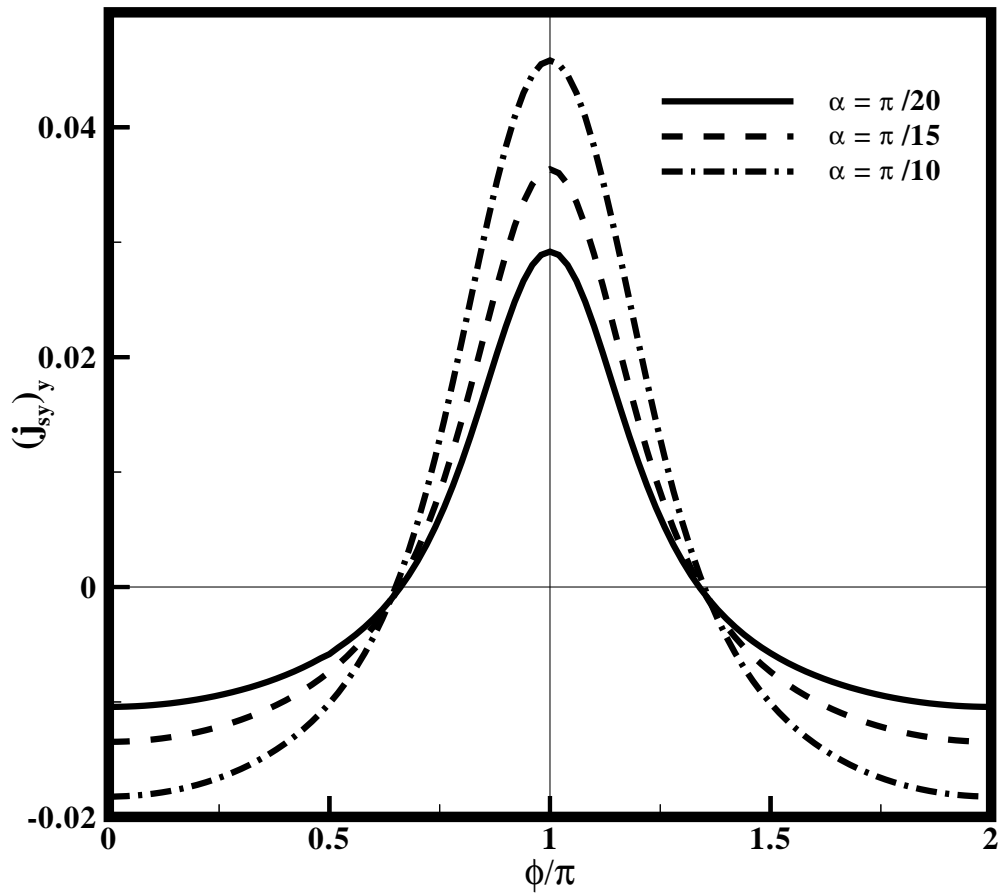}
\caption{Perpendicular component of the spin current ($s_{y}$)
versus the phase
difference $\protect%
\phi $ for the planar state (\ref{planar}), geometry (ii) and the
different misorientations. The tangential components ($x$ and $z$
directions) are absent.} \label{figb9}
\end{figure}
\newpage
\section{Weak link between $PrOs_4Sb_{12}$ supercondcuting banks}
\label{PrOsSb}
\subsection{Introduction} The ``$(p+h)-$wave" form of pairing
symmetry has been considered for the superconductivity in
$PrOs_4Sb_{12}$ compound, recently. In this chapter, a stationary
Josephson junction as a weak-link between $PrOs_4Sb_{12}$ triplet
superconductors is theoretically investigated. The Eilenberger
equation is solved for two distinct models of the order parameter
($A$ and $B-$phases). The spin and charge current-phase diagrams
are plotted and the effect of misorientation between crystals on
the Josephson, spontaneous and spin current is studied. It is
obtained that such experimental investigations of the
current-phase diagrams can be used to test the pairing symmetry in
the above-mentioned superconductors. In this chapter, it is shown
that this apparatus can be applied as a polarizer for the spin
current. Furthermore, it is observed that at some certain values
of the phase differences in which the charge current is zero, the
spin current exists while carriers of
both of charge and spin are the same (electrons).\\
The pairing symmetry of the recently discovered superconductor
compound $PrOs_4Sb_{12}$ is an interesting topic of research in
the  field of superconductivity \cite{Maki3,Maki1,Maki2}.
Superconductivity in this compound has been discovered in papers
\cite{Bauer1,Bauer2,Kotegawa} and two different phases ($A$ and
$B$) have been considered for this kind of superconductor in
papers \cite{Maki1,Nakajima}. Although people at first have
considered the spin-singlet ``$(s+g)-$wave" pairing symmetry to
this superconductor \cite{Maki1} but later it has been specified
that the spin-triplet is the real pairing symmetry of
$PrOs_4Sb_{12}$ complex \cite{Maki3,Tou2}. Authors of paper
\cite{Tou2}, using the knight shift in NMR measurement estimated
the spin-triplet pairing symmetry for the superconductivity in
$PrOs_4Sb_{12}$. Consequently, the ``$(p+h)-$wave" proposed for
the pairing symmetry of the superconductivity in $PrOs_4Sb_{12}$
compound, recently \cite{Maki3}. In this chapter, the
self-consistent equation for the gap vectors (BCS gap equation)
have been solved for the finite temperature numerically and for
the temperatures close to the zero and the critical temperature
analytically. For this compound and using the ``$(p+h)-$wave"
symmetry for the order parameter vector (gap function), the value
of the order parameter at the zero temperature has been obtained
for both $A$ and $B$-phases, in terms of the critical temperature
of this type of superconductivity ($T_c$). In addition, the gap
dependence on the temperature close to the zero and critical
temperatures have been obtained. Authors of paper \cite{Maki3}
have investigated the critical magnetic field and the temperature
dependence of critical
field, specific heat and heat conduction.\\
Also, the Josephson effect in the point contact between triplet
superconductors has been studied in paper \cite{Mahmoodi}. In this
chapter, the effect of misorientation on the charge transport has
been studied and a spontaneous current tangential to the interface
between the $f-$wave superconductors has been observed.
Additionally, the spin-current in the weak-link between the
$f-$wave superconductors has been investigated in the paper
\cite{Rashedi3}. The authors of paper \cite{Rashedi3}, have
proposed this kind of weak-link device as the filter for
polarization of the spin-current. These weak-link structures have
been used to
demonstrate the order parameter symmetry in paper \cite{Stefanakis}.\\
\begin{figure}
\includegraphics[width=6cm]{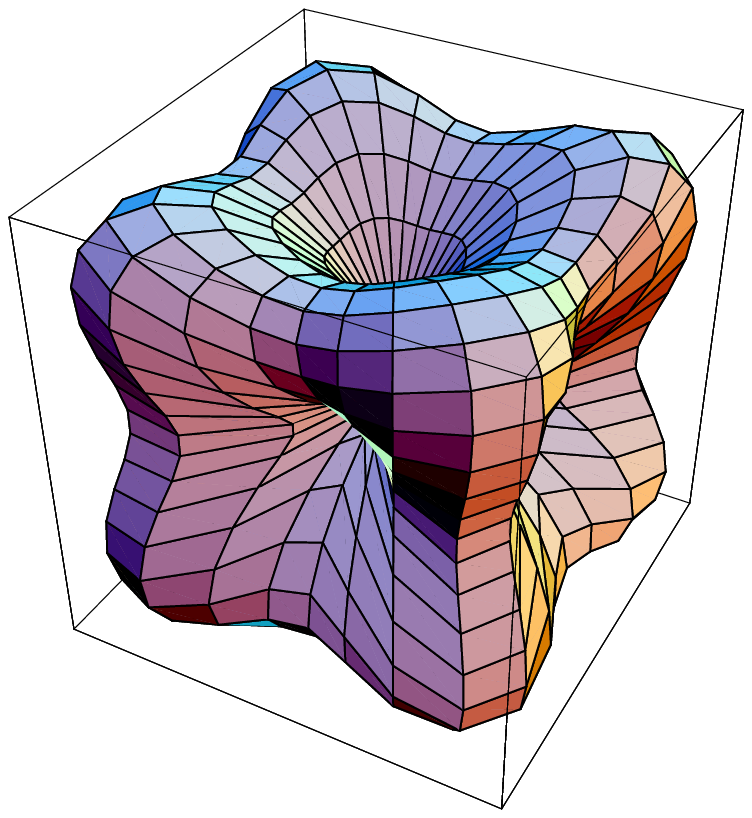}
\includegraphics[width=7.5cm]{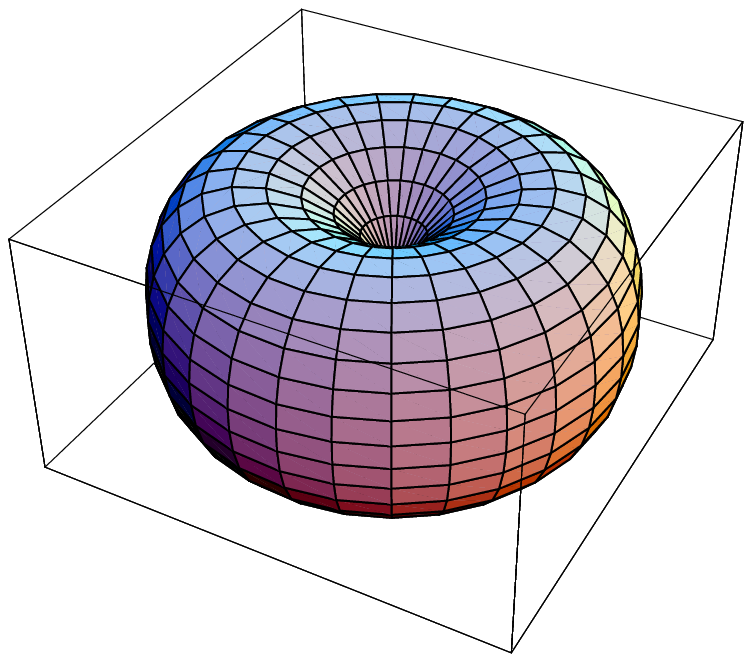}
\includegraphics[width=8cm]{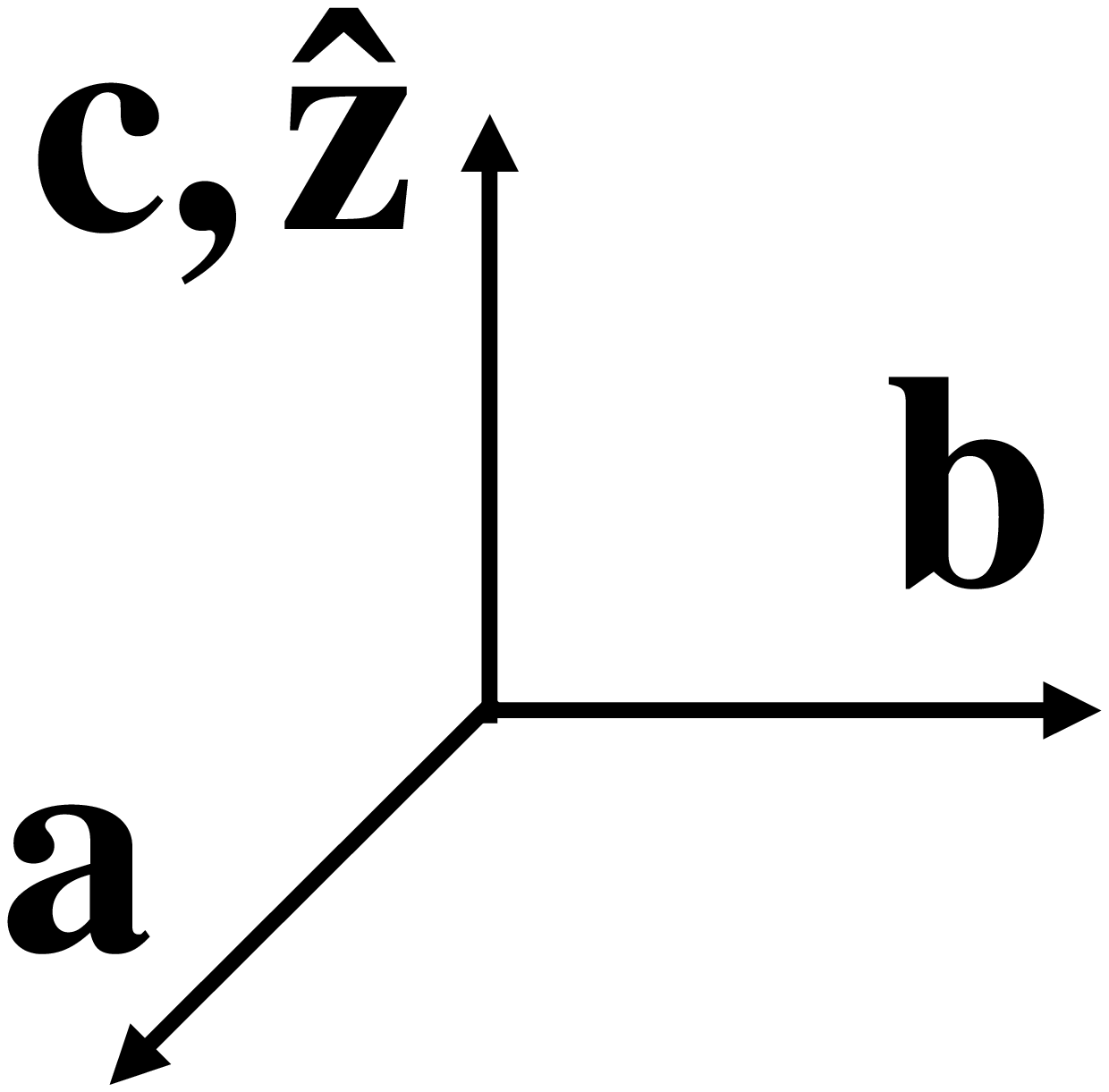}
\includegraphics[width=8cm]{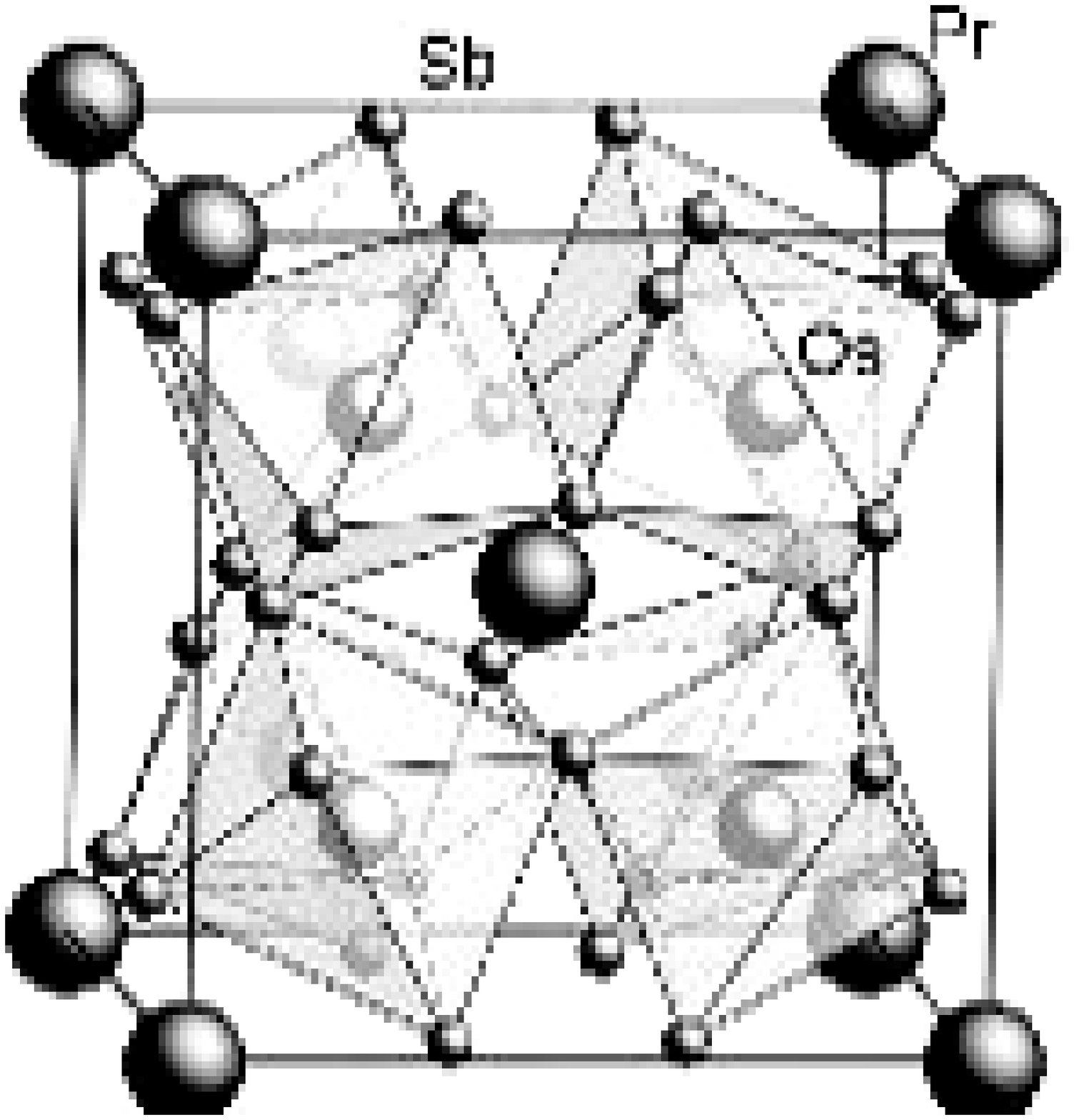}
\vspace{1cm} \caption{$A-$phase (left-above), $B-$phase
(right-above) order parameters, direction of $\mathbf{a}$,
$\mathbf{b}$, $\mathbf{c}$ and $\mathbf{\hat{z}}$ unit vectors
(left-below) and chemical structure of $PrOs_{4}Sb_{12}$ molecule
\cite{Maki1}. $A-$ and $B-$ phases are high field (high
temperature) and low field (low temperature) phases, respectively
\cite{Goryo}.}
\end{figure}\label{phases}
In this chapter, the ballistic Josephson weak-link via an
interface between two bulk of ``$(p+h)-$wave" superconductors with
different orientations of the crystallographic axes is
investigated. It is shown that the spin and charge current-phase
diagrams are totally different from the current-phase diagrams of
the junction between conventional ($s$-wave) superconductors
 \cite{Kulik}, high $T_{c}$ ($d$-wave) superconductors \cite{Coury}
and from the spin-current phase diagrams in the weak-link between
the $f$-wave superconductors \cite{Rashedi3}. In this weak-link
structure between the ``$(p+h)-$wave" superconductors, the
spontaneous current parallel to the interface as the
characteristic of unconventional superconductivity can be present.
The effect of misorientation on the spontaneous, Josephson and
spin currents for the different models of the paring symmetry
($A-$ and $B-$ phases in Fig.\ref{phases}) are investigated. It is
possible to find the value of the phase difference in which the
Josephson current is zero but the spontaneous current tangential
to the interface, which is produced by the interface, exists. In
some of configurations and at the zero phase difference, the
Josephson current is not zero and it has a finite value, this
finite value corresponds to a spontaneous phase difference which
is related to
the misorientation between the gap vectors.\\
Finally, It is observed that at the some certain values of the
phase differences $\phi$ in which the charge current is zero the
spin current exists and vise versa. In addition, in this
configuration in which the gap vectors are selected along the
$\mathbf{\hat{z}}$ direction and the unit vector perpendicular to
the interface is $\mathbf{\hat{y}}$ direction, only the spin
current of the $s_y$ can be present and the other terms of the
spin current are absent totally.
\begin{figure}[tbp]
\includegraphics[width=\columnwidth]{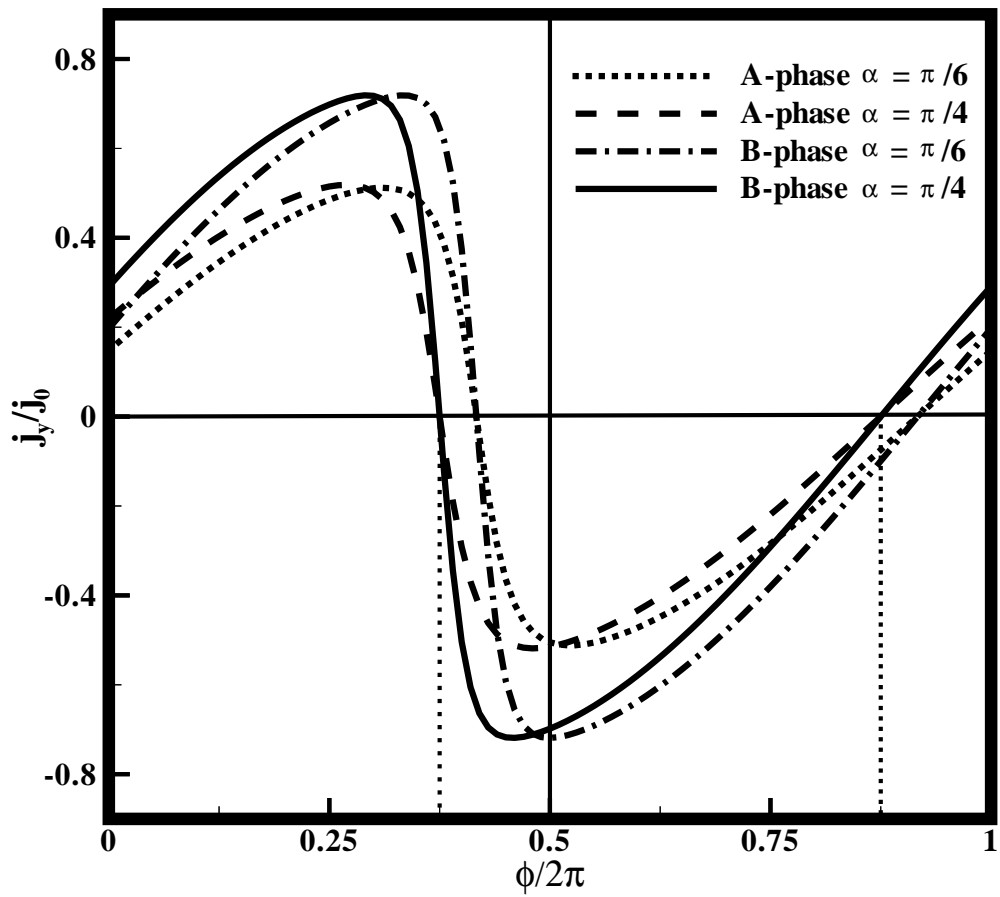}
\caption{Perpendicular component of current (Josephson current)
versus the phase difference $\protect\phi $ for $A$ and
$B-$phases, geometry (i), $T/T_{c}=0.08$ and the different
misorientations. Currents are given in units of
$j_{0}=\frac{\protect\pi }{2}eN(0)v_{F}\Delta _{0}(0).$}
\label{figc2}
\end{figure}
Consequently, this structure can be used as a filter for
polarization of the spin transport. Furthermore, our analytical
and numerical calculations have shown that the misorientation is
the origin of the spin current and in the absence of the
misorientation spin current is absent while the charge current
flows.\\
 In Sec.(\ref{subg}) the obtained formulas for the Green
functions will be discussed and an analysis of numerical results
will be done. \subsection{Discussions} \label{subg}
 In this chapter again we consider a model of a flat
interface $y=0$ between two misorientated ``$(p+h)-$wave"
superconducting half-spaces (Fig.\ref{figb1}) as a ballistic
Josephson junction.
\begin{figure}[tbp]
\includegraphics[width=\columnwidth]{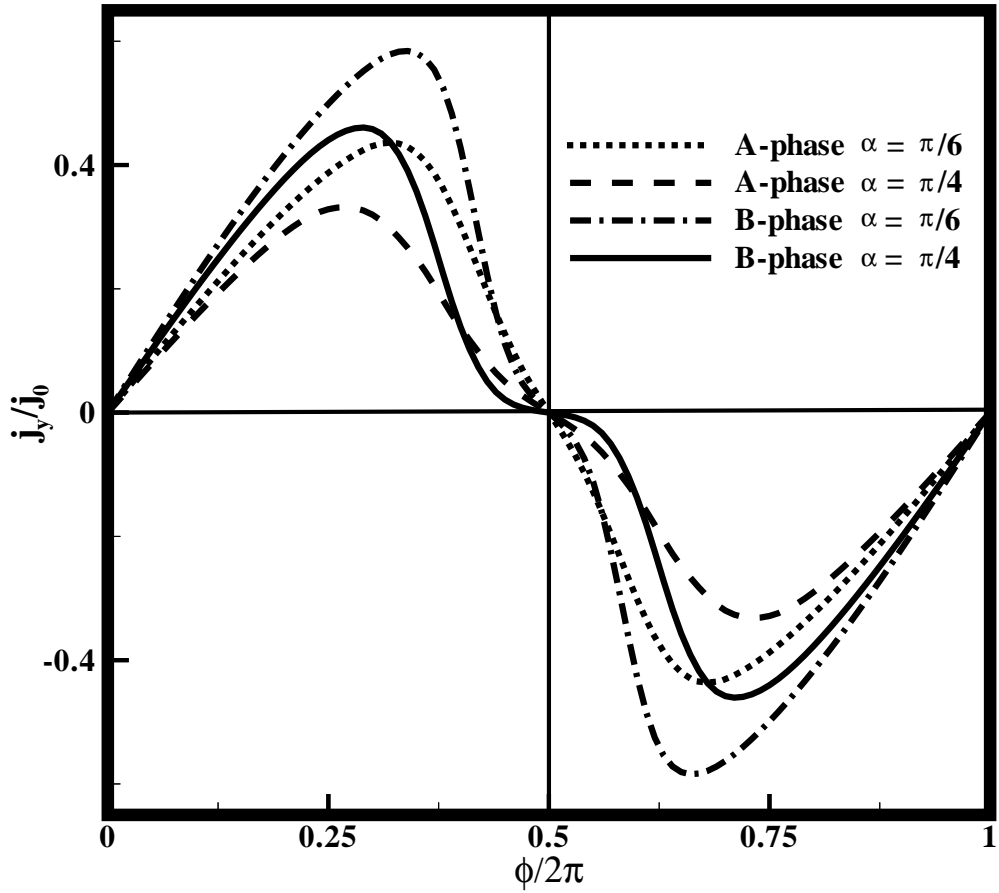}
\caption{Perpendicular component of current (Josephson current)
versus the phase difference $\protect\phi $ for $A$ and
$B-$phases, geometry (ii), $T/T_{c}=0.08$ and the different
misorientations.} \label{figc3}
\end{figure}
\begin{figure}[tbp]
\includegraphics[width=\columnwidth]{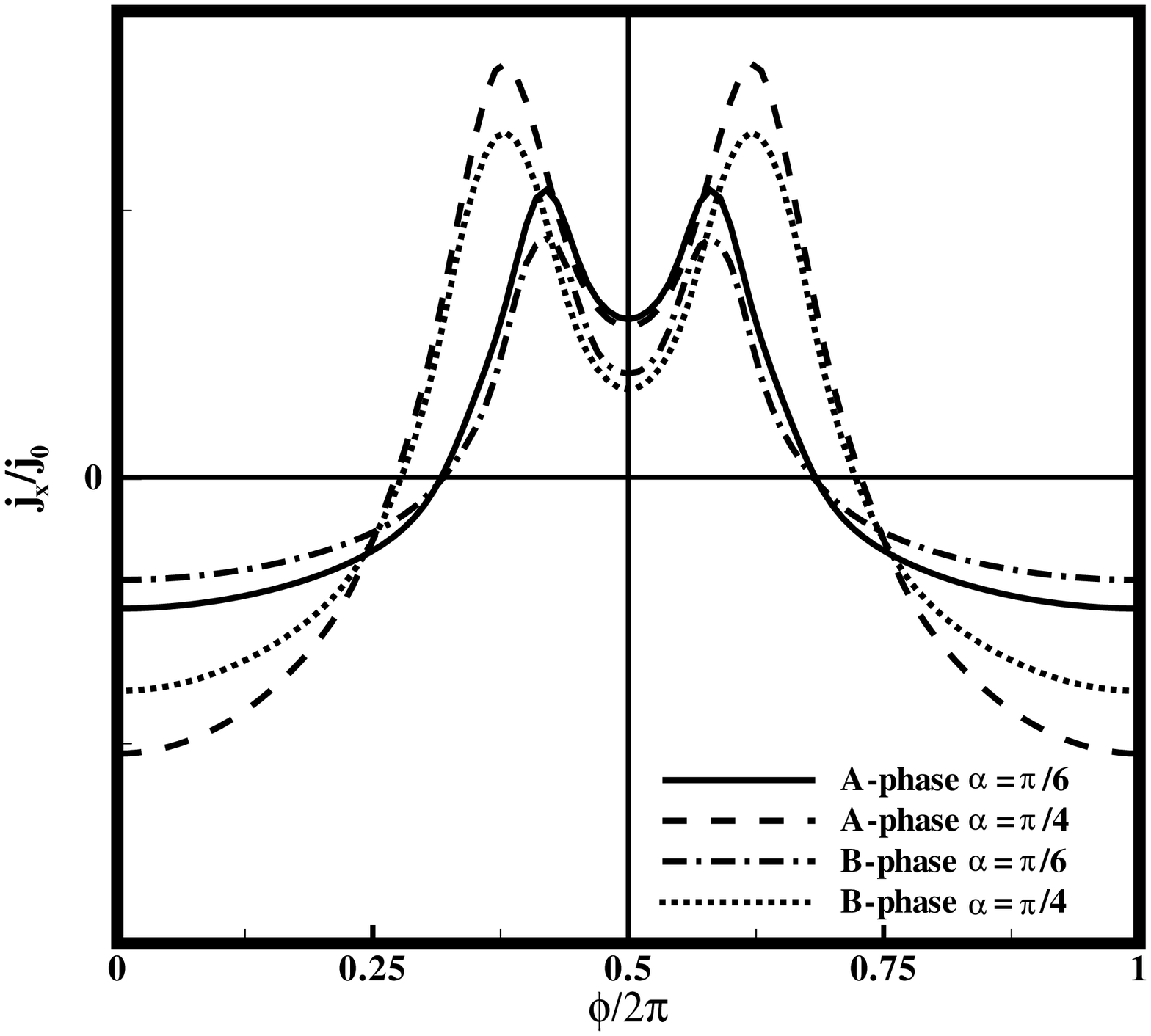}
\caption{Tangential (to the interface) component of current versus
the phase difference $\protect\phi $ for $A$ and $B-$phases,
geometry (ii), $T/T_{c}=0.08$ and the different
misorientations($x-$component).} \label{figc4}
\end{figure}
As same as the previous chapter, in our calculations, a simple
model of the constant order parameter up to the interface is
considered and the pair breaking and the scattering on the
interface are ignored. The solution of Eqs. (\ref{Eilenberger})
and (\ref{self-consistent}) allows us to calculate the charge and
spin current densities. Again, we assume that the order parameter
does not depend on coordinates and in each half-space it equals to its value (%
\ref{Bulk order parameter}) far from the interface in the left or
right bulks. For such a model, the current-phase dependence of a
Josephson junction can be calculated analytically. It enables us
to analyze the main features of current-phase dependence for the
different
models of the order parameter of ``%
$(p+h)-$wave'' superconductivity.
\begin{figure}[tbp]
\includegraphics[width=\columnwidth]{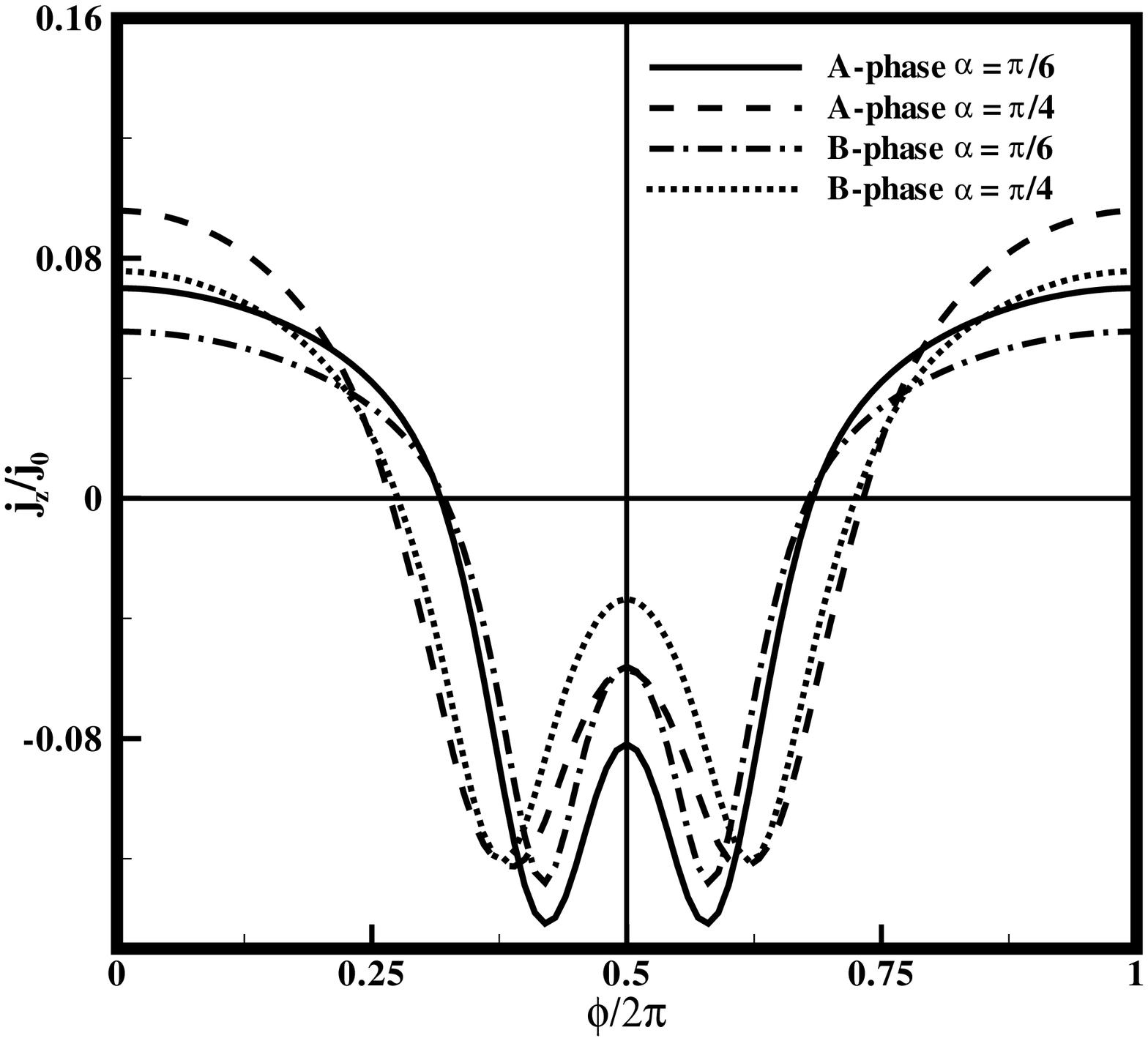}
\caption{Tangential (to the interface) component of current versus
the phase difference $\protect\phi $ for the both $A$ and
$B-$phases, geometry (ii), $T/T_{c}=0.08$ and the different
misorientations($z-$component).} \label{figc5}
\end{figure}
To illustrate the results obtained by computing the formula
(\ref{charge-term}), we plot the current-phase diagrams for the
different
models of the ``$(p+h)-$wave'' pairing symmetry (%
\ref{A-phase},\ref{B-phase}) and for two different geometries.
These geometries are corresponding to the different orientations
of the crystals in the right and left sides of the interface
(Fig.\ref{figb1}):\newline (i) The basal $ab$-plane in the right
side has been rotated around the $c$-axis by $\alpha $;
$\hat{\mathbf{c}}_{1}\Vert
\hat{\mathbf{c}}_{2}$.%
\newline
(ii) The $c$-axis in the right side has been rotated around the
$b$-axis by $\alpha $  ($y$%
-axis in Fig.\ref{figb1}); $\hat{\mathbf{b}}_{1}\Vert
\hat{\mathbf{b}}_{2}$.\newline Further calculations require a
certain model of the gap vector
(vector of order parameter) $%
\mathbf{d}$.

 In this chapter, two forms of the
``$(p+h)-$wave" unitary gap vector in $PrOs_4Sb_{12}$ are
considered. The first model to explain the properties of the
$A$-phase of $PrOs_4Sb_{12}$ is:
\begin{equation}
\mathbf{d}(T,\mathbf{\hat{k}})=\Delta _{0}(T)(k_x+ik_y)\frac{3}{2}
(1-\hat{k}_{x}^4 - \hat{k}_{y}^4 - \hat{k}_{z}^4)\hat{\mathbf{z}}
\label{A-phase}
\end{equation}
The coordinate axes
$\hat{\mathbf{x}},\hat{\mathbf{y}},\hat{\mathbf{z}}$
here and below are chosen along the crystallographic axes $\hat{\mathbf{a}},%
\hat{\mathbf{b}},\hat{\mathbf{c}}$ in the left side of
Fig.\ref{figb1}. The function $ \Delta _{0}=$ $\Delta _{0}\left(
T\right) $ in
\begin{figure}[tbp]
\includegraphics[width=\columnwidth]{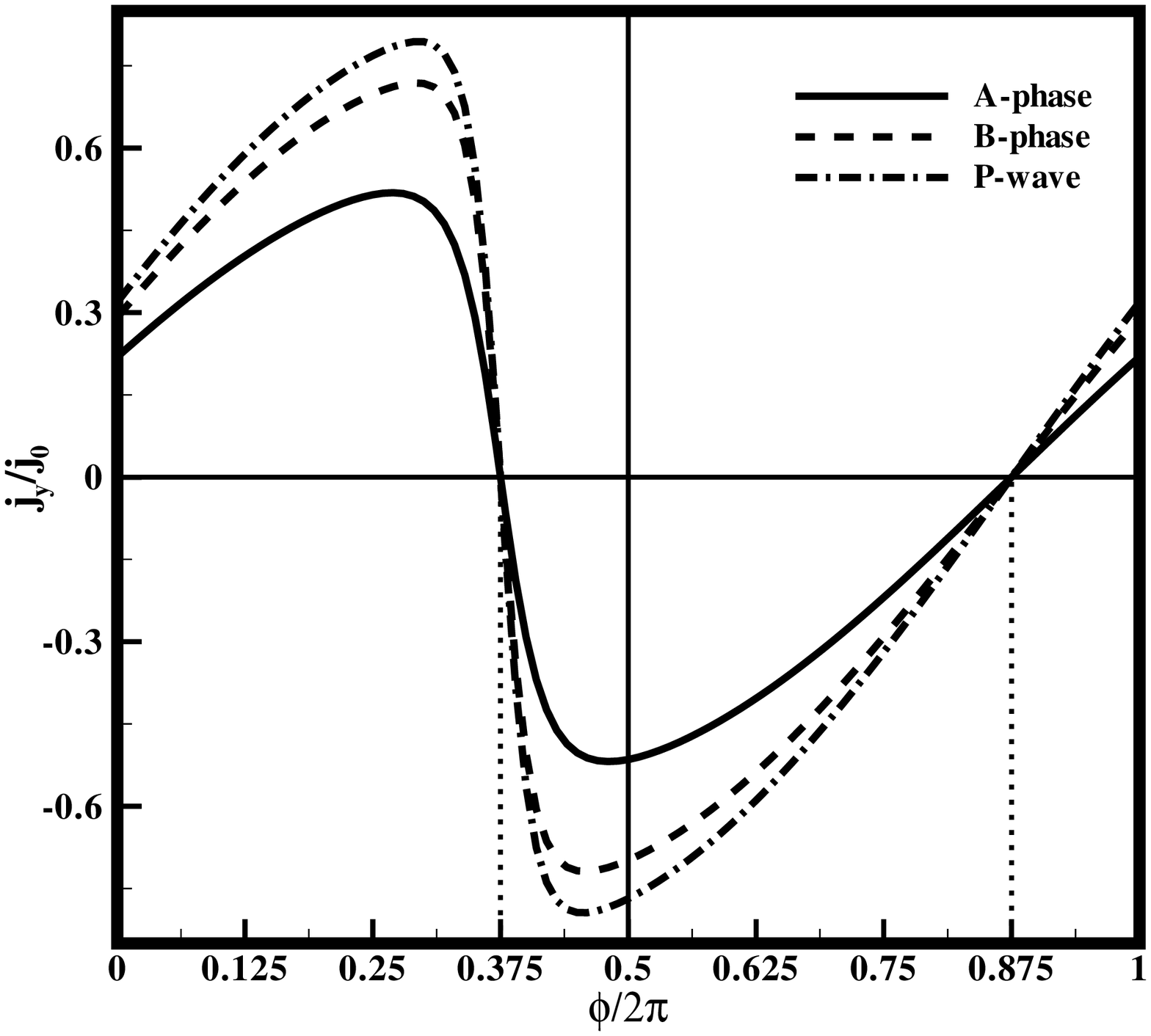}
\caption{Perpendicular component of current (Josephson current)
versus the phase difference $\protect\phi $ for geometry (i),
$\frac{T}{T_{c}}=0.08$, $\alpha=\frac{\pi}{4}$, $A-$phase,
$B-$phase and $p-$wave pairing symmetry.} \label{figc6}
\end{figure}
Eq. (\ref{A-phase}) and below describes the dependence of the gap
vector $\mathbf{d}$ on the temperature $T$. The second model to
describe the gap vector of the $B$-phase of $PrOs_4Sb_{12}$ is:
\begin{equation}
\mathbf{d}(T,\mathbf{\hat{k}})=\Delta _{0}(T)(k_x+ik_y)
(1-\hat{k}_{z}^4)\hat{\mathbf{z}}\label{B-phase}
\end{equation}
\begin{figure}[tbp]
\includegraphics[width=\columnwidth]{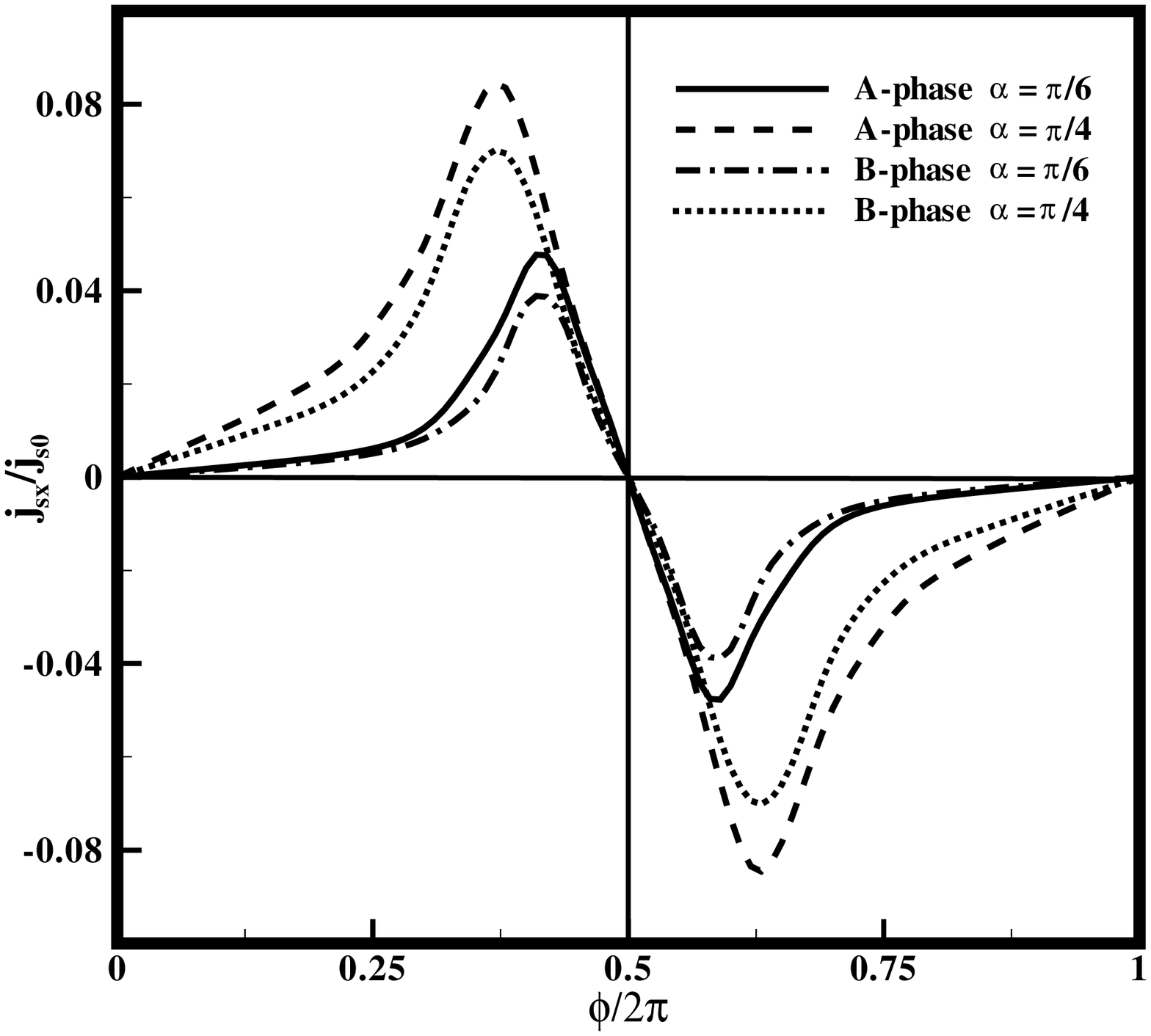}
\caption{Tangential component of spin ($s_y$) current
($x-$component) versus the phase difference $\protect\phi $ for
geometry (ii), $\frac{T}{T_{c}}=0.08$ and different
misorientations between the $A$ and $B-$phase of ``$(p+h)-$wave"
pairing symmetry.} \label{figc7}
\end{figure}
Our numerical calculations are done at the low temperatures,
$T/T_c=0.08$, and we have used the formulas $\ln(\Delta(T)/
\Delta(0)) = -\frac{7 \pi \zeta(3)}{8}(\frac{T}{\Delta(0)})^{3}$
for the $A-$phase and $\ln(\Delta(T)/ \Delta(0)) = -\frac{135 \pi
\zeta(3)}{512}(\frac{T}{\Delta(0)})^{3}$ for the $B-$phase, from
the paper \cite{Maki3}, for temperature dependence of the gap
functions $\Delta _{0}(T)$ at the low temperatures ($T<<T_{c}$).
Also, in the mentioned paper \cite{Maki3} the value of
$\Delta_{0}(T)$ in terms of the critical temperatures has been
calculated for both $A$ and $B-$ phases. They are
$\Delta(0)/T_{c}=2.34$ and $\Delta(0)/T_{c}=1.93$ for $A$ and $B-$
phases, respectively.
\begin{figure}[tbp]
\includegraphics[width=\columnwidth]{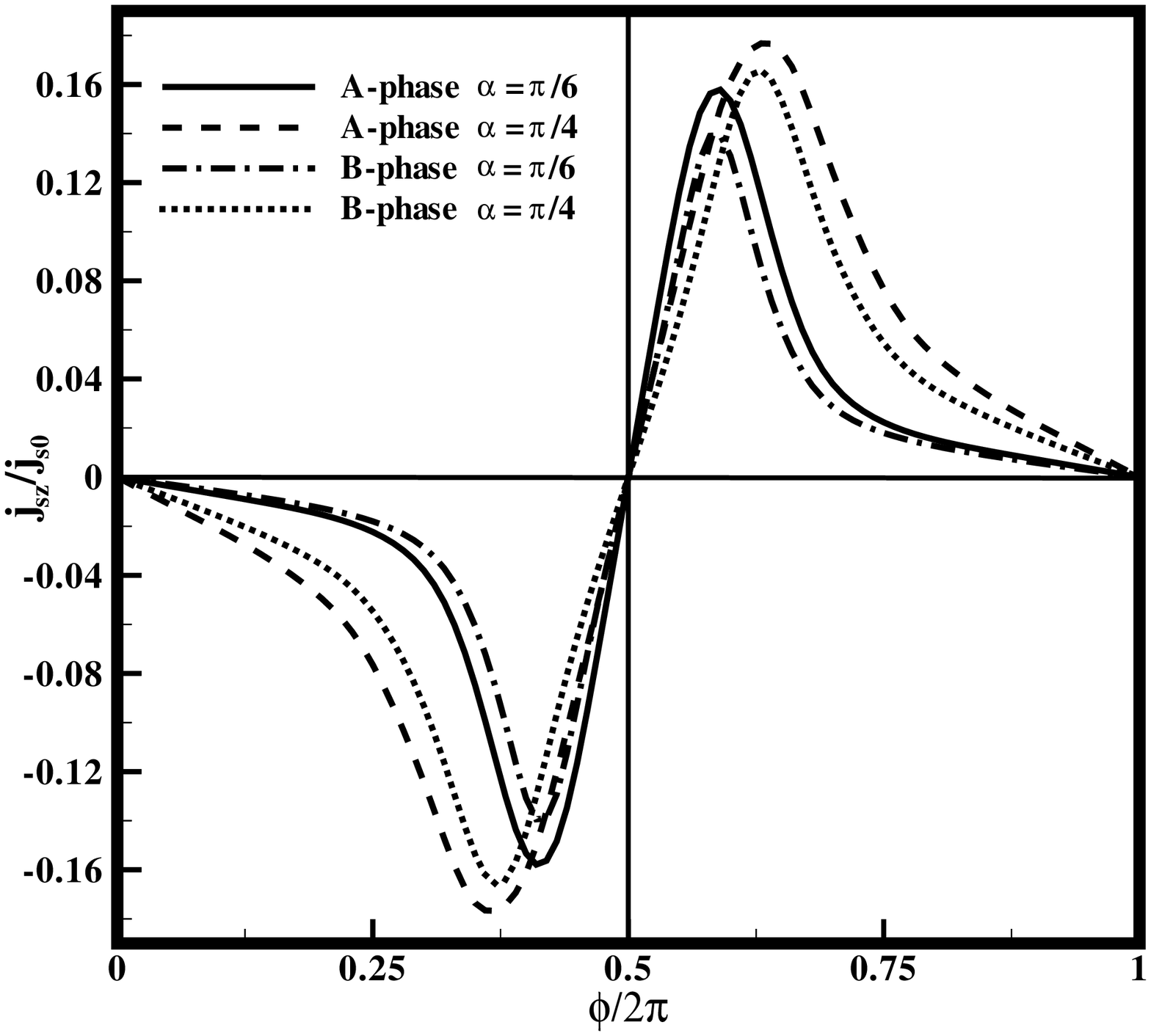}
\caption{Tangential component of spin ($s_y$) current
($z-$component) versus the phase difference $\protect\phi $ for
geometry (ii), $\frac{T}{T_{c}}=0.08$ and different
misorientations between the $A$ and $B-$phase of ``$(p+h)-$wave"
pairing symmetry.} \label{figc8}
\end{figure}
Using these two models of order parameters
(\ref{A-phase},\ref{B-phase}) and solution to the Eilenberger
equations (\ref{charge-term},\ref{spin-term}), we have calculated
the current density at the interface numerically.
These numerical results are listed below \cite{Rashedi4}:\\
1) In Fig.\ref{figc2}, the perpendicular component of current
(perpendicular to the interface) which is called Josephson current
is plotted for both $A$ and $B-$phases, geometry (i),
misorientations $\alpha=\pi/4$ and $\alpha=\pi/6$. It is observed
that the critical values of current for $B-$phase is greater than
$A-$phase. Also, in spite of junction between the conventional
superconductors and planar Josephson junction, at the $\phi=0$,
the current is not zero. The current is zero at the phase
difference value $\phi=\phi_{0}$, which depends on the
misorientation between the gap vectors. In Fig.\ref{figc2}, the
value of the spontaneous phase difference
$\phi_{0}$ is close to the misorientation $\alpha.$\\
2) In Fig.\ref{figc3}, Josephson current is plotted for both $A$
and $B-$phases, geometry (ii) and different misorientations.
Again, the maximum value of the current for the $B-$phase is
greater than $A-$phase. Increasing the misorientation between the
gap vectors, the maximum value of current decreases. It is
demonstrated that at the phase difference values $\phi=0$,
$\phi=\pi$ and $\phi=2\pi$, the Josephson current is zero while,
both of spontaneous and spin currents are not zero and have the
finite value. Increasing the misorientation between the gap
vectors decreases the derivative of the current with respect to
the phase difference ($\frac{dj_y}{d\phi}$) close to the
$\phi=\pi$.\\
3) In Figs.\ref{figc4},\ref{figc5}, the tangential components of
charge current ($x$ and $z-$components) in terms of the phase
difference $\phi$ are plotted.
\begin{figure}[tbp]
\includegraphics[width=\columnwidth]{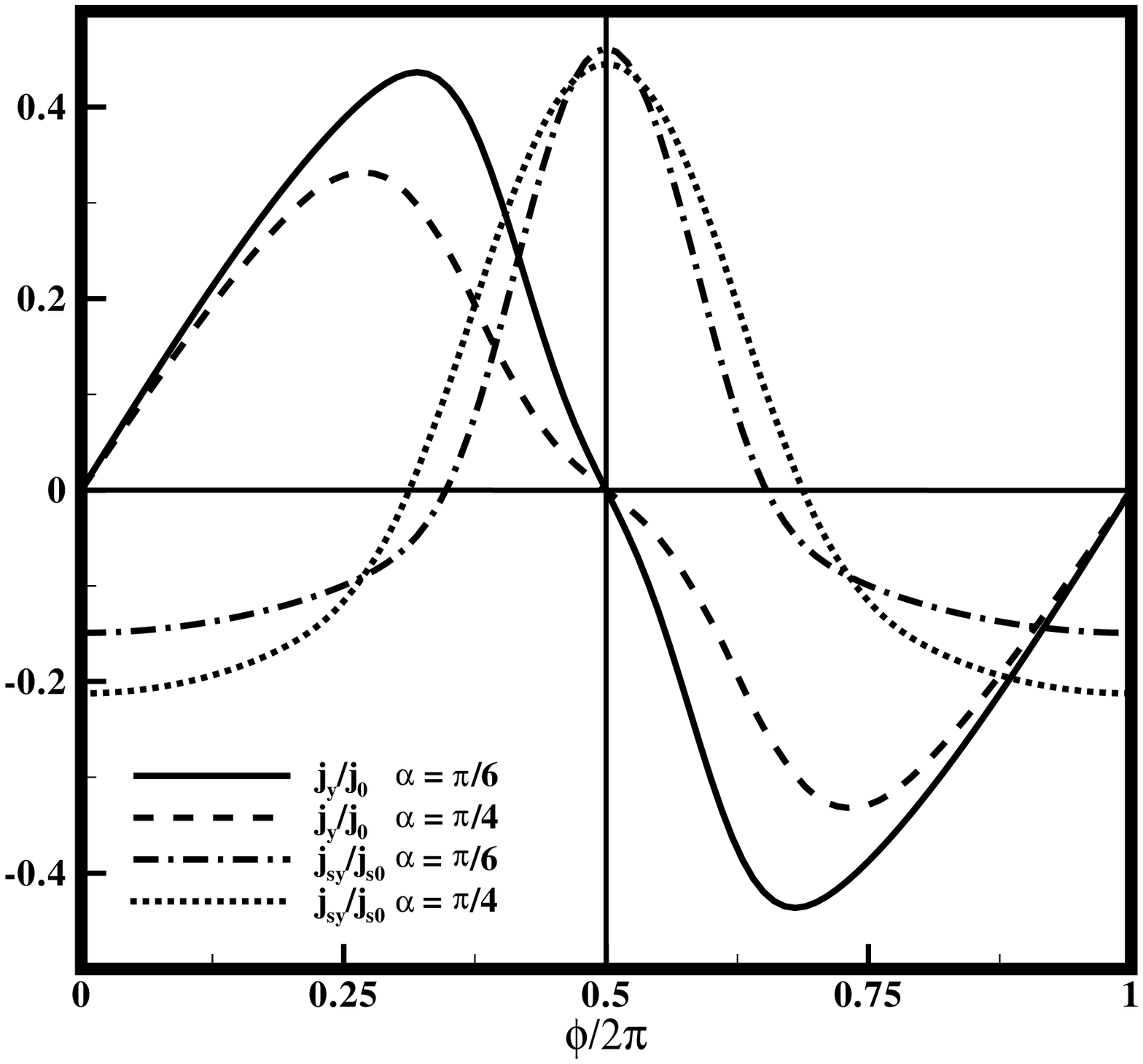}
\caption{Perpendicular component of charge and spin current ($j_y$
and $j_{sy}$) versus the phase difference $\protect\phi $ for
$A-$phase, geometry (ii), $\frac{T}{T_{c}}=0.08$,
$\alpha=\frac{\pi}{6}$ and $\alpha=\frac{\pi}{4}$.} \label{figc9}
\end{figure}
It is seen that at the $\phi=0$, $\phi=\pi$ and $\phi=2\pi$ in
which the Josephson current is zero the parallel spontaneous
currents have the finite values. Although the perpendicular
component of charge current (see Fig.\ref{figc3}) is an odd
function of the phase difference with respect to the line of
$\phi=\pi$ but the parallel charge currents are even
functions at the phase differences close to $\phi=\pi$ (see Fig.\ref{figc4} and Fig.\ref{figc5}).\\
4) In Fig.\ref{figc6}, the Josephson current in terms of the phase
difference is plotted for the case of $p-$wave, and $A$ and
$B-$phases of ``$(p+h)-$wave". The $p-$wave pairing symmetry as
the first candidate for the superconducting state in
$Sr_{2}RuO_{4}$ is as follows \cite{Rice}:
\begin{equation}
\mathbf{d}(T,\mathbf{\hat{k}})=\Delta
_{0}(T)(k_x+ik_y)\hat{\mathbf{z}}\label{p-wave}
\end{equation}
It is observed that the maximum value of Josephson current ($j_y$)
of junction between the $p-$wave superconductors, is greater than
the $B-$phase of ``$(p+h)-$wave" and the Josephson current of
second is greater than its $A-$phase counterpart. Also, the place
of the zero of the current is at the spontaneous phase difference
which is close to the misorientation $\phi_{0}=\alpha$ (look at the Fig.\ref{fig2}).\\
5) In Figs.\ref{figc7},\ref{figc8}, the tangential spin ($s_y$)
currents in terms of the phase difference is plotted for different
misorientations and geometry (ii). By increasing the
misorientation the maximum value of spin current increases. In
spite of the charge current for this state, the spin current at
the phase differences $\phi=0$, $\phi=\pi$ and $\phi=2\pi$ is
exactly zero (compare Figs.\ref{figc4} and \ref{figc5}
with Figs.\ref{figc7} and \ref{figc8}).\\
6) In Figs.\ref{figc9},\ref{figc10}, the perpendicular component
of charge and spin current ($j_y$ and $j_{sy}$) are plotted for
different misorientations and $A$ and $B-$phases respectively. An
interesting case in our observations is the finite value of the
perpendicular spin current at the $\phi=0$, $\phi=\pi$ and
$\phi=2\pi$ at which the perpendicular charge current ($j_y$) is
zero (see Figs.\ref{figc9} and \ref{figc10}).\\
7) In Fig.\ref{figc11}, the perpendicular component of the spin
current is plotted for $p-$wave, $A$ and $B-$phases of
``$(p+h)-$wave" pairing symmetries and for a specified value of
misorientation $\alpha=\frac{\pi}{4}$. In both Figs.\ref{figc6}
and \ref{figc11} the maximum value of current of junction between
the $p-$wave is greater than $B-$phase and $B-$phase has the
maximum value greater than junction between the $A-$phase. This
different character of the current-phase diagrams enables us to
distinguish between the three states.
\begin{figure}[tbp]
\includegraphics[width=\columnwidth]{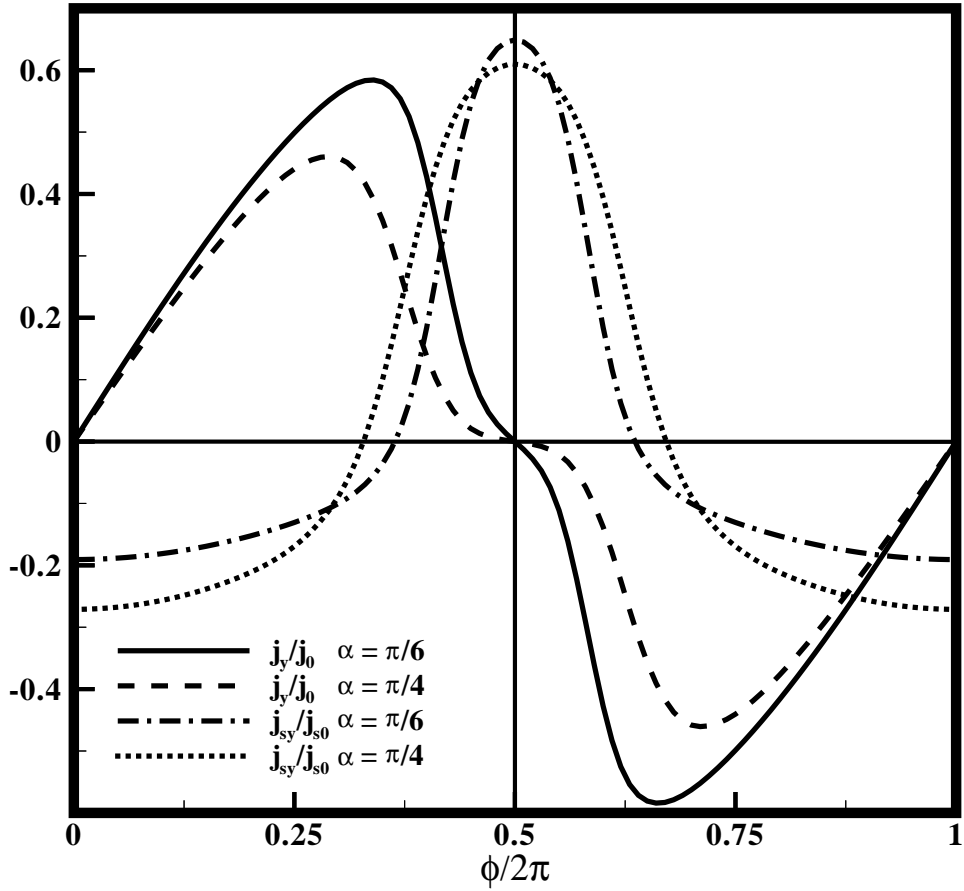}
\caption{Perpendicular component of charge and spin current ($j_y$
and $j_{sy}$) versus the phase difference $\protect\phi $ for
$B-$phase, geometry (ii), $\frac{T}{T_{c}}=0.08$,
$\alpha=\frac{\pi}{6}$ and $\alpha=\frac{\pi}{4}$.}\label{figc10}
\end{figure}
Also, It is observed that at the phase differences $\phi=0$,
$\phi=\pi$ and $\phi=2\pi$, the spin current has the finite value
and may have its maximum value. This is a counterpart of
Fig.\ref{figc3}, in that figure the charge currents are zero at
the mentioned value of
the phase difference but the spin current has the finite value.\\
Furthermore, our analytical and numerical calculations have shown
that the origin of the spin current is misorientation between the
gap vectors (cross product in Eq.\ref{spin-term}). Thus the spin
current in the weak link between geometry (i) misorientated
crystals is zero. Because the geometry (i) is a rotation by
$\alpha$ around the $\mathbf{\hat z}-$axis and both of the left
and right gap vectors are in the same direction and cross product
between them is zero.
\begin{figure}[tbp]
\includegraphics[width=\columnwidth]{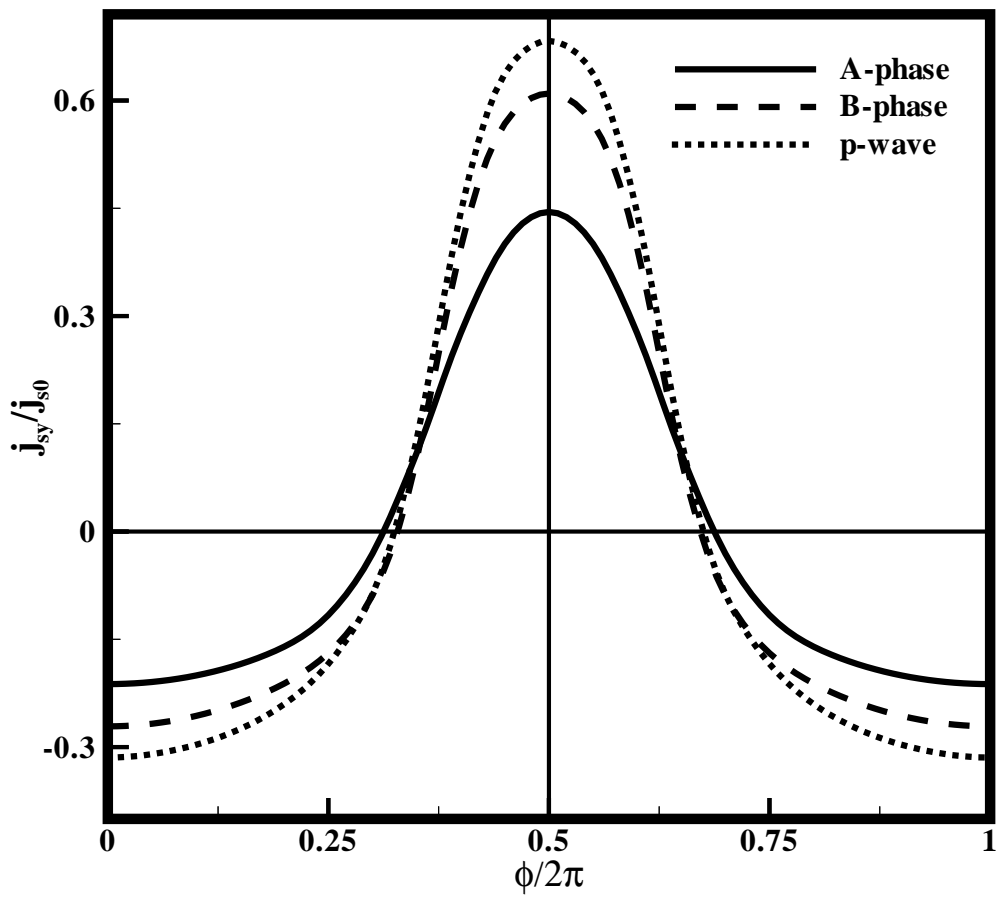}
\caption{Perpendicular component of spin current ($j_{sy}$) versus
the phase difference $\protect\phi $ for geometry (ii),
$\frac{T}{T_{c}}=0.08$, $\alpha=\frac{\pi}{4}$, $A-$phase,
$B-$phase and $p-$wave pairing symmetry.} \label{figc11}
\end{figure}
Also, it is shown that in this structure ($y-$direction is
perpendicular to the interface, gap vectors $c-$axis are selected
in the $z-$direction and rotation is done around the
$y-$direction) only the current of the $s_y$ flows and other terms
of the spin current are totally absent. So, this kind of weak-link
experiment can be used as the filter for polarization of spin
transport. Since the spin is a vector, the spin current is a
tensor and we have the current of spin $s_{y}$ in the three
$\mathbf{\hat{x}}$, $\mathbf{\hat{y}}$ and $\mathbf{\hat{z}}$
directions.
\newpage

\section{Weak link between nonunitary triplet superconductors}
\label{Nonunitary} \subsection{Introduction} A stationary
Josephson effect in a weak-link between misorientated nonunitary
triplet superconductors is investigated theoretically. The
non-self-consistent quasiclassical Eilenberger equation for this
system has been solved analytically. As an application of this
analytical calculation, the current-phase diagrams are plotted for
the junction between two nonunitary bipolar $f-$wave
superconducting banks. A spontaneous current parallel to the
interface between superconductors has been observed. Also, the
effect of misorientation between crystals on the Josephson and
spontaneous currents is studied. Such experimental investigations
of the current-phase diagrams can be used to test the pairing
symmetry in the above-mentioned superconductors.

In the recent years, the triplet superconductivity has become one
of the modern subjects for researchers in the field of
superconductivity \cite{Maeno,Mackenzie,Ueda}. Particularly, the
nonunitary spin triplet state in which, Cooper pairs may carry a
finite averaged intrinsic spin momentum, has attracted much
attention in the last decade \cite{Machida,Tou}. A triplet state
in the momentum inverse space $({\mathbf k})$ can be described by
the order parameter ${\hat\Delta}({\mathbf k})=i({\mathbf
d}({\mathbf k}) \cdot{\hat{\sigma }})\hat{\sigma }_y$ in a
2$\times$2 matrix form in which $\hat{\sigma }_j$s are 2$\times$2
Pauli matrices. The three dimensional complex vector ${\mathbf
d}(\mathbf k)$ demonstrates the triplet pairing state exactly. In
the nonunitary state, the product ${\hat\Delta}({\mathbf
k}){\hat\Delta}({\mathbf k})^{\dagger}={\mathbf d}({\mathbf
k})\cdot{\mathbf d}^{*}({\mathbf k})+i({\mathbf d}({\mathbf k})
\times{\mathbf d}^{*}({\mathbf k}))\cdot{\hat{\sigma }}$ is not a
multiple of unit matrix. Thus in a non-unitary state time reversal
symmetry necessarily is broken spontaneously and spontaneous
moment ${\mathbf m}({\mathbf k})=i{\mathbf d}({\mathbf k})
\times{\mathbf d}^{*}({\mathbf k})$ appears at each point
$(\mathbf k)$. In this case the macroscopic averaged moment
$<{\mathbf m}({\mathbf k})>$ integrated on the Fermi surface
non-vanishes. The variable ${\mathbf m}({\mathbf k})$ is related
to the net spin average of state ${\mathbf k}$ by
$\textbf{tr}[{\hat\Delta}({\mathbf k})^{\dag}{ \hat{\sigma }_j
}{\hat\Delta}({\mathbf k})]$, it is clear that the total spin
average over the Fermi surface can be nonzero. As an application,
the nonunitary bipolar state of $f-$wave pairing symmetry has been
considered for the $B-$phase of superconductivity in the $UPt_{3}$
compound which has been created in the low temperatures $T$ and
low values of $H$ \cite{Machida,Ohmi}. In this chapter, we want to
investigate the weak link between two nonunitary superconducting
bulks. This type of weak link structure can be used to demonstrate
and to test the pairing symmetry in the superconducting phase
\cite{Stefanakis}. Consequently, we generalize the formalism of
paper \cite{Mahmoodi} for the case of weak link between nonunitary
triplet superconducting bulks. In the paper \cite{Mahmoodi}, the
Josephson effect in the point contact between unitary $f-$wave
triplet superconductors has been studied. Also, the effect of
misorientation on the charge transport has been investigated and a
spontaneous current tangential to the
interface between the $f-$wave superconductors, has been observed.\\
In this chapter, the ballistic Josephson weak-link via an
interface between two bulks of nonunitary bipolar $f-$wave
superconductivity with different orientations of the
crystallographic axes is investigated. It is shown that the
current-phase diagrams are totally different from the
current-phase diagrams of the junction between conventional
($s$-wave) superconductors \cite{Kulik}, high $T_{c}$ ($d$-wave)
superconductors \cite{Coury} and unitary triplet ( axial and
planar) $f-$wave superconductors \cite{Mahmoodi}. This different
characters can be used to distinct between nonunitary bipolar
$f-$wave superconductivity from the other types of
superconductivity, roughly speaking. In this weak-link structure
between the nonunitary $f-$wave superconductors, the spontaneous
current parallel to the interface as a fingerprint of
unconventional superconductivity and spontaneous time reversal
symmetry breaking, has been observed. The effect of misorientation
on the spontaneous and Josephson currents is investigated. It is
possible to find the value of the phase difference in which the
Josephson current is zero but the spontaneous current tangential
to the interface, which is produced by the interface, is present.
In some configurations and at the zero phase difference, the
Josephson current is not zero generally and it has a finite value.
This finite value corresponds to a spontaneous phase difference
which is related to the misorientation between the gap vectors.\\
The arrangement of the rest of this chapter is as follows. In
Sec.(\ref{subh}) we describe our configuration, which has been
investigated. For a non-self-consistent model of the order
parameter, the quasiclassiacl Eilenberger equations
\cite{Eilenberger} are solved and suitable Green functions have
been obtained analytically. In Sec.(\ref{subi}) the obtained
formulas for the Green functions have been used for calculation
the current densities at the interface and an analysis of
numerical results will be done. \subsection{Quasiclassical
equations} \label{subh} We consider a model of a flat interface
$y=0$ between two misoriented nonunitary $f-$wave superconducting
half-spaces (Fig.\ref{figb1}) as a ballistic Josephson junction.
In order to calculate the current, we use  quaisclassical
Eilenberger equations \cite{Eilenberger} for Green functions
(\ref{Green function}) The Matsubara propagator $\breve{g}$ here,
like the unitary triplet superconductor case, can be written in
the form:
\begin{equation}
\breve{g}=\left(
\begin{array}{cc}
g_{1}+\mathbf{g}_{1}\cdot\mathbf{\hat{\sigma}} & \mathbf{g}_{2}\cdot\hat{%
\sigma } i\hat{\sigma}_{2} \\
i\hat{\sigma}_{2}\mathbf{g}_{3}\cdot\hat{\sigma } &
g_{4}-\hat{\sigma}_{2}\mathbf{g}_{4}\cdot\hat{\sigma
}\hat{\sigma}_{2}
\end{array}
\right) ,
\end{equation}
\label{Green's function} where, the matrix structure of the off-diagonal self energy $\breve{%
\Delta}$ in the Nambu space is
\begin{equation}
\breve{\Delta}=\left(
\begin{array}{cc}
0 & \mathbf{d}\cdot\hat{\sigma }i\hat{\sigma}_{2} \\
i\hat{\sigma}_{2}\mathbf{{d^{\ast }}\cdot\hat{\sigma}} & 0
\end{array}
\right) .
\end{equation}
\label{order parameter} In this chapter, the nonunitary states, for which $\mathbf{%
d\times d}^{\ast }\neq0,$ is investigated. Fundamentally, the gap
vector (order parameter) $\mathbf{d}$ has to be determined
numerically from the self-consistency equation \cite{Ueda}, while
in this chapter, we use a non-self-consistent model for the gap
vector which is much more suitable for the analytical
calculations. Solutions to Eq. (\ref{Eilenberger}) must satisfy
the conditions for Green functions and gap vector $\mathbf{d}$ in
the bulks of the superconductors far from the interface as follow:
\begin{equation}
\breve{g}=\frac{1}{\Omega_{n}}\left(
\begin{array}{cc}
\varepsilon(1-\mathbf{A_{n}}\cdot\mathbf{\hat{\sigma}})&
\left[i\mathbf{d_{n}}-\mathbf{d_{n}\times A_{n}}\right]\cdot\hat{\sigma } i\hat{\sigma}_{2} \\
i\hat{\sigma}_{2}\left[i\mathbf{d_{n}^{\ast}}+
\mathbf{d_{n}^{\ast}\times A_{n}}\right]\cdot\hat{\sigma }&
-\varepsilon\hat{\sigma}_{2}(1+\mathbf{A_{n}}\cdot\hat{\sigma
})\hat{\sigma}_{2}
\end{array}
\right)  \label{Bulk solution}
\end{equation}
where,
\begin{equation}
\hspace{-.4cm}\mathbf{A_{n}}=\frac{i\mathbf{d_{n}\times
d^{\ast}_{n}}}{\varepsilon^{2}+\mathbf{d_{n}\cdot d^{\ast}_{n}}
+\sqrt{(\varepsilon^{2}+\mathbf{d_{n}\cdot
d^{\ast}_{n}})^2+(\mathbf{d_{n}\times d^{\ast}_{n}})^{2}}}
\end{equation}
and,
\begin{equation}
\hspace{-.55cm}\Omega_{n}=\sqrt{\frac{2[(\varepsilon^{2}+\mathbf{d_{n}\cdot
d^{\ast}_{n}})^2+(\mathbf{d_{n}\times
d^{\ast}_{n}})^{2}]}{\varepsilon^{2}+\mathbf{d_{n}\cdot
d^{\ast}_{n}} +\sqrt{(\varepsilon^{2}+\mathbf{d_{n}\cdot
d^{\ast}_{n}})^2+(\mathbf{d_{n}\times d^{\ast}_{n}})^{2}}}}
\end{equation}
\begin{equation}
\mathbf{d}\left( \pm \infty \right)=\mathbf{d}_{2,1}\left(T,\mathbf{\hat{v%
}}_{F}\right) \exp \left( \mp \frac{i\phi }{2}\right)\label{Bulk
order parameter}
\end{equation}
where $\phi $ is the external phase difference between the order
parameters of the bulks and $n=1,2$ label the left and right half
spaces respectively. It is clear that, poles of Green function in
the energy space, are in
\begin{equation}
 \Omega_{2,1}=0
 \end{equation}
  consequently,
  \begin{equation}
((iE)^{2}+\mathbf{d_{2,1}\cdot
d^{\ast}_{2,1}})^2+(\mathbf{d_{2,1}\times d^{\ast}_{2,1}})^{2}=0
 \end{equation}
and
\begin{equation}
E=\pm\sqrt{\mathbf{d_{2,1}\cdot d^{\ast}_{2,1}}\mp
i\mathbf{d_{2,1}\times d^{\ast}_{2,1}}}
 \end{equation}
 in which $E$ is the
energy place of poles.
 Eq. (\ref{Eilenberger}) have to be supplemented by the
continuity conditions at the interface between superconductors.
For all quasiparticle trajectories, the Green functions satisfy
the boundary conditions both in the right and left bulks as well
as at
the interface.\\
The system of equations (\ref{Eilenberger}) and the
self-consistency equation \cite{Ueda} can be solved only
numerically.
\begin{figure}[tbp]
\includegraphics[width=\columnwidth]{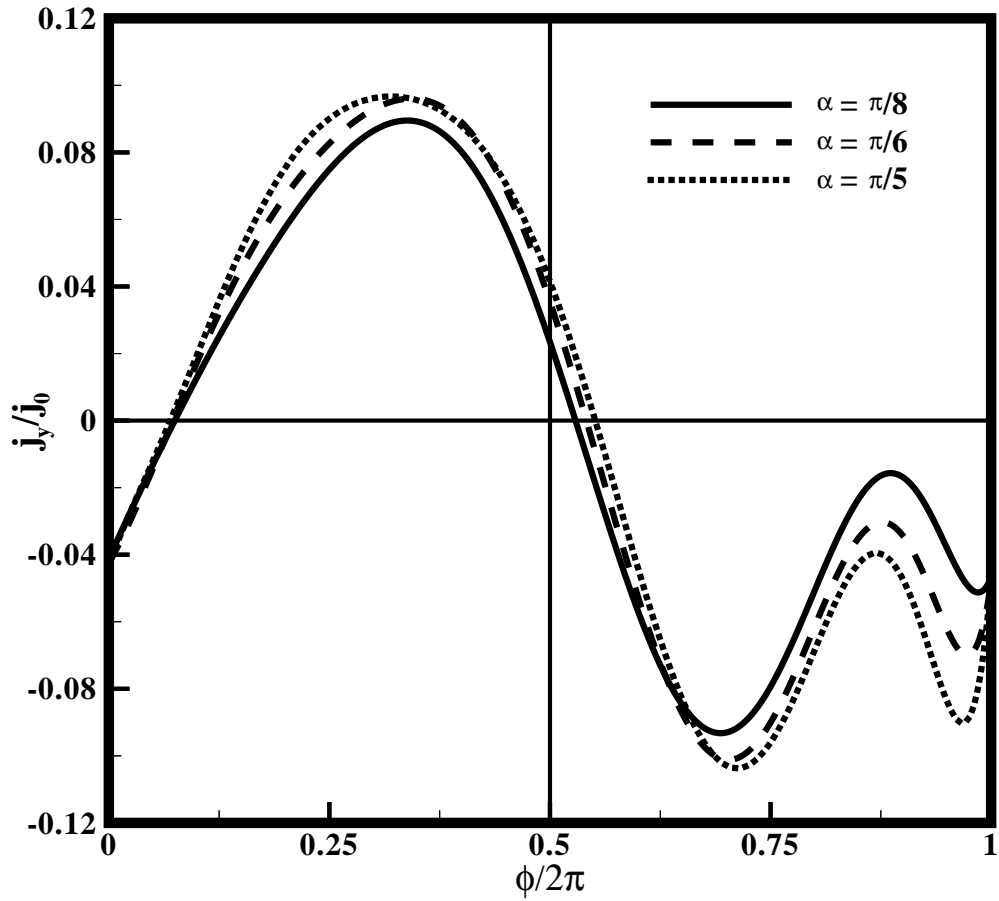}
\caption{Component of current normal to the interface (Josephson
current) versus the phase difference $\protect\phi $ for the
junction between nonunitary bipolar $f-$wave bulks ,
$T/T_{c}=0.15$, geometry (i) and different misorientations.
Currents are given in units of $j_{0}=\frac{\protect\pi
}{2}eN(0)v_{F}\Delta _{0}(0).$} \label{fig2}
\end{figure}
For unconventional superconductors such solution requires the
information of the interaction between the electrons in the Cooper
pairs and nature of unconventional superconductivity in novel
compounds which, in most cases are unknown. Also, It has been
shown that the absolute value of a self-consistent order parameter
is suppressed near the interface and at the distances of the order
of the coherence length, while its dependence on the direction in
the momentum space almost remains unaltered \cite{Barash}. This
suppression of the order parameter which keeps the current-phase
dependence unchanged but changes the amplitude value of the
current, doesn't influence the Josephson effect drastically. For
example, it has been verified in paper \cite{Coury} for the
junction between unconventional $d$-wave, in paper \cite{Barash}
for the case of unitary ``$f$-wave'' superconductors and in paper
\cite{Viljas} for pinholes in $^{3}He$ that, there is a good
qualitative agreement between self-consistent and
non-self-consistent results. Also, it has been observed that
results of the non-self-consistent model in \cite{Yip} are similar
to the experiment \cite{Backhaus}. Consequently, despite the fact
that this estimation cannot be applied directly for a quantitative
analyze of the real experiment, only a qualitative comparison of
calculated and experimental current-phase relations is possible.
In our calculations, a simple model of the constant order
parameter up to the interface is considered and the pair breaking
and the scattering on the interface are ignored. We believe that
under these strong assumptions our results describe the real
situation qualitatively. In the framework of such model, the
analytical expressions for the current can be obtained for a
certain form of the order parameter. \subsection{Analytical and
numerical results} \label{subi}
 The solution of Eq. (\ref{Eilenberger}) allows us to calculate the current densities.
 The expression for current is:
\begin{equation}
\mathbf{j}\left( \mathbf{r}\right) =2i\pi eTN\left( 0\right)
\sum_{m}\left\langle \mathbf{v}_{F}g_{1}\left( \mathbf{\hat{v}}_{F},\mathbf{r%
},\varepsilon _{m}\right) \right\rangle
\end{equation}
where, $\left\langle ...\right\rangle $ stands for averaging over
the directions of an electron momentum on the Fermi surface
${\mathbf{\hat{v}}}_{F}$ and $N\left( 0\right) $ is the electron
density of states at the Fermi level of energy.
\begin{figure}[tbp]
\includegraphics[width=\columnwidth]{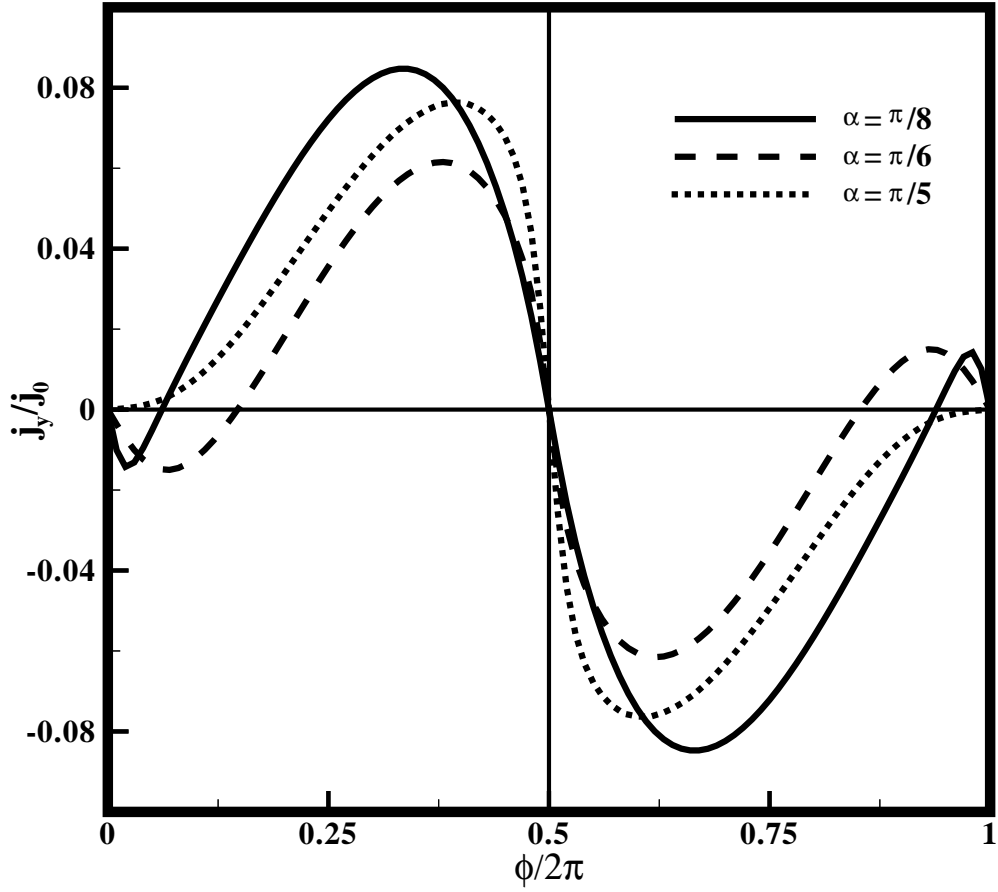}
\caption{Component of current normal to the interface (Josephson
current) versus the phase difference $\protect\phi $ for the
junction between nonunitary bipolar $f-$wave bulks ,
$T/T_{c}=0.15$, geometry (ii) and different misorientations.}
\label{fig3}
\end{figure}
We assume that the order parameter is constant in space and in each half-space it equals to its value (%
\ref{Bulk order parameter}) far form the interface in the left or
right bulks. For such a model, the current-phase dependence of a
Josephson junction can be calculated analytically. It enables us
to analyze the main features of current-phase dependence for a
certain model of the order parameter of nonunitary $f-$wave
superconductivity (bipolar). The Eilenberger equations
(\ref{Eilenberger}) for Green functions $\breve{g}$, which are
supplemented by the condition of continuity of solutions across
the interface, $y=0$, and the boundary conditions at the bulks,
are solved for a non-self-consistent model of the order parameter
analytically. In a ballistic case the system of equations for functions $g_{i}$ and $%
{\mathbf g}_{i}$ can be decomposed on independent blocks of
equations. The set of equations which enables us to find the Green
function $g_{1}$ is:
\begin{eqnarray}
v_{F}\hat{{\mathbf k}}\nabla g_{1}=i\left({\mathbf d}\cdot{\mathbf
g}_{3}-{\mathbf d}^{\ast }\cdot{\mathbf g}_{2}
\right);  \label{anon} \\
v_{F}\hat{{\mathbf k}}\nabla {\mathbf g}_{-}=-2\left( {\mathbf
d\times g}_{3}+{\mathbf d
}^{\ast }{\mathbf \times g}_{2}\right);  \label{bnon} \\
v_{F}\hat{{\mathbf k}}\nabla {\mathbf g}_{2}=-2\varepsilon
_{m}{\mathbf g}
_{2}+2ig_{1}{\mathbf d}+{\mathbf d}\times {\mathbf g}_{-};  \label{dnon}\\
v_{F}\hat{{\mathbf k}}\nabla {\mathbf g}_{3}=2\varepsilon
_{m}{\mathbf g} _{3}-2ig_{1}{\mathbf d}^{\ast }+{\mathbf d}^{\ast
}\times {\mathbf g}_{-}; \label{cnon}
\end{eqnarray}
where ${\mathbf g}_{-}={\mathbf g}_{1}-{\mathbf g}_{4}.$ The Eqs.
(\ref{anon})-(\ref{dnon}) can be solved by integrating over
ballistic trajectories  of electrons in the\ right and left
half-spaces. The general solution satisfying the boundary
conditions (\ref{Bulk solution}) at infinity is
\begin{equation}
g_{1}^{\left( n\right) }=\frac{\varepsilon _{m}}{\Omega
_{n}}+a_{n}e^{-2s\Omega _{n}t};  \label{e}
\end{equation}
\begin{equation}
{\mathbf g}_{-}^{\left( n\right) }=-2\frac{\varepsilon
_{m}}{\Omega _{n}}\mathbf{A_{n}}+{\mathbf
C}_{n}e^{-2s\Omega_{n}t}; \label{f}
\end{equation}
\begin{equation}
\hspace{-.3cm}{\mathbf g}_{2}^{\left( n\right) }=\frac{i{\mathbf
d}_{n}-{\mathbf d}_{n}\times\mathbf{A_{n}}}{\Omega
_{n}}-\frac{2ia_{n}{\mathbf d}_{n}+{\mathbf d}_{n}\times {\mathbf
C}_{n}}{2s\eta \Omega _{n}-2\varepsilon _{m}}e^{-2s\Omega _{n}t};
\label{g}
\end{equation}
\begin{equation}
\hspace{-.3cm}{\mathbf g}_{3}^{\left( n\right) }=\frac{i {\mathbf
d}_{n}^{\ast }+{\mathbf
d}_{n}^{\ast}\times\mathbf{A_{n}}}{\Omega _{n}}+\frac{2ia_{n}{\mathbf d}_{n}^{\ast }-{\mathbf d}%
_{n}^{\ast }\times {\mathbf C}_{n}}{2s\eta \Omega
_{n}+2\varepsilon _{m}}e^{-2s\Omega _{n}t}; \label{h}
\end{equation}
where $t$ is time of flight along the trajectory, $sgn\left(
t\right) =sgn\left( y\right) =s$ and $\eta =sgn\left(
v_{y}\right).$  By matching the solutions (\ref{e}-\ref{h}) at the
interface $\left(y=0, t=0\right) $, we find constants $a_{n}$ and
${\mathbf C_{n}}.$ Indices $n=1,2$ label the left and right
half-spaces respectively. The function $g_{1}\left( 0\right)
=g_{1}^{\left( 1\right) }\left( -0\right) =g_{1}^{\left( 2\right)
}\left( +0\right) ,$ which is a diagonal term of Green matrix and
determines the current density at the interface, $y=0$, is as
follows:
\begin{equation}
g_{1}\left( 0\right)=\frac{\eta(\mathbf{d_2\cdot
d_2}(\eta\Omega_1+\varepsilon)^2-\mathbf{d_1\cdot
d_1}(\eta\Omega_2-\varepsilon)^2+B)}
{[\mathbf{d_2}(\eta\Omega_1+\varepsilon)+\mathbf{d_1}(\eta\Omega_2-\varepsilon)]^{2}}\label{charge-term-nonunitary}
\end{equation}
where $B=i\mathbf{d_1\times
d_2}\cdot(\mathbf{A_1+A_2})(\eta\Omega_2-\varepsilon)(\eta\Omega_1+\varepsilon).$
We consider a rotation $\breve{R}$ only in the right
superconductor (see, Fig.\ref{figb1}), i.e.,
$\mathbf{d}_{2}(\hat{\mathbf{k}})=\breve{R}\mathbf{d}_{1}(\breve{R}^{-1}\hat{%
\mathbf{k}});$ $\hat{\mathbf{k}}$ is the unit vector in the
momentum space.
\begin{figure}[tbp]
\includegraphics[width=\columnwidth]{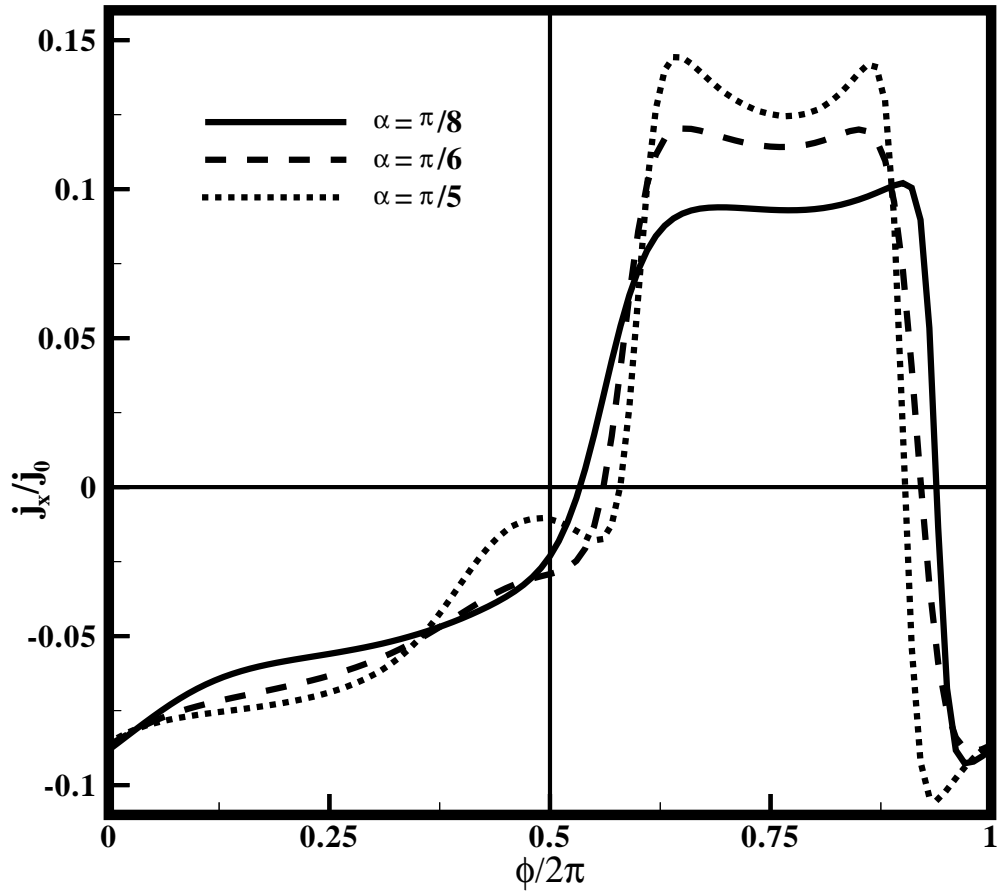}
\caption{The $x-$component of charge current tangential to the
interface versus the phase difference $\protect\phi $ for the
junction between nonunitary bipolar $f-$wave superconducting
bulks, $T/T_{c}=0.15$,geometry (i) and the different
misorientations.} \label{fig4}
\end{figure}
The crystallographic $c$-axis in the left half-space is selected
parallel to the partition between the superconductors (along
$z$-axis in Fig.\ref{figb1}). To illustrate the results obtained
by computing the formula (\ref{charge-term-nonunitary}), we plot
the current-phase diagrams for two different geometries. These
geometries are corresponding to the different orientations of the
crystals in the right and left sides of the interface
(Fig.\ref{figb1}):\newline (i) The basal $ab$-plane in the right
side has been rotated around the $c$-axis by $\alpha $;
$\hat{\mathbf{c}}_{1}\Vert \hat{\mathbf{c}}_{2}$.
\newline
(ii) The $c$-axis in the right side has been rotated around the
$b$-axis by $\alpha $; $\hat{\mathbf{b}}_{1}\Vert
\hat{\mathbf{b}}_{2}$.\newline Further calculations require a
certain model of the gap vector
(order parameter) $%
\mathbf{d}$.\\
In this chapter, the nonunitary $f-$wave gap vector in the
$B-$phase (low temperature $T$ and low field $H$) of
superconductivity in $UPt_{3}$ compound has been considered. This
nonunitary bipolar state which explains the weak spin-orbit
coupling in $UPt_{3}$ is \cite{Machida}: \begin{equation}
\mathbf{d}(T,\mathbf{v}_{F})=\Delta
_{0}(T)k_{z}(\hat{\mathbf{x}}\left( k_{x}^{2}-k_{y}^{2}\right)
+\hat{\mathbf{y}}2ik_{x}k_{y}), \label{Model} \end{equation} The
coordinate axes
$\hat{\mathbf{x}},\hat{\mathbf{y}},\hat{\mathbf{z}}$ are selected
along the crystallographic axes
$\hat{\mathbf{a}},\hat{\mathbf{b}},\hat{\mathbf{c}}$ in the left
side of Fig.\ref{figb1}. The function $\Delta _{0}=$$\Delta
_{0}\left( T\right) $ describes the dependence of the gap vector
on the temperature $T$ (our numerical calculations are done at the
low value of temperature $T/T_{c}=0.1$). Using this model of the
order parameter (\ref{Model}) and solution to the Eilenberger
equations (\ref{charge-term-nonunitary}), we have calculated the
current density at the interface numerically. These numerical
results are listed below \cite{Rashedi5}:\newline 1) The
nonunitary property of Green's matrix diagonal term consists of
two part. The explicit part which is in the $B$ mathematical
expression in Eq. (\ref{charge-term-nonunitary}) and the implicit
part in the $\Omega _{1,2}$ and $\mathbf{d}_{1,2}$ terms. These
$\Omega _{1,2}$ and $\mathbf{d}_{1,2}$ terms are different from
their unitary counterparts. In the mathematical expression for
$\Omega _{1,2}$ the nonunitary mathematical terms
$\mathbf{A}_{1,2}$ are presented. The explicit part will be
present only in the presence of misorientation between gap
vectors,\newline $B=i\mathbf{d}_{1}\times \mathbf{d}_{2}\cdot
(\mathbf{A}_{1}\mathbf{+A}_{2})(\eta \Omega _{2}-\varepsilon
)(\eta \Omega _{1}+\varepsilon )$, but the implicit part will be
present always. So, in the absence of misorientation
$(\mathbf{d}_{1}\mathbf{\Vert d}_{2})$, although the implicit part
of nonunitary exists but the explicit part is absent.
\begin{figure}[tbp]
\includegraphics[width=\columnwidth]{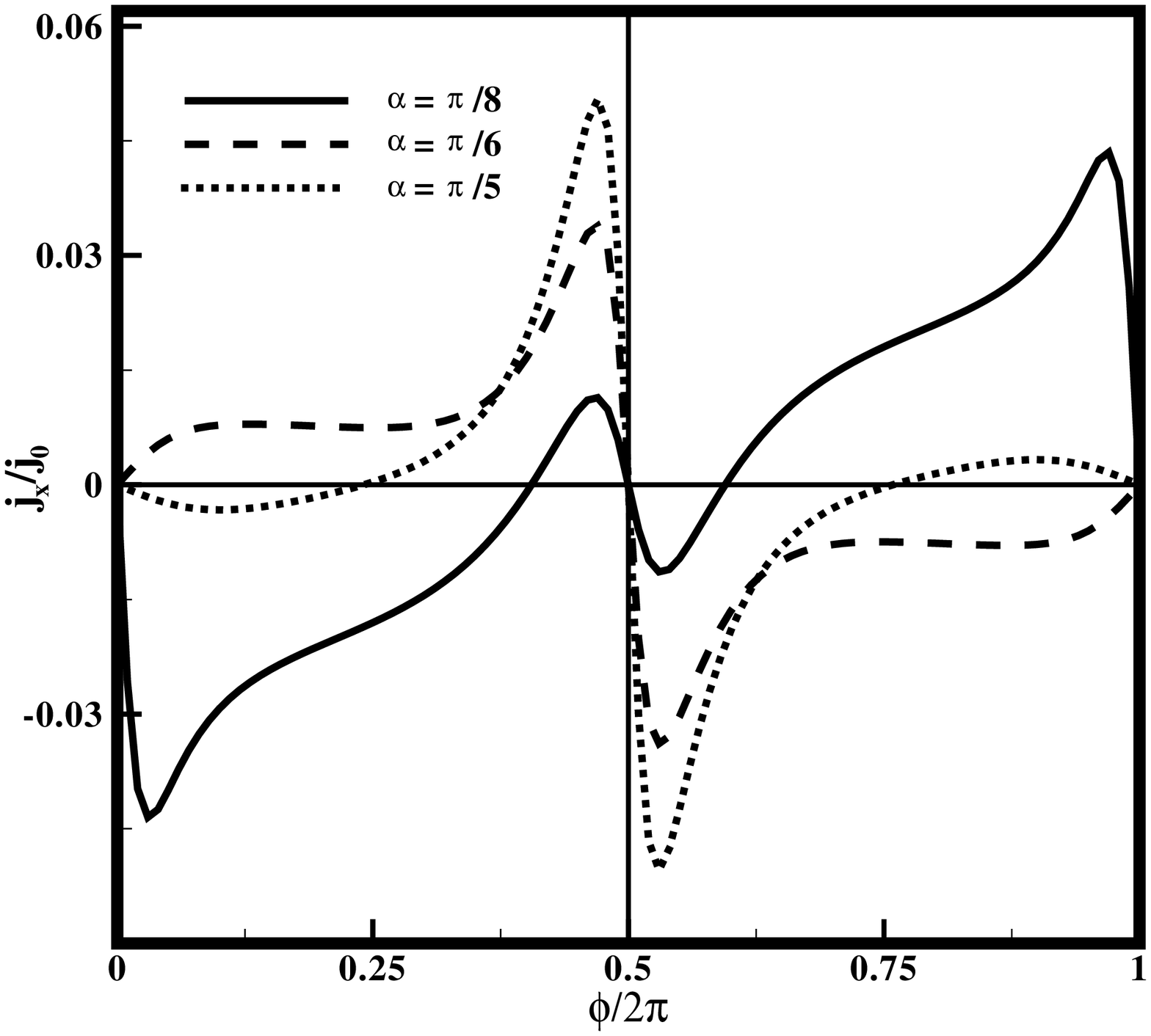}
\caption{Current tangential to the interface versus the phase
difference $\protect\phi $ for the junction between nonunitary
bipolar $f-$wave superconducting bulks, $T/T_{c}=0.15$, geometry
(ii) and the different misorientations ($x$component) .}
\label{fig5}
\end{figure}
This means that, in the absence of misorientation current-phase
diagrams for planar unitary and nonunitary bipolar systems are the
same but the maximum values is different slightly.\newline 2) A
component of current parallel to the interface $j_{z}$ for
geometry (i) is zero as same as the unitary case \cite{Mahmoodi}
while the other parallel component $j_{x}$ has a finite value (see
Fig.\ref{fig4}). This later case is a difference between unitary
and nonunitary cases. Because in the junction between unitary
$f-$wave superconducting bulks all parallel components of the
current ($j_{x}$ and $j_{z}$) for geometry (i) are absent
\cite{Mahmoodi}.\newline 3) In Figs.\ref{fig2},\ref{fig3}, the
Josephson current $j_{y}$ is plotted for certain nonunitary model
of $f-$wave and different geometries. Figs.\ref {fig2},\ref{fig3}
are plotted for the geometry (i) and geometry (ii) respectively.
They are completely unusual and totally different from their
unitary counterparts which have been obtained in
\cite{Mahmoodi}.\newline 4) In Fig.\ref{fig2} for geometry (i), it
is observed that by increasing the misorientation, some small
oscillations, as the result of non-unitary property of the order
parameter, appear in the current-phase diagrams. Also, the
Josphson current at the zero external phase difference $\phi =0$
is not zero but it has a finite value. The Josephson current will
be zero at the some finite values of the phase difference.\newline
5) In Fig.\ref{fig3} for geometry (ii), it is observed that by
increasing the misorientation the new zeros in current-phase
diagrams appear and the maximum value of current will be changed
non-monotonically.
\begin{figure}[tbp]
\includegraphics[width=\columnwidth]{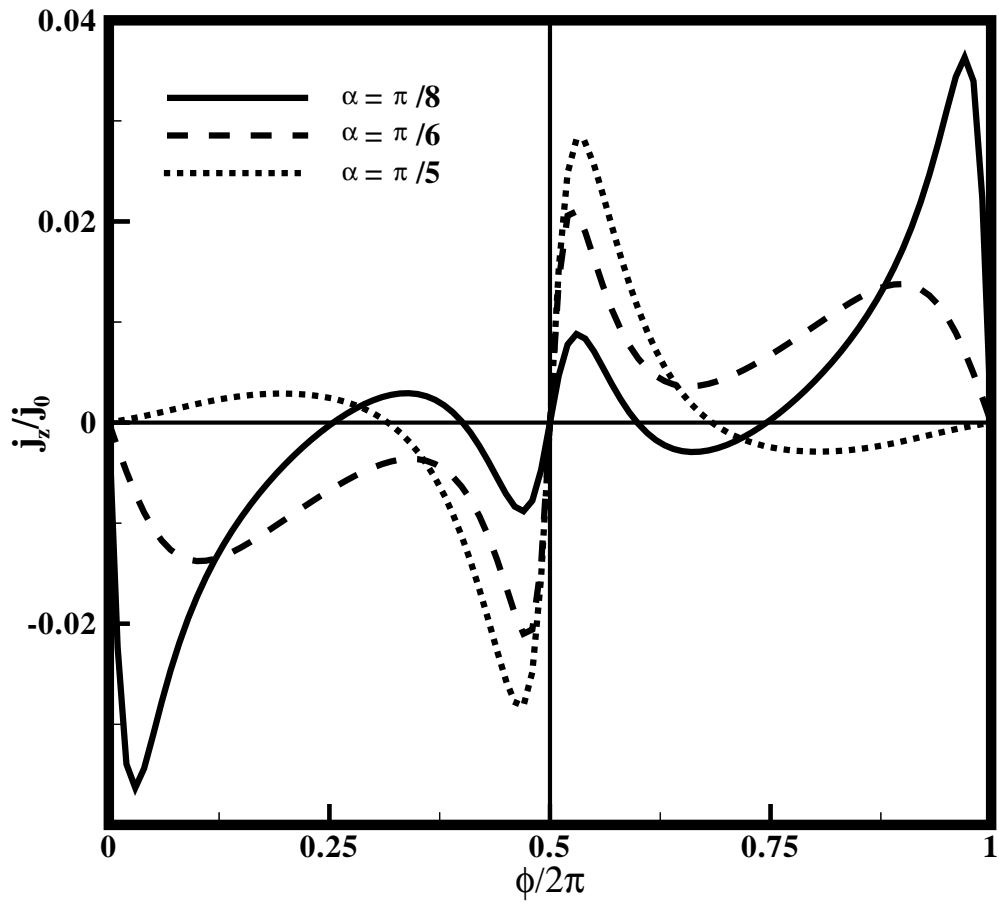}
\caption{Current tangential to the interface versus the phase
difference $\protect\phi $ for the junction between nonunitary
bipolar $f-$wave superconducting bulks, $T/T_{c}=0.15$, geometry
(ii) and the different misorientations ($z-$component).}
\label{fig6}
\end{figure}
In spite of the Fig\ref{fig2} for geometry (i), the Josephson
current at the phase differences $\phi =0$, $\phi =\pi $ and $\phi
=2\pi $ is zero exactly.\newline 6) The current-phase diagram for
geometry (i) and $x-$component (Fig.\ref {fig4}) is totally
unusual. By increasing the misorientation, the maximum value of
current increases. The components of current parallel to the
interface for geometry (ii) are plotted in Fig.\ref{fig5} and
Fig\ref{fig6}. All of terms at the phase differences $\phi =0$,
$\phi =\pi $ and $\phi =2\pi $ are zero. The maximum value of the
current-phase diagrams is not a monotonic function of the
misorientation.
\newpage
\section{Conclusions} \label{Conclusions}
At first, in the Chapter $3$ of this thesis, we have studied the
stationary Josephson effect in the ballistic point contact with
transport current on the banks in the model $S-c-S$ taking into
account the reflection of electrons from the contact. The contact
is subject to the phase difference $\phi $ and the transport
current tangential to the boundary of the contact. As it was shown
in \cite{KOSh}, in the ideal transparent point-contact at $\phi
=\pi $ and near the orifice the tangential current flows in the
opposite direction to the transport current, and there are two
anti-symmetric vortex-like structures. It is observed that, by
decreasing the transparency $D_{c}<D<1$ the vortex-like current is
destroyed. The critical value of $D=D_{c}\left( T\right) $ depends
on the temperature $T$ and $D_{c}\left( T\rightarrow
0\right) \rightarrow 1,$ $D_{c}\left( T\rightarrow T_{c}\right) \rightarrow {%
\frac{1}{2}}$, so that we can never find a vortex for transparency
values lower than $\frac{1}{2}$ . This anomalous temperature
behavior of the vortices is the result of non-monotonic dependence
of the interference current on the temperature.\\
Then, we have studied the spin current in the ballistic Josephson
junction in the model of an ideal transparent interface between
two misorientated $f$-wave superconductors which are subject to a
phase difference $\phi $, in Chapter (\ref{SpinCharge}). Our
analysis has shown that the misorientation and different models of
triplet gap vectors influence the spin current \cite{Rashedi3}.
This has been shown for the charge current in the paper
\cite{Mahmoodi}. Misorientation between the left and right gap
vectors changes strongly the critical values of both of the spin
current and charge current. It has been obtained that the spin
current is the result of the misorientation between the gap
vectors. Furthermore, as an interesting and new result, it is
observed that the different models of the gap vectors and
geometries can be applied for the polarization of the spin
transport. Another result of these calculations is the state in
which the currents select one of the two possible directions
(perpendicular and parallel to the interface) to flow. This
property can be used as a switch to control the direction of the
charge and spin current. Finally, it is observed that at some
certain values of the phase difference $\phi $, the charge-current
vanishes while the spin-current flows, although the carriers of
both spin and charge are the same (electrons). The spatial
variation of the phase of the order parameter plays a role as the
origin of the charge current and, similarly, due to a spatial
difference of the gap vectors in two half-spaces causes spin
currents. This is because there is a position-dependent phase
difference between ``spin up'' and ``spin down'' Cooper pairs and,
although the total charge current vanishes, there can be a net
transfer of the spin. Therefore, in our system, there is a
discontinuous jump between the gap vectors and, consequently the
spin currents should generally be present. For instance, if
spin-up states and spin-down states have a velocity in the
opposite directions, the charge currents cancel each other whereas
the spin current is being transported. Mathematically speaking,
$\mathbf{{j_{charge}}={j_{\uparrow }}+{j_{\downarrow }},
{j_{spin}}={j_{\uparrow }}-{j_{\downarrow }}}$, so it is possible
to find the state in which one of these current terms is zero and
the other term has a finite value \cite{Maekawa}. In addition, the
spin imbalance which is the result of the different density of
states for ``spin-up'' and ``spin-down'' can be other reason of
spin current \cite{Sun}. In conclusion, the spin current in the
absence
of the charge current can be observed.\\
Also, the weak link between two misorientated $PrOs_4Sb_{12}$
superconducting bulks has been investigated in Chapter
(\ref{PrOsSb}) of this thesis. We have considered the
transportation of the spin and charge in the ballistic Josephson
junction between two $PrOs_4Sb_{12}$ crystals with ``$(p+h)-$wave"
pairing symmetry which are subject to a phase difference $\phi $.
In this case as same as the case of $f-$wave junctions in Chapter
(\ref{SpinCharge}), the different misorientations and different
models of the gap vectors influence the spin and charge currents
and it is shown that the misorientation of the superconductors
leads to a spontaneous phase difference that corresponds to the
zero Josephson current. This phase difference depends on the
misorientation angle. A spontaneous charge current tangential to
the interface which is not equal to zero in the absence of the
Josephson current is observed in this junction. It has been
obtained that the spin current is the result of the misorientation
between the gap vectors which is the characteristic of unitary
triplet superconductors. Again in this case as same as the case of
$f-$wave superconductors, it is observed that the certain model of
the gap vectors and geometries can be applied to polarize of the
spin transport. Finally, it is observed that at some certain
values of the phase difference $\phi $, the charge-current
vanishes while the spin-current flows. The reason for this case is
the above-mentioned discussion for $f-$wave. The spin and charge
currents can be used to recognize the $A-$phase and $B-$phase of
``$(p+h)-$wave" and the pure $p-$wave pairing symmetry and can be
used to determine the pairing symmetry. Particularly, this
proposed experiment can be used to demonstrate the ``$(p+h)-$wave"
pairing symmetry for which many
doubts and challenges exist.\\
Finally, in Chapter (\ref{Conclusions}), we have theoretically
studied the charge currents in the ballistic Josephson junction in
the model of an ideal transparent interface between two
misoriented $UPt_{3}$ crystals with nonunitary bipolar $f-$wave
superconducting bulks which are subject to a phase difference
$\phi $. Our analysis has shown that misorientation between the
gap vectors create a current parallel to the interface and the
different misorientations between gap vectors influence the
spontaneous parallel and normal Josephson currents. These have
been shown for the currents in the point contact between two bulks
of unitary axial and planar $f$-wave superconductor in
\cite{Mahmoodi} separately. Also, It is shown that the
misorientation of the superconductors leads to a spontaneous phase
difference that corresponds to the zero Josephson current and to
the minimum of the weak link energy in the presence of the finite
spontaneous current. This phase difference depends on the
misorientation angle. Again in this case, the tangential
spontaneous charge current is not equal to zero in the absence of
the Josephson current. The difference between junction behavior of
unitary planar and nonunitary bipolar superconductivity can be
used to distinct between them. This experiment can be used to test
the pairing symmetry and recognize the different phases of
$UPt_{3}$ which has two unitary and a nonunitary triplet pairing
symmetry.
\end{document}